\documentclass[%
preprint,
preprintnumbers,
superscriptaddress,
nofootinbib,
 amsmath,amssymb,
 aps,
]{revtex4-2}

\usepackage[pdftex]{graphicx}
\usepackage{epsfig} 
\usepackage{dcolumn}
\usepackage{bm}
\usepackage[pdfencoding=unicode,psdextra]{hyperref}
\usepackage{natbib}
\usepackage{color}
\usepackage[normalem]{ulem}
\usepackage{upgreek}
\usepackage{overpic}
\usepackage{tabularx}
\usepackage{cleveref}
\usepackage[mathlines]{lineno}
\usepackage[utf8]{inputenc}
\usepackage{orcidlink}
\usepackage{float}
\crefname{figure}{Fig.}{Figs.}
\crefname{equation}{Eq.}{Eqs.}
\crefname{section}{Sec.}{Secs.}
\crefname{appendix}{Appendix}{Appendices}

\begin{document}

\title{Ambiguities in the Partial-Wave Analysis of the Photoproduction of Pairs of Pseudoscalar Mesons}
\author{J.~Guo\,\orcidlink{0000-0003-2936-0088}}
\affiliation{Carnegie Mellon University, Pittsburgh, Pennsylvania 15213, USA}
\author{E.~Barriga\orcidlink{0000-0003-3415-617X}} \affiliation{Florida State University, Tallahassee, Florida 32306, USA}
\author{K.~Scheuer\orcidlink{0009-0000-4604-9617}} \affiliation{William \& Mary, Williamsburg, Virginia 23185, USA}
\author{A.~Austregesilo\orcidlink{0000-0002-9291-4429}} \affiliation{Thomas Jefferson National Accelerator Facility, Newport News, Virginia 23606, USA}
\author{P.~Eugenio\orcidlink{0000-0002-0588-0129}} \affiliation{Florida State University, Tallahassee, Florida 32306, USA}
\author{D.~I.~Glazier\orcidlink{0000-0002-8929-6332}}\affiliation{University of Glasgow, Glasgow, G12 8QQ, UK}
\author{B.~Grube\orcidlink{0000-0001-8473-0454}} 
\affiliation{Thomas Jefferson National Accelerator Facility, Newport News, Virginia 23606, USA}
\author{W.~Imoehl\orcidlink{0000-0002-1554-1016}} \affiliation{Carnegie Mellon University, Pittsburgh, Pennsylvania 15213, USA}
\author{C.~A.~Meyer\orcidlink{0000-0001-7599-3973}}
\affiliation{Carnegie Mellon University, Pittsburgh, Pennsylvania 15213, USA}
\author{A.~Ostrovidov\orcidlink{0000-0001-6415-6061}} \affiliation{Florida State University, Tallahassee, Florida 32306, USA}
\author{J.~R.~Stevens\orcidlink{0000-0002-0816-200X} } \affiliation{William \& Mary, Williamsburg, Virginia 23185, USA}
\date{ \today }

\begin{abstract}
Applying the technique of partial-wave analysis, there are cases where more than one set of underlying complex-valued amplitudes can describe the measured observables. These ambiguities can sometimes be resolved using additional information, but assumptions are often required. It is known that the partial-wave analysis of two-pseudoscalar meson systems produced in photoproduction with a linearly polarized photon beam is free from discrete ambiguities stemming from the Barrelet zeros when the nucleon spin is ignored. In this article, we show that continuous ambiguities are possible for certain wave sets, even though the discrete ambiguities do not appear. We also explore ways to resolve these ambiguities and determine the maximal amount of information that can be obtained from analyses that suffer from these continuous ambiguities.  

\end{abstract}

\maketitle

\section{Introduction}

According to the quark model proposed by Gell-Mann~\cite{GELLMANN1964214} and Zweig~\cite{Zweig:1964ruk}, conventional mesons are bound states of quark-antiquark pairs. However, gluonic degrees of freedom are allowed within Quantum Chromodynamics (QCD), the fundamental theory of strong interaction. A possible realization is hybrid mesons~\cite{Meyer:2015eta}, which are composed of a quark-antiquark pair and excited gluonic field configurations that contribute to quantum numbers. The experimental observation of such states would be a direct confirmation of QCD in the low-energy region. Therefore, investigating the spectrum of exotic mesons,\footnote{Exotic mesons are any states beyond quark-antiquark configurations.} along with their production and decay mechanisms, is crucial to understand the non-perturbative behavior of QCD.

In the search for exotic mesons, two-pseudoscalar meson final states have been an important area of study and the analyses of these simple final states remain a key part in particular for the program to search for hybrid mesons. The GlueX experiment~\cite{GlueX:2020idb} at Jefferson Lab has been designed to search for hybrid mesons in photoproduction reactions using a beam of linearly polarized photons with an energy of about 9 GeV. Recent work includes the $\pi^{+}\pi^{-}$ final state in the region of the $\rho(770)$~\cite{GlueX:2023fcq}, while other analyses are looking at the $\eta\pi$ ~\cite{GlueX:2025kma} and $\eta^{\prime}\pi$~\cite{Albrecht:2024qdh} final states.

It is well known that in the partial-wave analysis of two-pseudoscalar final states~\cite{Gersten:1969, Sadovsky:1991hm}, multiple sets of complex-valued amplitudes may describe the observables equally well. These mathematical ambiguities can arise due to under-constrained equation systems that determine the partial-wave amplitudes or due to periodicity of the involved functions. An extensive study of ambiguities of two-pseudoscalar final states produced by spinless beam particles has been carried out by Chung~\cite{Chung:1997qd} in terms of Barrelet zeros. A recent study of discrete ambiguities in linearly polarized photoproduction has been carried out by the JPAC collaboration~\cite{JointPhysicsAnalysisCenter:2023gku}.

In partial-wave analysis, a standard approach involves initiating the fitting process with a basic wave set, incorporating dominant partial-wave amplitudes based on existing knowledge. This initial set is then refined by progressively adding more amplitudes to assess its validity.  In this paper, we extend the ambiguity study to both reflectivities for linearly polarized photoproduction of pseudoscalar meson pairs and demonstrate that continuous mathematical ambiguities appear when a particular type of simple wave set, commonly employed as a starting point in partial-wave analysis, is used with both reflectivities. In \cref{sec:formalism}, we present the theoretical formalism that describes this process. Subsequently, in \cref{sec:analysis}, we develop a matrix representation of the moment formalism and establish the criteria to identify mathematical ambiguities. In \cref{sec:case}, we analyze three different wave sets using numerical simulations, demonstrating that continuous ambiguities can appear when both reflectivities are populated. In \cref{sec:discussion}, we discuss the maximal amount of information that can be retrieval in the case of continuous ambiguities, explore potential external constraints to resolve these ambiguities, and discuss challenges related to model selection and experimental effects. Finally, the summary and conclusions are given in \cref{sec:conclusion}

\section{Formalism}
\label{sec:formalism}
\begin{figure}[ht!]\centering
\includegraphics[width=0.35\textwidth]{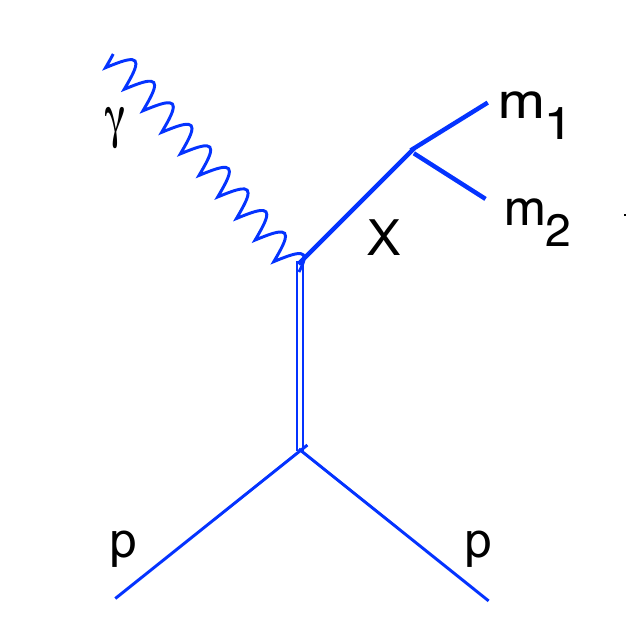}
\caption[]{\label{fig:t-chan}Photoproduction of a meson resonance $X$, which decays into a two-meson final state, off a proton target. At high energy, the production occurs through $t$-channel exchange.}
\end{figure}

We consider photoproduction reactions as depicted in \cref{fig:t-chan}, where some intermediate meson resonance $X$ is produced off a proton target through a $t$-channel process, i.e. 
\begin{eqnarray}
    \vec{\gamma}p & \rightarrow & Xp .
\end{eqnarray}
The intermediate state $X$ subsequently decays into two pseudoscalar mesons, i.e.
\begin{eqnarray}
    X & \rightarrow & m_{1}\,m_{2} \, .
\end{eqnarray}
Due to the combined requirements of angular momentum and parity conservation in the decay, the intermediate states $X$ are restricted to the natural-parity series with $J^{P}=0^{+}, 1^{-}, 2^{+}, ...$ 

For completeness, here we summarise the formalism derived in Ref.~\cite{Mathieu:2019fts}. Following the convention, the angle $\Phi$ describes the direction of the photon's linear polarization relative to the production plane of the $Xp$ system. The decay of $X$ into two pseudoscalar mesons is described by the angles $\theta$ and $\phi$ of the particle $m_1$ in the helicity frame of $X$. The decay angular distribution is typically expressed in terms of spherical harmonics $Y_{\ell}^{m}(\Omega)$, where $\Omega=(\theta, \phi)$. Given the two-pseudoscalar invariant mass $\omega$ and the squared four-momentum transfer $t$ from the photon to the target proton, the intensity distribution, i.e. the density function of the number of events $N$, can be expanded in terms of partial-wave amplitudes:
\begin{equation}
\begin{aligned}
    \label{eq:intensity}
    \mathcal{I}(\Omega, \Phi;\omega,t) &= \frac{dN}{d\omega \ dt \ d\Omega \ d\Phi} \\
    &= \sum_{\substack{\ell m \\ \ell^{\prime} m^{\prime}}}^{\infty} \sum_{\substack{\lambda,\lambda'=\pm1 \\ \lambda_1,\lambda_2=\pm1/2}} Y_{\ell}^m(\Omega) \mathcal{T}_{m\lambda;\lambda_1 \lambda_2}^{\ell}(\omega,t) \ \rho_{\lambda\lambda'}^{\gamma}(\Phi) \ \mathcal{T}_{m' \lambda';\lambda_1 \lambda_2}^{\ell' *}(\omega,t) \ Y_{\ell'}^{m' *}(\Omega) \, .
\end{aligned}
\end{equation}
Here, the $\mathcal{T}_{m\lambda;\lambda_1 \lambda_2}^{\ell}(\omega,t)$ are the partial-wave amplitudes for an intermediate state with a spin equal to the orbital angular momentum $\ell$ between the two pseudoscalar mesons and a spin projection $m$. The target proton and the recoil proton have helicities $\lambda_1, \lambda_2 =\pm 1/2$, respectively, while $\lambda, \lambda' = \pm 1$ represent the helicity of the beam photon. The photon polarization is described by the spin-density matrix $\rho_{\lambda,\lambda^{\prime}}^{\gamma}(\Phi)$. 
For linearly polarized photons, we can decompose $\rho_{\lambda,\lambda^{\prime}}^{\gamma}(\Phi)$ and hence $\mathcal{I}$ into three terms as in Ref.~\cite{Mathieu:2019fts}:
\begin{equation}
    \mathcal{I}(\Omega, \Phi;\omega,t) = \mathcal{I}_0(\Omega;\omega,t) - \mathcal{I}_1(\Omega;\omega,t)P_{\gamma} \cos{(2\Phi)} - \mathcal{I}_2(\Omega;\omega,t)P_{\gamma} \sin{(2\Phi)} ,
    \label{eq:intensity_formula}
\end{equation}
where $P_{\gamma}$ is the degree of linear polarization of the photon. The intensity components are
\begin{equation}
    \mathcal{I}_i(\Omega;\omega,t) = \sum_{\substack{\ell m \\ \ell^{\prime} m^{\prime}}}^{\infty} Y_{\ell}^m(\Omega) \  ^{i}\rho_{m,m'}^{\ell,\ell'}(\omega, t) \ Y_{\ell'}^{m' *}(\Omega); \ \ \ \ i=0,1,2
\label{eq:intensity_sdme}
\end{equation}
and can be expressed in terms of the components
\begin{align}
    ^0\rho_{m,m'}^{\ell,\ell'}(\omega,t) =& \frac{1}{2} \sum_{\substack{\lambda=\pm1 \\ \lambda_1,\lambda_2=\pm1/2}}  \mathcal{T}_{m\, \lambda;\lambda_1\, \lambda_2}^{\ell}(\omega,t) \ \mathcal{T}_{m'\,  \lambda;\lambda_1\, \lambda_2}^{\ell' *}(\omega,t) \label{eq:sdme0}\\
    ^1\rho_{m,m'}^{\ell,\ell'}(\omega,t) =& \frac{1}{2} \sum_{\substack{\lambda=\pm1 \\ \lambda_1,\lambda_2=\pm1/2}}  \mathcal{T}_{m\, -\lambda;\lambda_1\, \lambda_2}^{\ell}(\omega,t) \ \mathcal{T}_{m'\, \lambda;\lambda_1\, \lambda_2}^{\ell' *}(\omega,t) \label{eq:sdme1}\\
    ^2\rho_{m,m'}^{\ell,\ell'}(\omega,t) =& \frac{i}{2} \sum_{\substack{\lambda=\pm1 \\ \lambda_1,\lambda_2=\pm1/2}} \lambda \ \mathcal{T}_{m\, -\lambda;\lambda_1\, \lambda_2}^{\ell}(\omega,t) \ \mathcal{T}_{m'\, \lambda;\lambda_1\, \lambda_2}^{\ell' *}(\omega,t)
    \label{eq:sdme2} \, 
\end{align}
of the spin-density matrix of $X$.

The intensity components $\mathcal{I}_i$ in \cref{eq:intensity_formula} are orthogonal to each other in the $\Phi$ space, and we can expand them individually into spherical harmonics:
\begin{align}
    \mathcal{I}_0(\Omega) &= \sum_{LM}^{\infty}\sqrt{\frac{2L+1}{4\pi}}H_0(L,M)Y_L^M(\Omega) \\
    \mathcal{I}_{1,2}(\Omega) &= -\sum_{LM}^{\infty}\sqrt{\frac{2L+1}{4\pi}}H_{1,2}(L,M)Y_L^M(\Omega) .
\end{align}
Here, $H_i(L,M)$ are the moments, where $H_0$ describes unpolarized photoproduction and $H_{1,2}$ describe linearly polarized photoproduction. For a given intensity distribution, the moment decomposition is unique. The contraction property of the spherical harmonics relates the moments to the partial-wave amplitudes via the spin-density matrix elements:
\begin{equation}
\begin{aligned}
    H_0(L,M) =& \sum_{\substack{\ell m \\ \ell' m'}}^{\infty}  C_{\ell m;\ell'm'}^{L M} \ {^0}\rho_{m,m'}^{\ell,\ell'}\\
    H_{1,2}(L,M) =& -\sum_{\substack{\ell m \\ \ell' m'}}^{\infty} C_{\ell m;\ell'm'}^{L M} \ {^{1,2}}\rho_{m,m'}^{\ell,\ell'} \ ,
    \label{eq:moments_sdmes}
\end{aligned}
\end{equation}
where
\begin{equation}    
 C_{\ell m;\ell'm'}^{L M} \equiv \sqrt{\frac{2\ell'+1} {2\ell+1}} \langle \ell' \ 0, L \ 0 \ | \ \ell \ 0\rangle \langle \ell' \ m', L \ M \ | \ \ell \ m\rangle .
\label{eq:coef_c}
\end{equation}
Here, $\langle j_1 \ m_1, j_2 \ m_2 \ | \ J \ M\rangle$ are the Clebsch-Gordan coefficients. From \cref{eq:moments_sdmes} it is clear that $H_i(L,M)$ vanishes unless $|\ell'-L| \leq \ell \leq \ell' +L $ and $m=m'+M$.\footnote{Note that in addition $\langle \ell' \ 0, L \ 0 \ | \ell \ 0 \rangle$ requires $\ell' + L +\ell = \text{even}$.} Using parity conservation and the symmetry properties of the spherical harmonics and the Clebsch-Gordan coefficients, one can show that $H_{0,1}(L,M)$ are real-valued whereas $H_2(L,M)$ is purely imaginary and that moments with opposite $M$ values are related by
\begin{equation}
    \begin{aligned}
        H_{0,1}(L,-M) =& (-1)^{M}\, H_{0,1}(L,M) \\
        H_{2}(L,-M) =& -(-1)^{M}\, H_{2}(L,M) .
\end{aligned}   
\end{equation}
We hence need to measure only the moments with $M \geq 0$.
%

The reflectivity operator is defined as a $180^{\circ}$ rotation about the normal to the production plane followed by a parity transformation~\cite{PhysRevD.11.633}. Because of parity conservation, the partial-wave amplitudes in \cref{eq:sdme0,eq:sdme1,eq:sdme2} can be expressed as eigenstates of the reflectivity operator, where amplitudes with positive reflectivity $\epsilon = +1$ do not change sign under this transformation and those with $\epsilon = -1$ do. Following Ref.~\cite{Mathieu:2019fts}, the partial-wave amplitudes of the reflectivity eigenstates can be written in terms of the amplitudes for the two photon helicity states with $\lambda=\pm 1$, i.e.
\begin{eqnarray}
    ^{(\epsilon)}\mathcal{T}_{m;\lambda_1 \lambda_2}^{\ell} & = & \frac{1}{2}\biggl[ \mathcal{T}_{m+1;\lambda_1\lambda_2}^{\ell} - \epsilon (-1)^{m} \mathcal{T}^{\ell}_{-m-1;\lambda_1\lambda_2} \biggr].
\label{eq:refl_def}
\end{eqnarray}
Due to parity conservation, partial-wave amplitudes with opposite reflectivities do not interfere, leading to an incoherent sum over $\epsilon$ in the intensity.

In the high-energy limit, the reflectivity $\epsilon$ corresponds to the naturality of the exchange particle. Due to parity, the amplitudes with opposite helicities of target and recoil proton are related by~\cite{Mathieu:2019fts}
\begin{equation}
    ^{(\epsilon)}\mathcal{T}_{m;-\lambda_1 -\lambda_2}^{\ell} = \epsilon (-1)^{\lambda_1-\lambda_2} \ ^{(\epsilon)}\mathcal{T}_{m;\lambda_1 \lambda_2}^{\ell} .
\label{eq:parity_relation}
\end{equation}
Hence, we have two incoherent amplitudes for a given $\epsilon, \ell$, and $m$: a proton spin-non-flip amplitude\footnote{We use short-hand notation $\pm$ for $\lambda_{1,2} = \pm 1/2$ here.}
\begin{equation}
    ^{(\epsilon)}\mathcal{T}_{m; + +}^{\ell} \equiv [\ell]_{m;k=0}^{(\epsilon)}
\end{equation}
and a proton spin-flip amplitude
\begin{equation}
    ^{(\epsilon)}\mathcal{T}_{m; + -}^{\ell} \equiv [\ell]_{m;k=1}^{(\epsilon)} .
\end{equation}
When performing the partial-wave analysis, we assume that one of these two amplitudes is dominant and ignore the index $k$ as in Ref.~\cite{Chung:1997qd}.

Finally, we can relate the moments $H_i(L,M)$ to the partial-wave amplitudes in the reflectivity basis:
\begin{align}
    H_0(L,M) = \sum_{\substack{\ell m \\ \ell' m'}}^{\infty} C_{\ell m;\ell'm'}^{L M} & \ \sum_{\epsilon=\pm}\bigg( [\ell]_m^{(\epsilon)}[\ell']_{m'}^{(\epsilon)*} + (-1)^{m-m'} [\ell]_{-m}^{(\epsilon)}[\ell']_{-m'}^{(\epsilon)*} \bigg) \label{eq:moments0}\\
    H_1(L,M) = \sum_{\substack{\ell m \\ \ell' m'}}^{\infty} C_{\ell m;\ell'm'}^{L M} & \ \sum_{\epsilon=\pm} \epsilon \ \bigg( (-1)^m [\ell]_{-m}^{(\epsilon)}[\ell']_{m'}^{(\epsilon)*} + (-1)^{m'} [\ell]_{m}^{(\epsilon)}[\ell']_{-m'}^{(\epsilon)*} \bigg) \label{eq:moments1}\\
    H_2(L,M) = i\sum_{\substack{\ell m \\ \ell' m'}}^{\infty} C_{\ell m;\ell'm'}^{L M} & \ \sum_{\epsilon=\pm} \epsilon \ \bigg( (-1)^m [\ell]_{-m}^{(\epsilon)}[\ell']_{m'}^{(\epsilon)*} - (-1)^{m'} [\ell]_{m}^{(\epsilon)}[\ell']_{-m'}^{(\epsilon)*} \bigg),
    \label{eq:moments2}
\end{align}
where $C_{\ell m;\ell'm'}^{L M}$ is defined in \cref{eq:coef_c}. The moments hence do not separate the contributions from different reflectivities.

\section{Analysis}
\label{sec:analysis}
The presence of mathematical ambiguities in the partial-wave amplitudes is determined by whether, for a given set of moments $\{H_i(L,M)\}$, the system of equations in \cref{eq:moments0,eq:moments1,eq:moments2} has a unique solution for the set of amplitudes $\{[\ell]_m^{(\epsilon)}\}$. In the case of spinless beam particles, it is known that discrete ambiguities arise due to the intrinsic property of polynomial equations to have more than one root~\cite{Chung:1997qd}. However, the JPAC collaboration has shown in Ref.~\cite{JointPhysicsAnalysisCenter:2023gku} that for photoproduction of two-pseudoscalar final states with a linearly polarized photon beam the ambiguities from Barrelet zeros~\cite{Barrelet1972} do not exist when the spin of the target nucleon is ignored. When both reflectivities are populated, we find that another type of ambiguity may arise that leads to a continuous set of solutions. Continuous ambiguities appear when the equation system in \cref{eq:moments0,eq:moments1,eq:moments2} has fewer linearly independent equations than the number of unknown parameters in the set of partial-wave amplitudes. In other words, the system of polynomial equations is underdetermined leading to an infinite number of possible solutions. 

As shown in \cref{appen:pair_amplitude}, the moments in \cref{eq:moments0,eq:moments1,eq:moments2} can be written as linear combinations of real-valued terms of the form
\begin{equation}
    \begin{aligned}
    \relax [\ell]_m^{(\epsilon)}[\ell']_{m'}^{(\epsilon)*} + [\ell]_m^{(\epsilon) *}[\ell']_{m'}^{(\epsilon)} &= \begin{cases}
        2\left( r_{\ell m}^{(\epsilon)} \right)^2, &\text{if} \ \ell=\ell',m=m' \\
        2 \, r_{\ell m}^{(\epsilon)} \, r_{\ell' m'}^{(\epsilon)} \, \cos(\eta_{\ell m}^{(\epsilon)}-\eta_{\ell' m'}^{(\epsilon)}), &\text{if} \ \ell \neq \ell' \text{or} \ m\neq m' .
    \end{cases}
    \end{aligned}
    \label{eq:PairOfAmplitudes}
\end{equation}
Here, $r_{\ell, m}^{(\epsilon)}$ and $\eta_{\ell, m}^{(\epsilon)}$ are magnitude and phase, respectively, of the complex amplitude $[\ell]_m^{(\epsilon)}$. Using \cref{eq:PairOfAmplitudes}, we can cast \cref{eq:moments0,eq:moments1,eq:moments2} into matrix form
\begin{equation}
\mathbf{H} = \mathbf{A f}.
\label{eq:matrix_h=af}
\end{equation}
Here,
\begin{equation}
\mathbf{f} = \left[..., \operatorname{Re}([\ell]_m^{(\epsilon)}[\ell']_{m'}^{(\epsilon)*}),...\right]^T
\label{eq:f-term}
\end{equation}
is the vector of all $\sum_\epsilon N^{(\epsilon)}_\text{wave} (N^{(\epsilon)}_\text{wave} + 1) / 2$ different bilinear terms of partial-wave amplitudes for a given set of $N_\text{waves} = N^{(+)}_\text{waves} + N^{(-)}_\text{waves}$ partial waves and 
\begin{equation}
\mathbf{H} = [\ldots, H_0(L, M), \ldots,  H_1(L, M), \ldots, H_2(L, M), \ldots]^T
\end{equation}
is the vector of all unique, non-zero moments. The matrix $\mathbf{A}$ contains the coefficients that relate $\mathbf{f}$ to $\mathbf{H}$.

The set of $N_{\text{wave}}$ partial-wave amplitudes is in principle defined by $2N_{\text{wave}}$ real parameters consisting of the magnitudes $r_{l,m}^{(\epsilon)}$ and the phases $\eta_{l,m}^{(\epsilon)}$. However, we can fix one phase parameter to $0$ for each reflectivity because only the phase differences between each pair of amplitudes matter, not their absolute phases. Therefore, the number of free real parameters is 
\begin{equation}
    N_{\text{par}}=
    \begin{cases}
        2N_{\text{wave}}-1, &\text{if only one reflectivity is included} \\
        2N_{\text{wave}}-2, &\text{if both reflectivities are included} .
    \end{cases}
\end{equation}

As a simple example, consider a wave set that contains only $S$-wave amplitudes. In this case, $\mathbf{H}$ contains $2$ elements, i.e.
\begin{equation}
    \mathbf{H} = \left[ H_{0}(0,0) , H_{1}(0,0) \right]^{T} \, ,
\end{equation}
because $H_{2}(L,0)$ is zero and \cref{eq:matrix_h=af} becomes
\begin{equation}
    \begin{pmatrix}
        H_{0}(0,0) \\
        H_{1}(0,0)
    \end{pmatrix}
    =
    \left(
    \begin{array}{cr} 
        2 & 2 \\
        2 & -2
    \end{array}
    \right)
    \begin{pmatrix}
        \operatorname{Re}[S_0^+ (S_0^{+})^*)] \\
        \operatorname{Re}[S_0^- (S_0^{-})^*)] \\
    \end{pmatrix}.
\end{equation}
Here and in the rest of the text, we use the spectroscopic notation where $[\ell] = S, P, D, \ldots$ for $\ell =0,1,2,\ldots$. 

The rank of the coefficient matrix $\mathbf{A}$ is equal to the number of linearly independent equations that relate $\mathbf{H}$ and $\mathbf{f}$. When all the elements of $\mathbf{f}$ are non-zero, i.e. when all the partial-wave amplitudes are non-zero, $\text{rank}(\mathbf{A})$ is the number of equations that we have to solve for the amplitudes. In practice, we usually remove amplitudes that have a sufficiently small contribution to guarantee all the amplitudes included are non-zero. If $\text{rank}(\mathbf{A})<N_{\text{par}}$, the equation system is underdetermined leading to continuous ambiguities. If rank$(\mathbf{A}) \geq N_{\text{par}}$, we have a sufficient number of equations to solve the system with two possible outcomes: a unique solution or no solution, since there are no discrete ambiguities in the form of Barrelet zeros, if we ignore the trivial ambiguous solution, where the sign of all phases is flipped simultaneously. In the case of no solution, this means that the wave set cannot describe the data well. In practice, it usually corresponds to bad fit results, which can be a hint to optimize the wave set.


\section{Case Studies}
\label{sec:case}
Mathematical ambiguities in the partial-wave analysis of two-pseudoscalar meson systems produced with a linearly polarized photon beam were already studied by the JPAC collaboration in Ref.~\cite{JointPhysicsAnalysisCenter:2023gku}. 
In this section, we will show that continuous ambiguities can appear when both reflectivities are populated. 

In examining several wave sets, we will limit our studies to amplitudes with $\ell \leq 2$ since the analyses of GlueX data seem to indicate that contributions from partial waves with $\ell > 2$ are small. For each wave set, we perform input-output studies based on Monte Carlo simulations to show whether mathematical ambiguities exist or not. The pseudo-data sets are generated by weighting phase-space distributed events with the intensity distribution given by
\begin{equation}
\begin{aligned}
    I\left(\Phi, \Omega; \omega,t \right)  =  2 \kappa  \Biggl\{ \Biggr. & \left(1-P_\gamma\right)\biggl[\biggl|\sum_{\ell,m} [\ell]_{m}^{-} \operatorname{Re}[Z_{\ell}^{m}(\Phi, \Omega)]\biggr|^2+\biggl|\sum_{\ell,m} [\ell]_{m}^{+} \operatorname{Im}[Z_{\ell}^{m}(\Phi, \Omega)]\biggr|^2\biggr] \\ 
    + & \Biggl.\left(1+P_\gamma\right)\biggl[\biggl|\sum_{\ell,m} [\ell]_{m}^{+} \operatorname{Re}[Z_{\ell}^{m}(\Phi, \Omega)]\biggr|^2+\biggl|\sum_{\ell,m} [\ell]_{m}^{-} \operatorname{Im}[Z_{\ell}^{m}(\Phi, \Omega)]\biggr|^2\biggr]\Biggr\} \, ,
\end{aligned}
\label{eq:Zlm}
\end{equation}
which is obtained from \cref{eq:intensity_formula,eq:intensity_sdme,eq:sdme0,eq:sdme1,eq:sdme2}~\cite{Mathieu:2019fts}. Here, $Z_{\ell}^m(\Phi,\Omega) = e^{-i\Phi}Y_{\ell}^m(\Omega)$ and $\kappa$ is a normalization coefficient. The degree of polarization $P_{\gamma}$ is set to 0.35 according to the GlueX experiment. We generate $10^6$ events for each wave set and normalize the sum of the squared magnitudes of all amplitudes to 1. In addition, experimental effects like acceptance, resolution, and systematic effects are not considered here since we want to discuss only the mathematical ambiguities with a perfect model.

From the pseudo-data, the partial-wave amplitudes are extracted by performing unbinned extended maximum likelihood fits of \cref{eq:Zlm} using the AmpTools software framework~\cite{amptools}. For each data sample, 100 fit attempts are performed by minimizing the negative log likelihood function $-2\log\mathcal{L}$ with the \texttt{MIGRAD} algorithm starting from randomly initialized values for each amplitude.

Initially, we consider the wave set $\{S_0^+, D_{+1}^+, D_0^+, D_{-1}^+\}$, as used in Ref.~\cite{JointPhysicsAnalysisCenter:2023gku}, which includes only positive-reflectivity amplitudes. Our study confirms the JPAC result that this wave set has no ambiguous solutions. Next, we extend the study to include both reflectivities by examining the wave set $\{P_{+1}^{\pm}, P_{-1}^{\pm}\}$, which we find to exhibit continuous ambiguities. Finally, we consider the comprehensive wave set consisting of all $S$ and $P$-waves, $\{S_0^{\pm}, P_{+1}^{\pm}, P_{0}^{\pm}, P_{-1}^{\pm}\}$. For this wave set, we find no continuous or discrete ambiguities.

\subsection{$S_{0}^{+}$ and $D_{+1,0,-1}^{+}$ wave set}
\label{sec:SD_japc}
We start from the wave set \{$S_{0}^{+}, D_{+1}^{+}, D_{0}^{+}, D_{-1}^{+}$\} used in Ref.~\cite{JointPhysicsAnalysisCenter:2023gku} and show that we can replicate the results for the case of a single reflectivity. The wave set contains only positive-reflectivity partial-wave amplitudes and has $N_{\text{par}}=7$ free real parameters: 4 magnitudes and 3 phase differences. From \cref{eq:moments0,eq:moments1,eq:moments2}, we obtain 18 non-zero moments, i.e.
\begin{equation}
\begin{aligned}
    \mathbf{H} = \bigg[&H_{0}(0,0), \ H_{0}(2,0), \ H_{0}(2,1), \ H_{0}(2,2), \ H_{0}(4,0), \ H_{0}(4,1),\\
                  &H_{0}(4,2), \ H_{1}(0,0), \ H_{1}(2,0), \ H_{1}(2,1), \ H_{1}(2,2), H_{1}(4,0),\\
                  &\ H_{1}(4,1), \ H_{1}(4,2), \ H_{2}(2,1)/i, \ H_{2}(2,2)/i, \ H_{2}(4,1)/i, \ H_{2}(4,2)/i \bigg]^{T}.
\end{aligned}
\end{equation}

In \cref{appen:SPD_moments}, we list all non-zero moments expressed in terms of $S$, $P$, and $D$-waves with all $m$-projections and both reflectivities. The matrix $\mathbf{A}$ for the related bilinear terms of partial-wave amplitudes
\begin{equation}
\begin{aligned}
    \mathbf{f} = \bigg[&\operatorname{Re}[S_0^+(S_0^{+})^{*}], \ \operatorname{Re}[S_0^+(D_0^{+})^{*}], \ \operatorname{Re}[S_0^+(D_{+1}^{+})^*], \ \operatorname{Re}[S_0^+(D_{-1}^{+})^*], \ \operatorname{Re}[D_0^+(D_0^{+})^*],\\ 
    &\operatorname{Re}[D_0^+(D_{+1}^{+})^*], \ \operatorname{Re}[D_0^+(D_{-1}^{+})^*], \ \operatorname{Re}[D_{+1}^+(D_{+1}^{+})^*], \ \operatorname{Re}[D_{+1}^+(D_{-1}^{+})^*], \ \operatorname{Re}[D_{-1}^+(D_{-1}^{+})^*] \bigg]^T
\end{aligned}
\end{equation}
is given by the coefficients of the respective terms in the moment expressions in \cref{appen:SPD_moments}. Since $\mathbf{A}$ has rank $10>N_{\text{par}}$, this wave set does not have mathematical ambiguities, as expected.

\begin{figure}[H]
\centering
\includegraphics[width=0.45\textwidth]{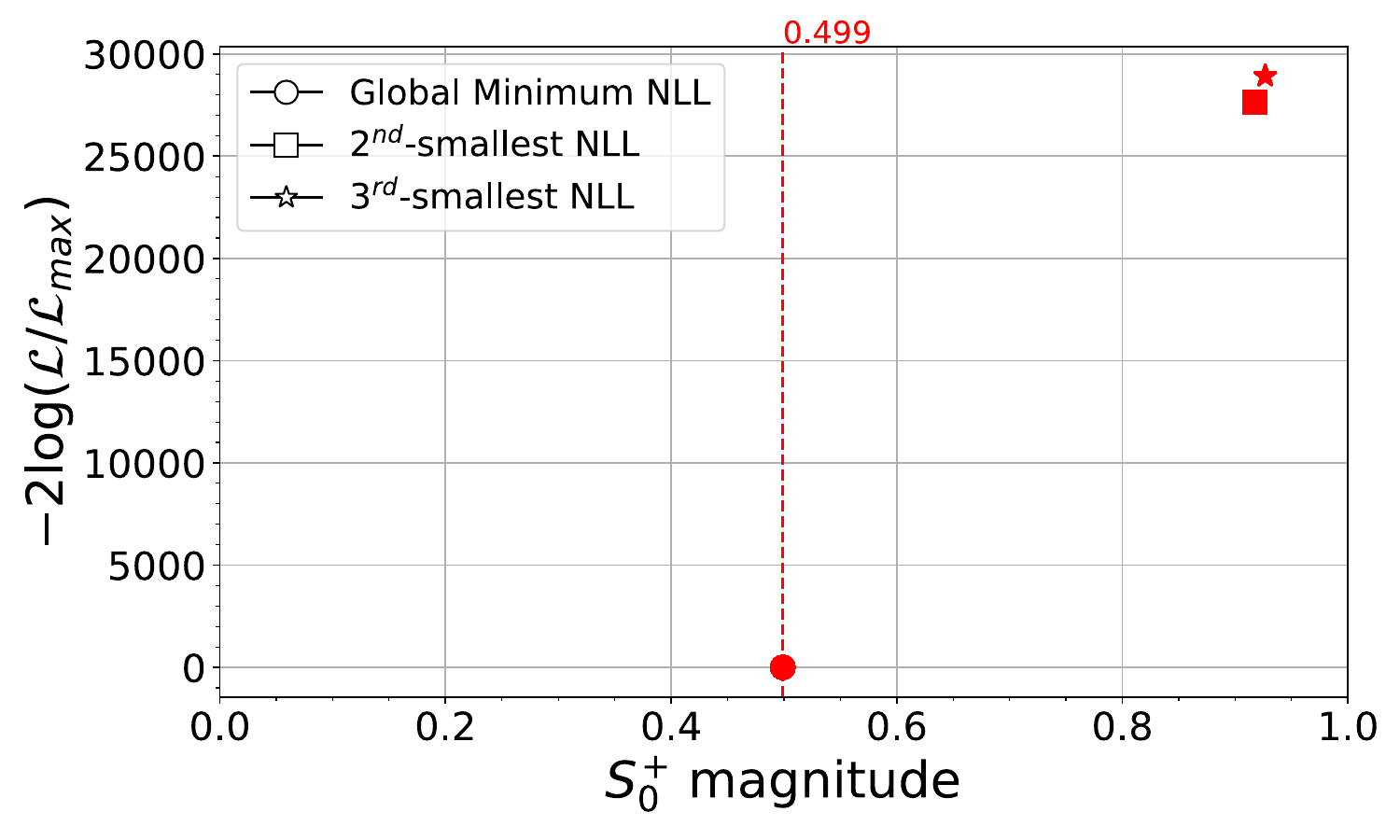}
\includegraphics[width=0.45\textwidth]{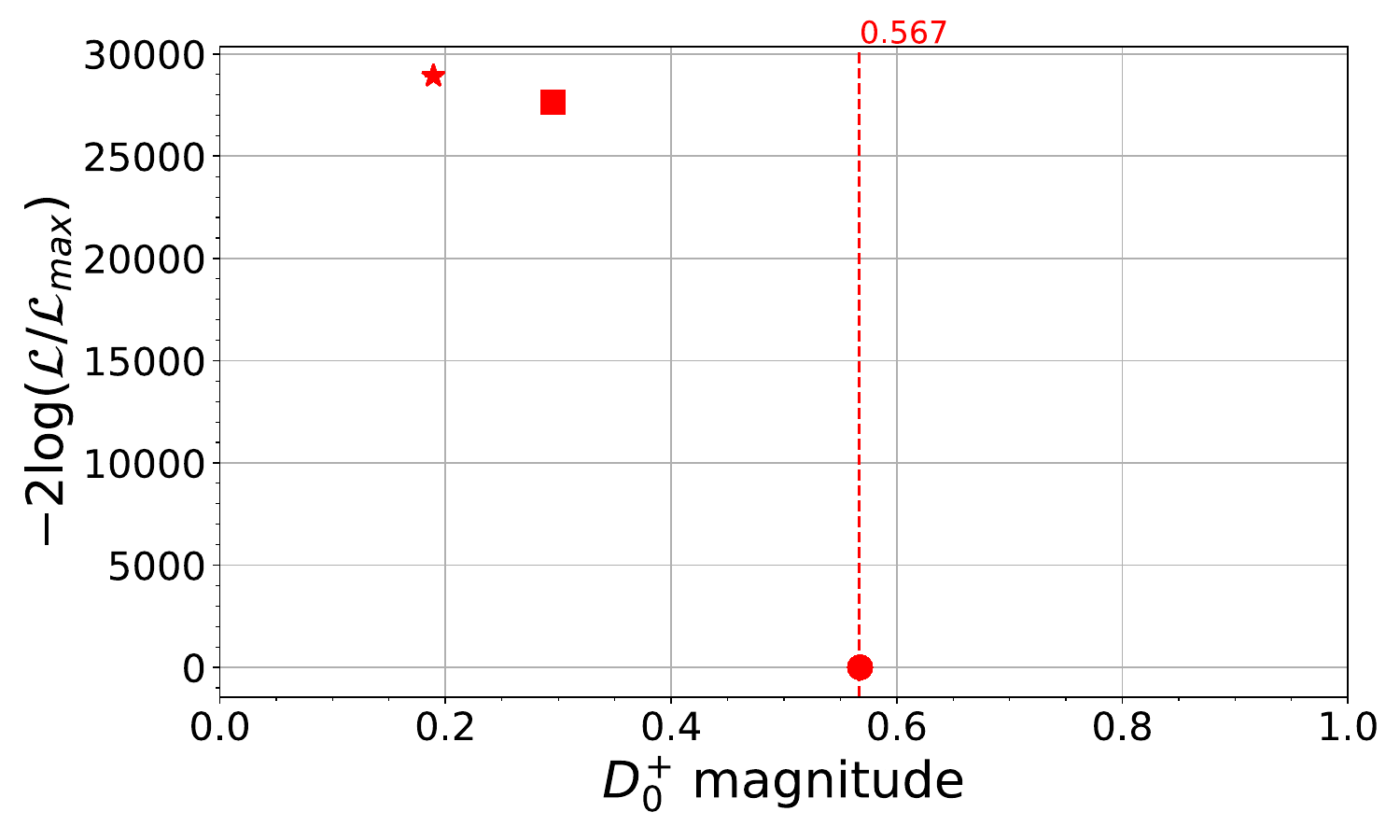}
\includegraphics[width=0.45\textwidth]{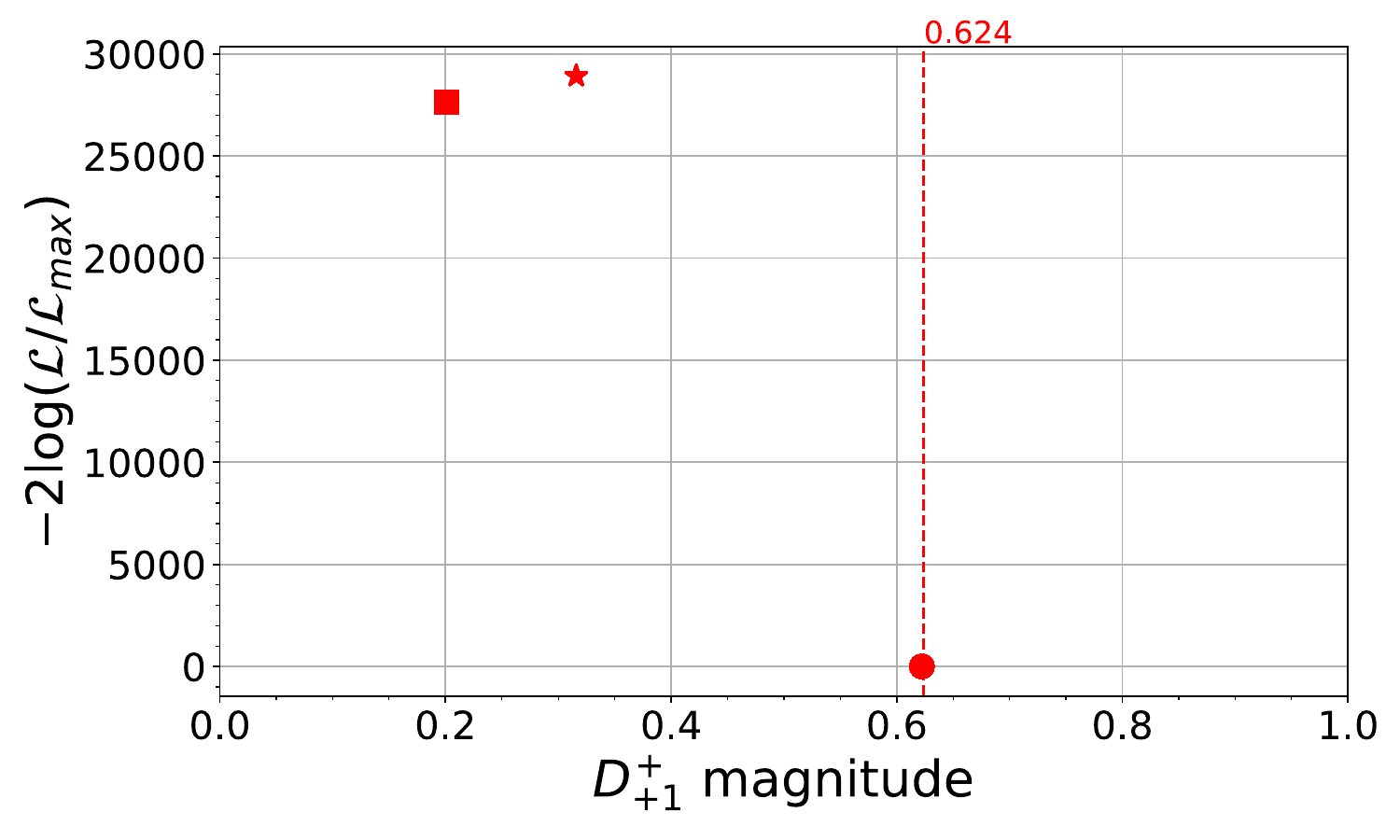}
\includegraphics[width=0.45\textwidth]{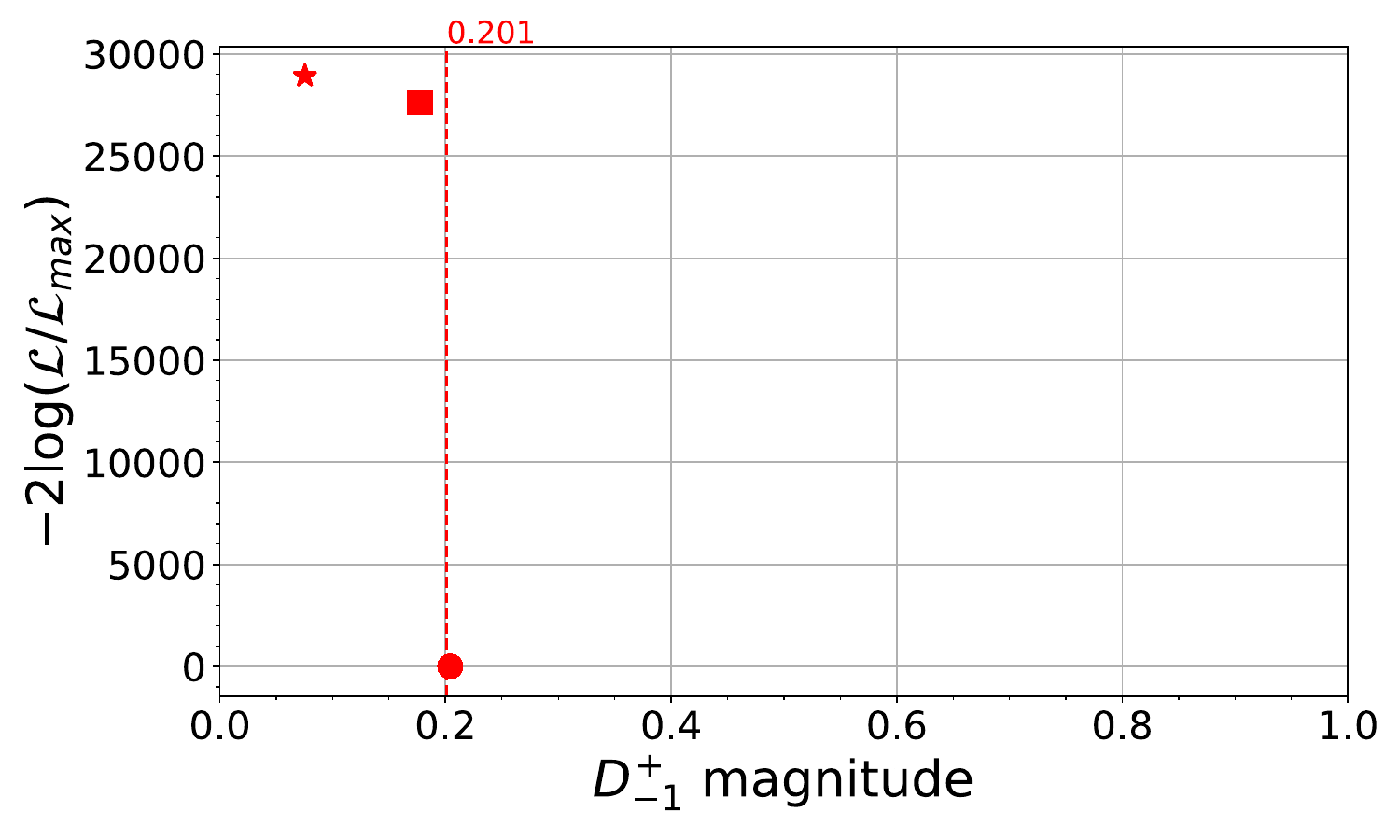}
\includegraphics[width=0.45\textwidth]{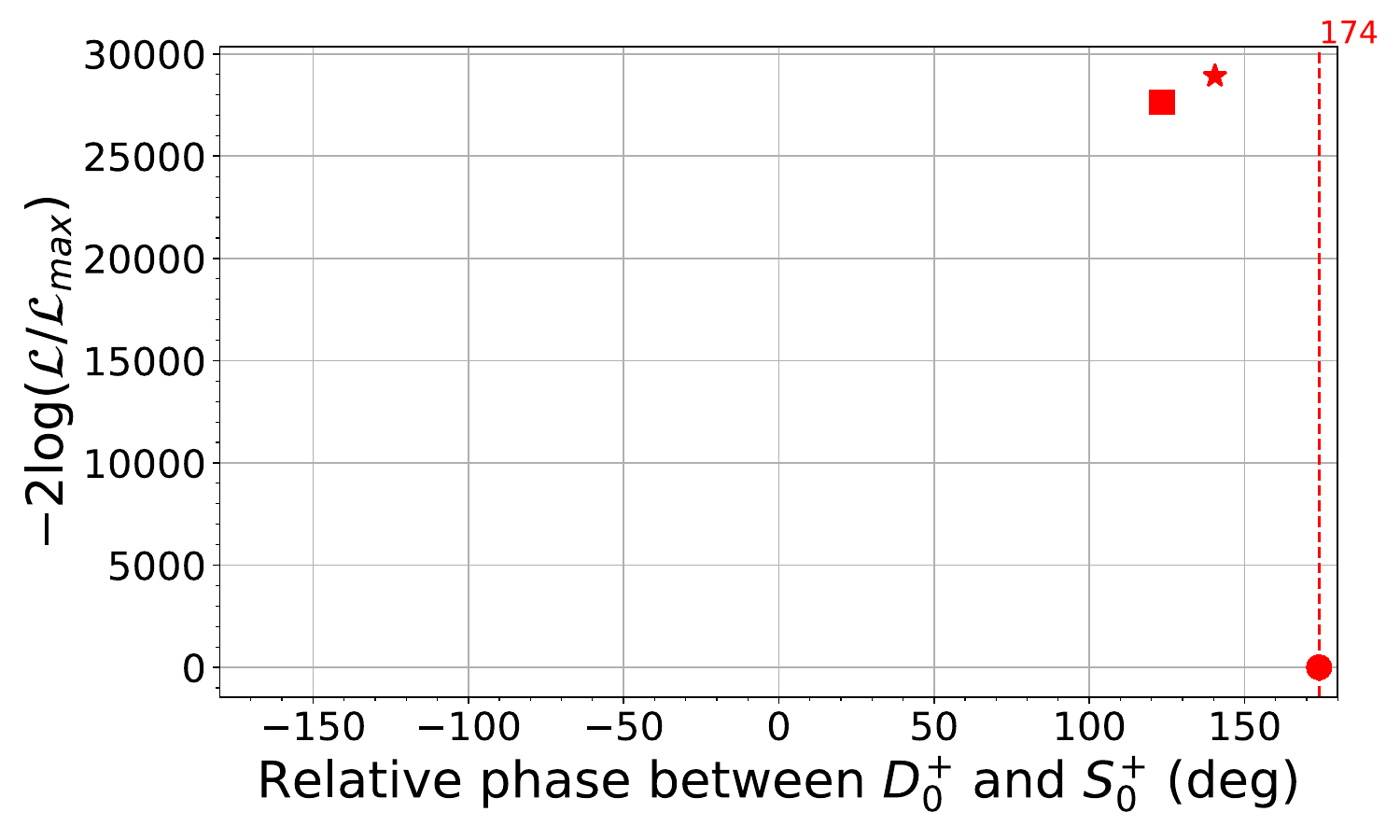}
\includegraphics[width=0.45\textwidth]{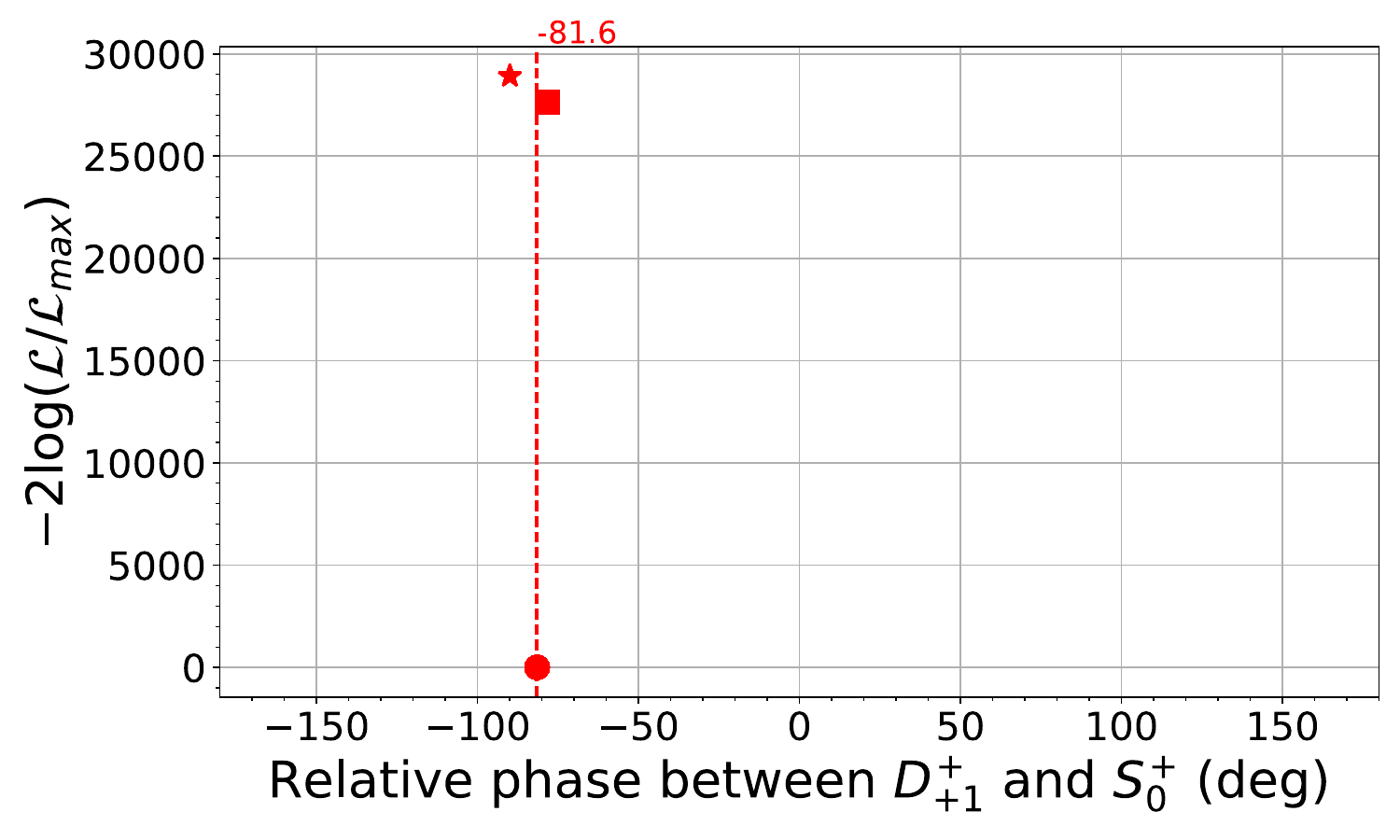}
\includegraphics[width=0.45\textwidth]{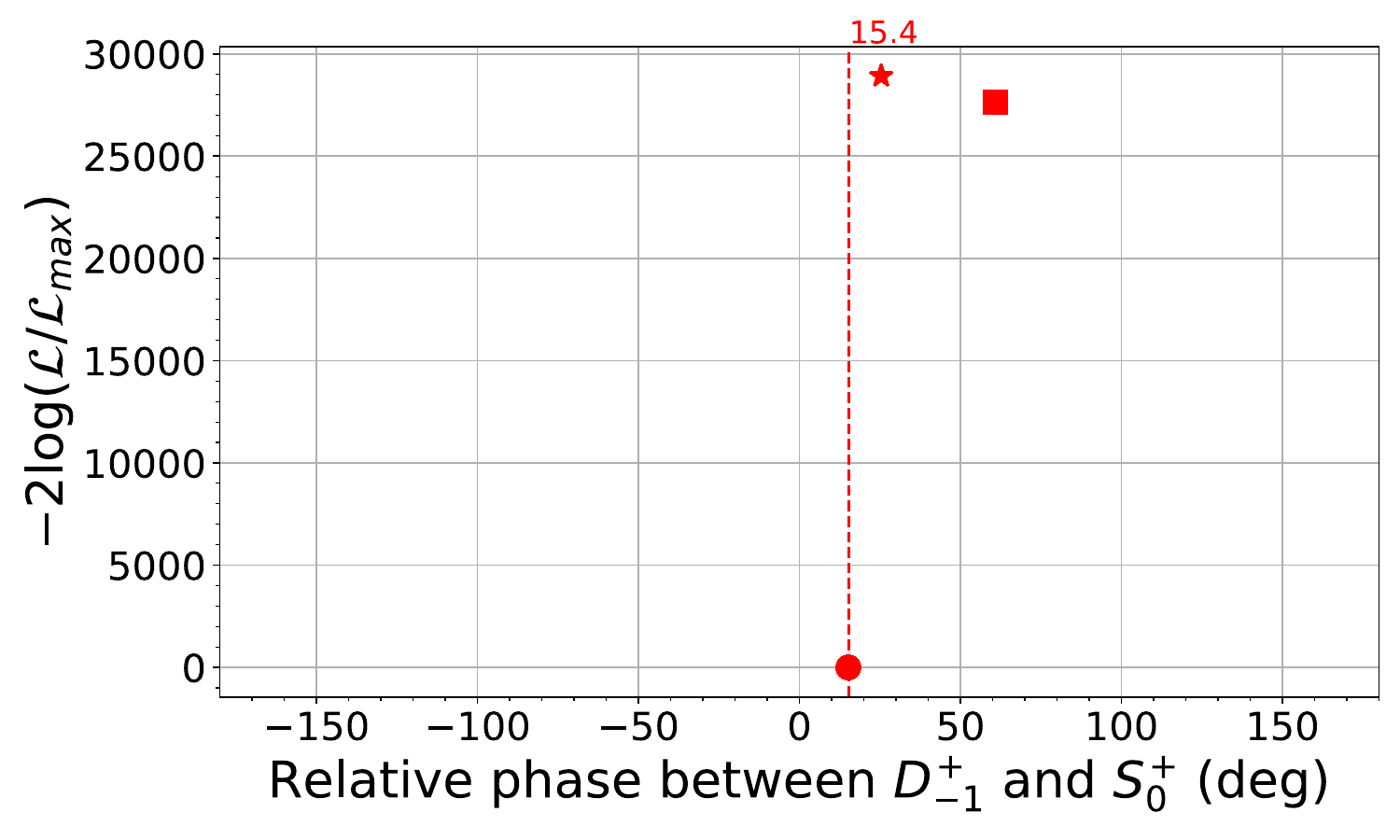}
\caption[]{\label{fig:SD_pos}Results of 100 fits of $10^6$ events generated with the wave set \{$S_{0}^{+}, D_{+1}^{+}, D_{0}^{+}, D_{-1}^{+}$\}, performed using random start values. The upper four plots show the negative log likelihood (NLL) differences with respect to the lowest NLL of the global minimum versus the partial-wave magnitudes of the $S^+_0$, $D^+_0$, $D^+_{+1}$, and $D^+_{-1}$ waves. The lower three plots show the NLL difference versus the phases of the $D^+_0$, $D^+_{+1}$, and $D^+_{-1}$ waves relative to the $S^+_0$ wave. The circles represent the parameter values at the global minimum, the other symbols at local minima. The statistical uncertainties are present but are too small to be visible at this scale. The dashed lines indicate the input values used to generate the pseudo-data.}
\end{figure}

For the Monte Carlo input-output study, the pseudo-data are generated using the same values for the partial-wave amplitudes as in Ref.~\cite{JointPhysicsAnalysisCenter:2023gku}. \cref{fig:SD_pos} shows the results of the PWA fits. For clarity, we flip the sign of the trivial ambiguous solution back in \cref{fig:SD_pos}. Out of the 100 fit attempts, 66 converge\footnote{Convergence requires a successful minimization status from \texttt{MINUIT} and a positive-definite error matrix.} to the same global minimum, whereas the remaining ones converge to two local minima with significantly higher values of the negative log likelihood (NLL). Only the partial-wave amplitude values obtained from the global minimum are consistent with the input values. 

\begin{figure}[!hb]
    \centering
    \includegraphics[width=0.5\textwidth]{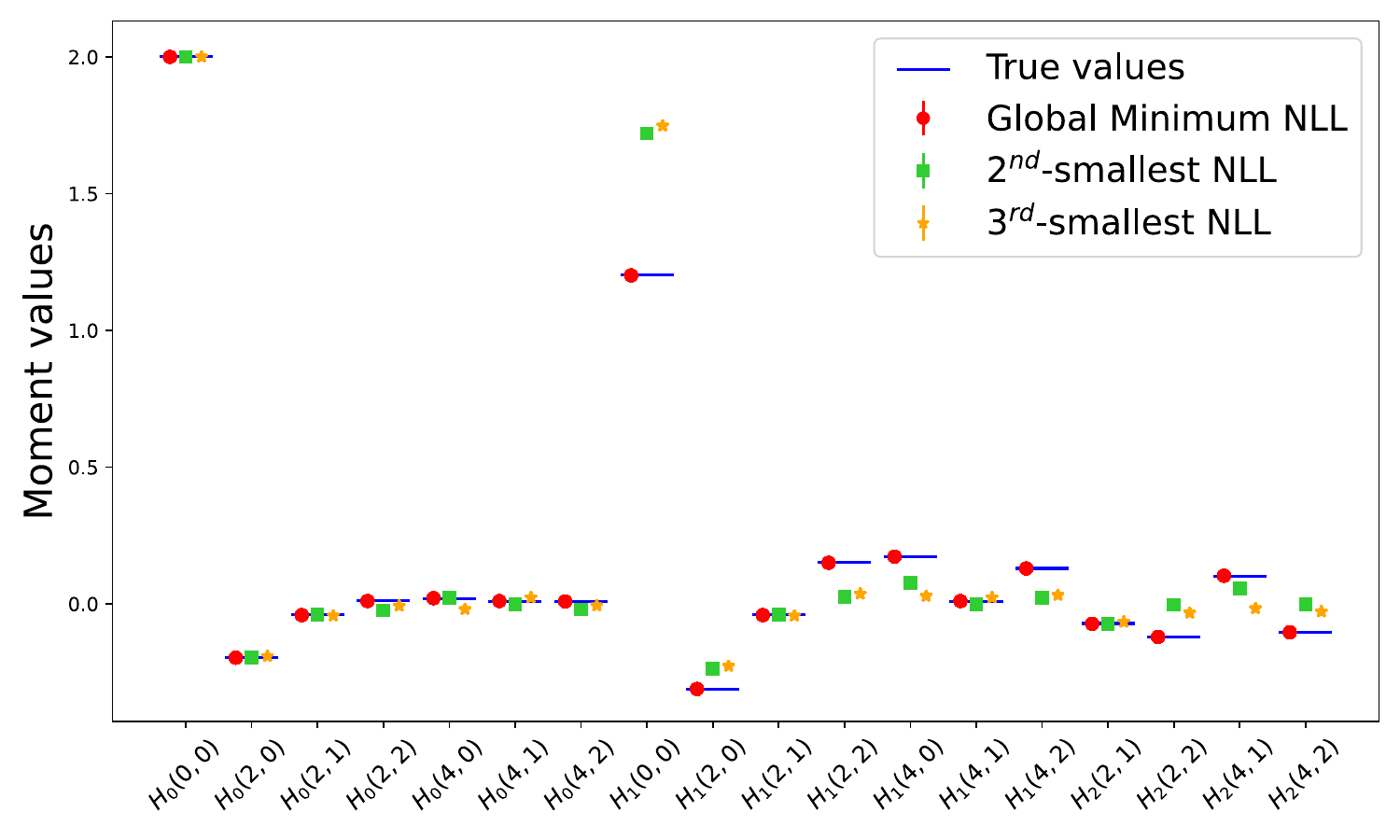}
    \caption{The moments calculated from 100 fit results are presented as colored markers, with statistical uncertainties that are smaller than the marker size. These are compared against the expected values, shown as blue lines. Marker shapes represent different minima of the NLL function, as detailed in \cref{fig:SD_pos}.}
    \label{fig:moments_SD_jpac}
\end{figure}

Since moments represent a unique decomposition of the angular distribution, we compare in \cref{fig:moments_SD_jpac} the moments calculated from the maximum-likelihood estimates of the partial-wave amplitudes using \cref{eq:moments0,eq:moments1,eq:moments2} with the true values to demonstrate that the amplitude values found at the global minimum indeed correspond to the generated angular distribution, whereas the moments from local minima deviate strongly from the true values. Overall, the results are consistent with the conclusion in Ref.~\cite{JointPhysicsAnalysisCenter:2023gku} and indicate that the wave set \{$S_{0}^{+}, D_{+1}^{+}, D_{0}^{+}, D_{-1}^{+}$\} does not have mathematical ambiguities.

\subsection{The $P_{+1}^{\pm}$ and $P_{-1}^{\pm}$ wave set} 
\label{sec:case_P+-}
To extend the study to cases with both reflectivities, we examine the simple wave set \{$P_{+1}^{\pm}, P_{-1}^{\pm}$\}, which includes just the $P$ waves with $m$-projections $\pm 1$ and both reflectivities. This represents one of the simplest wave sets we used to study the photoproduction of vector mesons like $\rho(770)$. From \cref{eq:moments0,eq:moments1,eq:moments2}, we get 7 non-zero moments and \cref{eq:matrix_h=af} becomes
\begin{equation}
\begin{pmatrix}
    2 & 0  & 2 & 0 & 2 & 2 \\
    -\frac{2}{5} & 0 & -\frac{2}{5} & 0 & -\frac{2}{5} & -\frac{2}{5} \\
    0 & -\frac{2\sqrt{6}}{5} & 0 & -\frac{2\sqrt{6}}{5} & 0 & 0 \\
    0 & -4 & 0 & 4 & 0 & 0 \\
    0 & \frac{4}{5} & 0 & -\frac{4}{5} & 0 & 0 \\
    \frac{\sqrt{6}}{5} & 0 & -\frac{\sqrt{6}}{5} & 0 & \frac{\sqrt{6}}{5} & -\frac{\sqrt{6}}{5} \\
    -\frac{\sqrt{6}}{5} & 0 & \frac{\sqrt{6}}{5} & 0 & \frac{\sqrt{6}}{5} & -\frac{\sqrt{6}}{5}
\end{pmatrix}
\begin{pmatrix}
    \operatorname{Re}[P_{+1}^+ (P_{+1}^{+})^*] \\ \operatorname{Re}[P_{+1}^+ (P_{-1}^{+})^*] \\ \operatorname{Re}[P_{+1}^- (P_{+1}^{-})^*] \\ \operatorname{Re}[P_{+1}^- (P_{-1}^{-})^*] \\ \operatorname{Re}[P_{-1}^+ (P_{-1}^{+})^*] \\ \operatorname{Re}[P_{-1}^- (P_{-1}^{-})^*]
\end{pmatrix}
=\begin{pmatrix}
    H_{0}(0,0) \\  H_{0}(2,0) \\ H_{0}(2,2) \\ H_{1}(0,0) \\ H_{1}(2,0) \\ H_{1}(2,2) \\ H_{2}(2,2)/i
\end{pmatrix}.
\label{eq:P+1-1_moments}
\end{equation}
We see that the rows for $H_{0}(0,0)$ and $H_{0}(2,0)$, as well as the ones for $H_{1}(0,0)$ and $H_{1}(2,0)$ are trivially related by a factor of $-\frac{1}{5}$. Thus, $\mathbf{A}$ has a rank of only 5. However, we have 6 free real parameters: 4 magnitudes and 2 phase differences. It is impossible to uniquely solve a system of 5 equations for 6 unknown parameters. Consequently, an infinite set of amplitude values that correspond to the same intensity distribution solves \cref{eq:P+1-1_moments} leading to continuous ambiguities in the partial-wave analysis. 

\begin{figure}[H]
\centering
\includegraphics[width=0.45\textwidth]{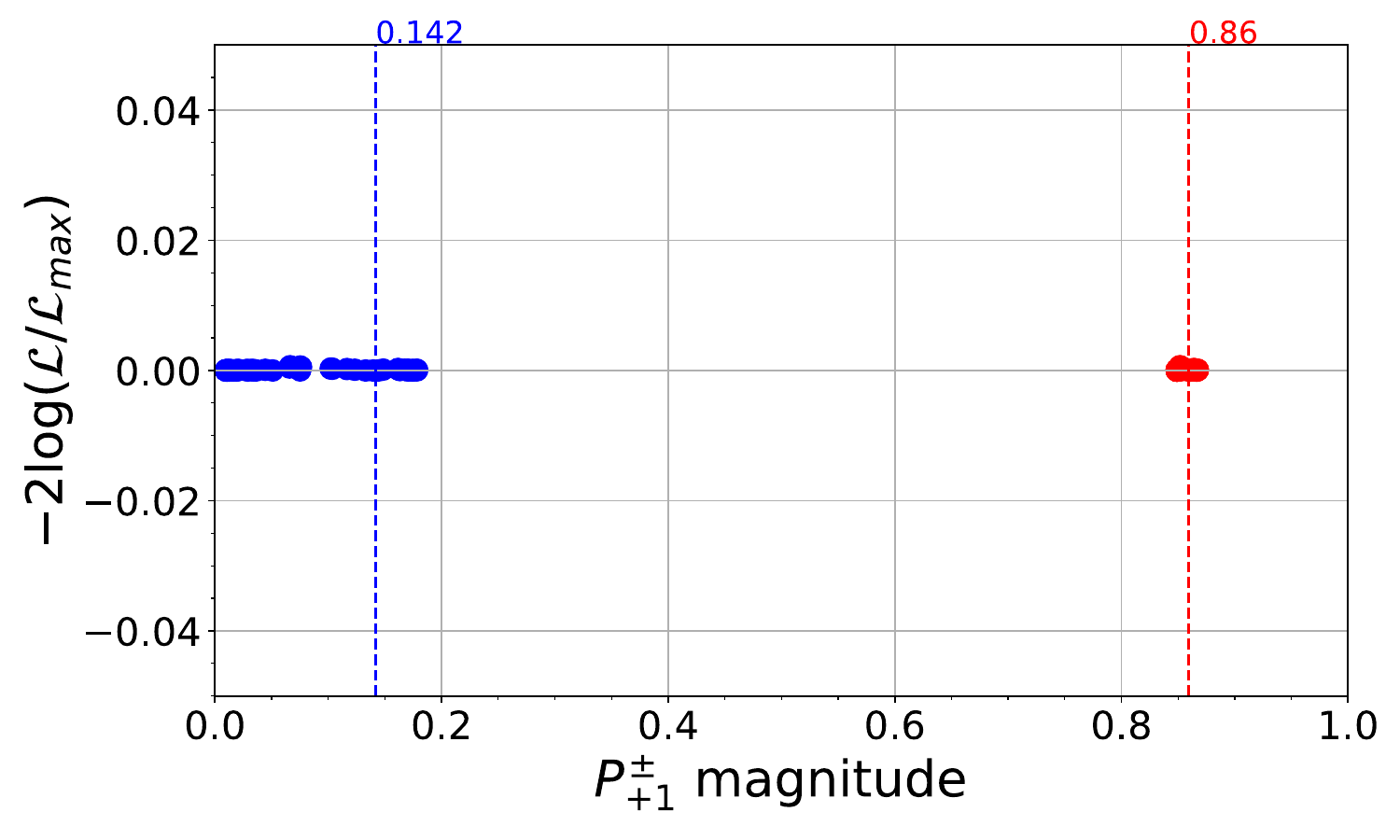}
\includegraphics[width=0.45\textwidth]{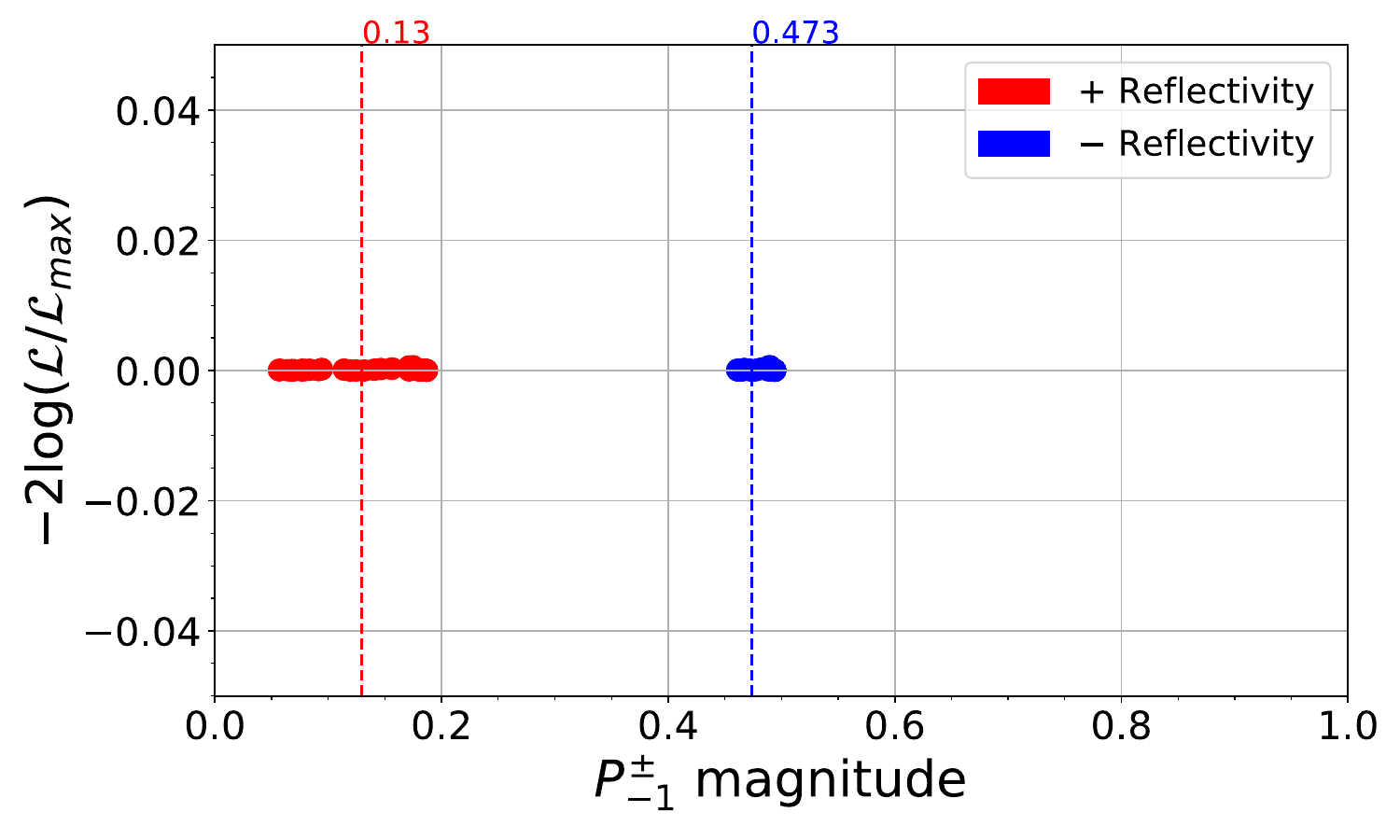}
\includegraphics[width=0.45\textwidth]{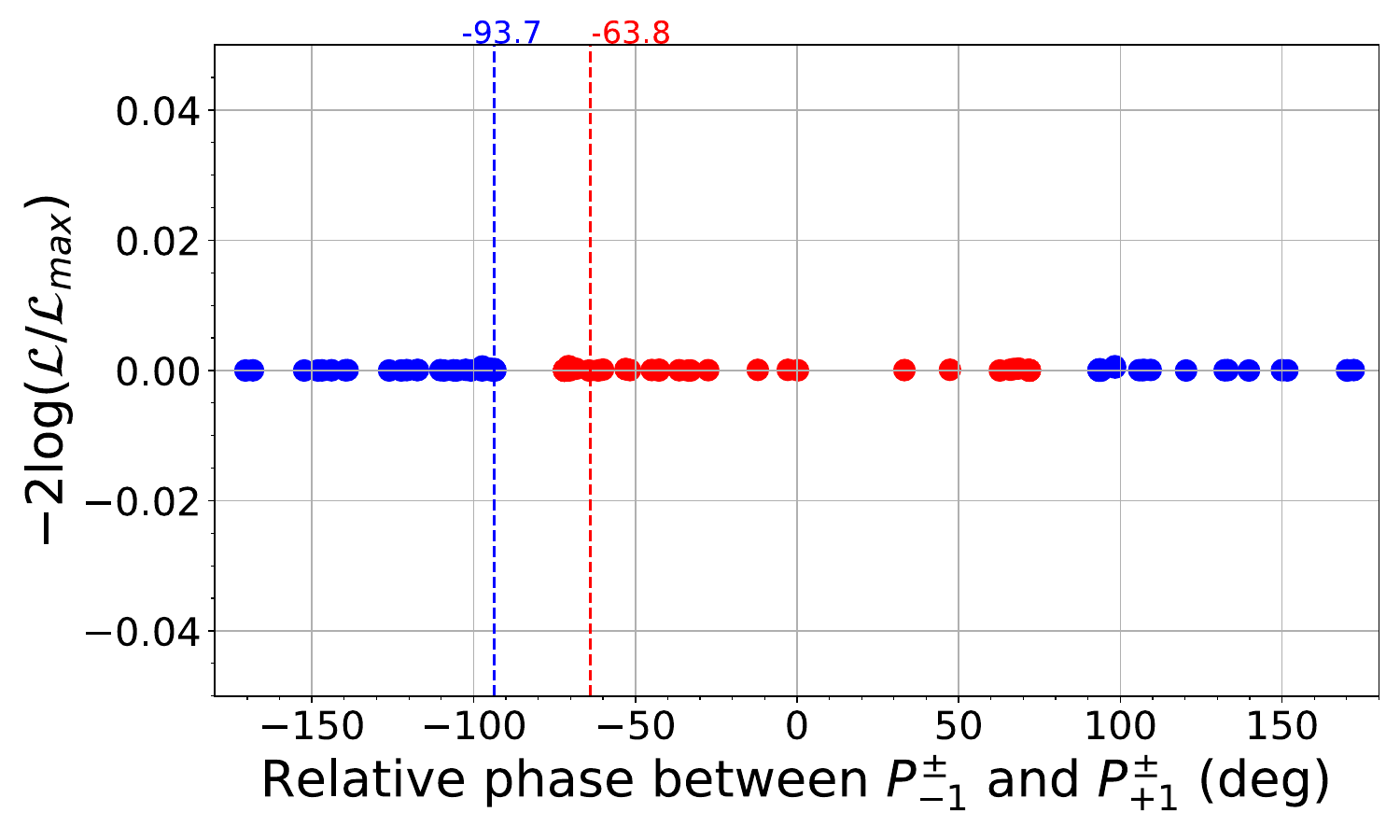}
\caption[]{\label{fig:P+1-1} Results of the $60$ converged out of 100 attempted fits to $10^6$ events generated with the wave set \{$P_{+1}^{\pm}, P_{-1}^{\pm}$\}, performed using random start values. The upper two plots show the NLL differences with respect to the lowest NLL of the global minimum versus the partial-wave magnitudes (left for $P^{\pm}_{+1}$ and right for $P^{\pm}_{-1}$). The lower one shows the NLL differences versus the phase difference between $P^{\pm}_{+1}$ and $P^{\pm}_{-1}$. Positive reflectivity is shown in red and negative reflectivity in blue; uncertainties are not shown. The dashed lines indicate the input values used to generate the pseudo-data.}
\end{figure}

For the Monte Carlo input-output study, we generate pseudo-data using the normalized amplitudes and relative phases shown as dashed lines in \cref{fig:P+1-1} together with the results of the partial-wave analysis fits. We find that the NLL values of the 60 converged fits are nearly identical within $10^{-3}$ units, whereas the estimated amplitude values scatter over large ranges. The ranges of each amplitude magnitude are determined by the related moment values. This is consistent with the expectation that all solutions are equivalent due to the continuous ambiguity. Since these solutions are all equivalent and indistinguishable, the estimated uncertainties are unreliable and are therefore not shown in \cref{fig:P+1-1}. The symmetric pattern of the phase values corresponds to a trivial sign ambiguity because the complex conjugate of a solution is also a solution.

\begin{figure}[H]
    \centering
    \includegraphics[width=0.5\textwidth]{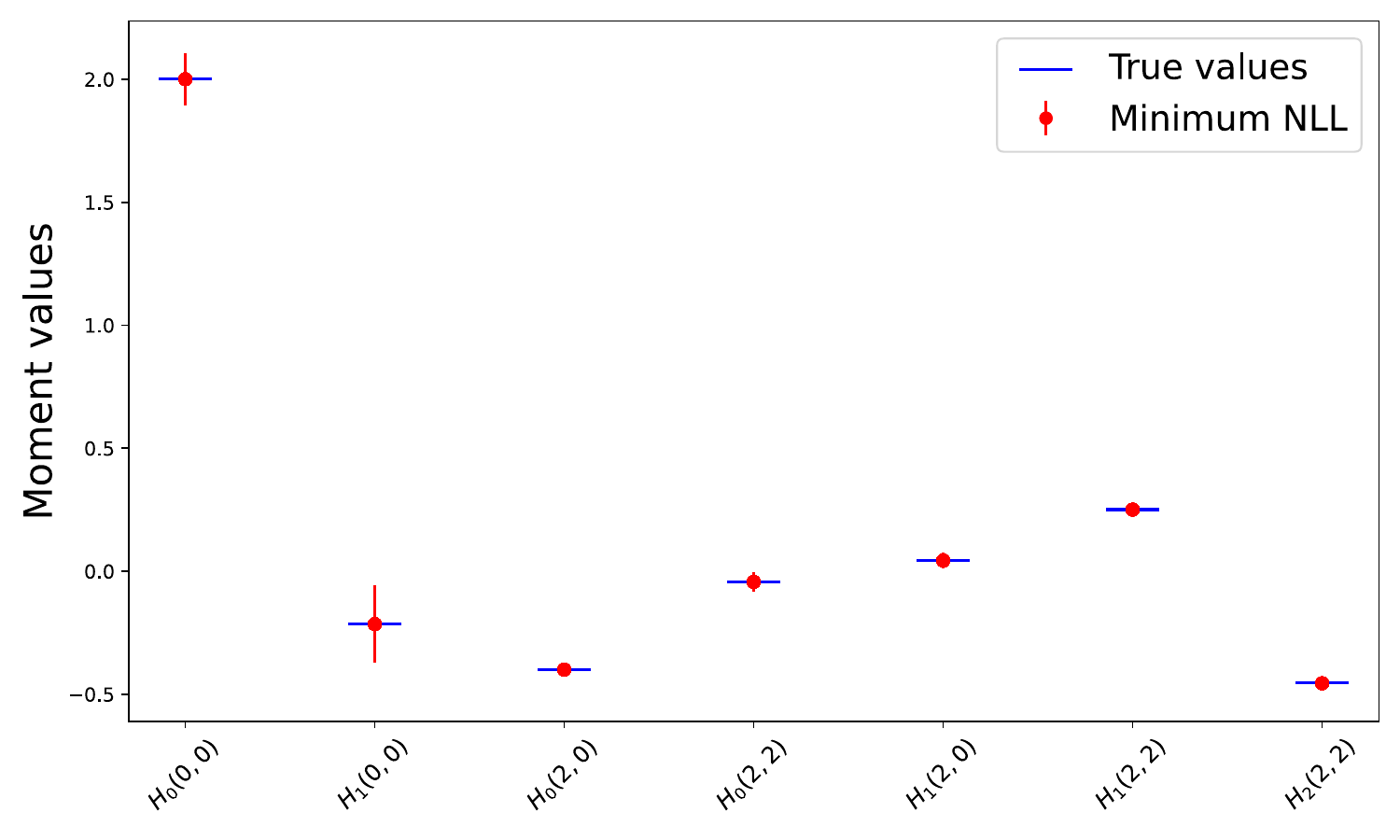}
    \caption{Comparison of the moments calculated from the 60 fit results shown in \cref{fig:P+1-1} using \cref{eq:P+1-1_moments} (red points with uncertainties) with the expected values (blue lines).}
    \label{fig:moments_P+1-1}
\end{figure}

 In \cref{fig:moments_P+1-1}, we compare the moments calculated from the estimated partial-wave amplitudes shown in \cref{fig:P+1-1} using \cref{eq:P+1-1_moments} with the true moment values used to generate the pseudo-data. As expected, all estimated moments are consistent with the true values. This demonstrates that the different partial-wave amplitude values estimated in the various fit attempts correspond indeed to the same angular distribution.


\subsection{The $S_{0}^{\pm}$ and $P_{+1,0,-1}^{\pm}$ wave set}
\label{sec:SP-simulation}
Finally, we consider the wave set $\{S_0^{\pm}, P_{+1}^{\pm}, P_{0}^{\pm}, P_{-1}^{\pm}\}$, which contains all $S$ and $P$-waves with all $m$-projections and both reflectivities. This wave set can be used, for example, to study the interference of $\rho(770)$ production with $\pi^+\pi^-$ $S$-wave production. The 8 partial-wave amplitudes correspond to $N_{\text{par}}=14$ free real parameters. From \cref{eq:moments0,eq:moments1,eq:moments2}, we obtain 15 non-zero moments, i.e.
\begin{equation}
\begin{aligned}
\mathbf{H} = \bigg[&H_{0}(0,0), \  H_{0}(1,0), \ H_{0}(1,1), \ H_{0}(2,0), \  H_{0}(2,1), \\
                &H_{0}(2,2), \ H_{1}(0,0), \ H_{1}(1,0), \ H_{1}(1,1), \ H_{1}(2,0), \\
                &H_{1}(2,1), \ H_{1}(2,2), \ H_{2}(1,1)/i, \ H_{2}(2,1)/i, \ H_{2}(2,2)/i \bigg]^{T}.
\label{eq:SP_moments}
\end{aligned}
\end{equation}
The matrix $\mathbf{A}$ for the bilinear terms of partial-wave amplitudes
\begin{equation}
\begin{aligned}
    \mathbf{f} = \bigg[&\operatorname{Re}[S_0^+(S_0^{+})^{*}], \ \operatorname{Re}[S_0^+(P_0^{+})^{*}], \ \operatorname{Re}[S_0^+(P_{+1}^{+})^*], \ \operatorname{Re}[S_0^+(P_{-1}^{+})^*], \ \operatorname{Re}[P_0^+(P_0^{+})^*], \\ 
    &\operatorname{Re}[P_0^+(P_{+1}^{+})^*], \ \operatorname{Re}[P_0^+(P_{-1}^{+})^*], \ \operatorname{Re}[P_{+1}^+(P_{+1}^{+})^*], \ \operatorname{Re}[P_{+1}^+(P_{-1}^{+})^*], \ \operatorname{Re}[P_{-1}^+(P_{-1}^{+})^*], \\
    &\operatorname{Re}[S_0^-(S_0^{-})^{*}], \ \operatorname{Re}[S_0^-(P_0^{-})^{*}], \ \operatorname{Re}[S_0^-(P_{+1}^{-})^*], \ \operatorname{Re}[S_0^-(P_{-1}^{-})^*], \ \operatorname{Re}[P_0^-(P_0^{-})^*], \\ 
    &\operatorname{Re}[P_0^-(P_{+1}^{-})^*], \ \operatorname{Re}[P_0^-(P_{-1}^{-})^*], \ \operatorname{Re}[P_{+1}^-(P_{+1}^{-})^*], \ \operatorname{Re}[P_{+1}^-(P_{-1}^{-})^*], \ \operatorname{Re}[P_{-1}^-(P_{-1}^{-})^*]\bigg]^T
\end{aligned}
\end{equation}
is given by the coefficients of the corresponding amplitude terms in \cref{appen:SPD_moments}. This matrix has a rank of $15>N_{\text{par}}$. Therefore, the partial-wave amplitudes are well constrained, and continuous ambiguities are ruled out. Considering that discrete ambiguities from Barrelet zeros do not exist, the wave set does not have mathematical ambiguities.

\begin{figure}[H]
\centering
\includegraphics[width=0.45\textwidth]{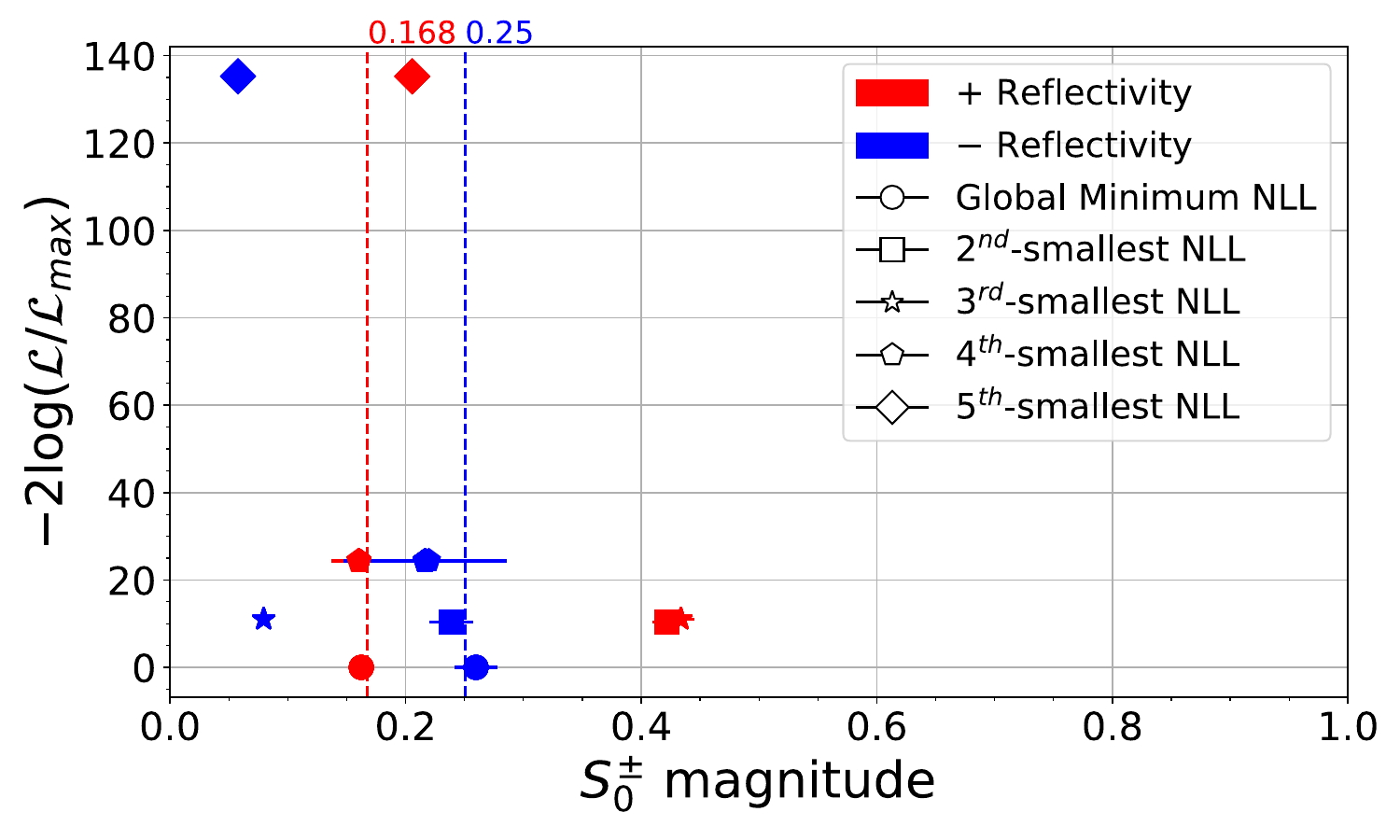}
\includegraphics[width=0.45\textwidth]{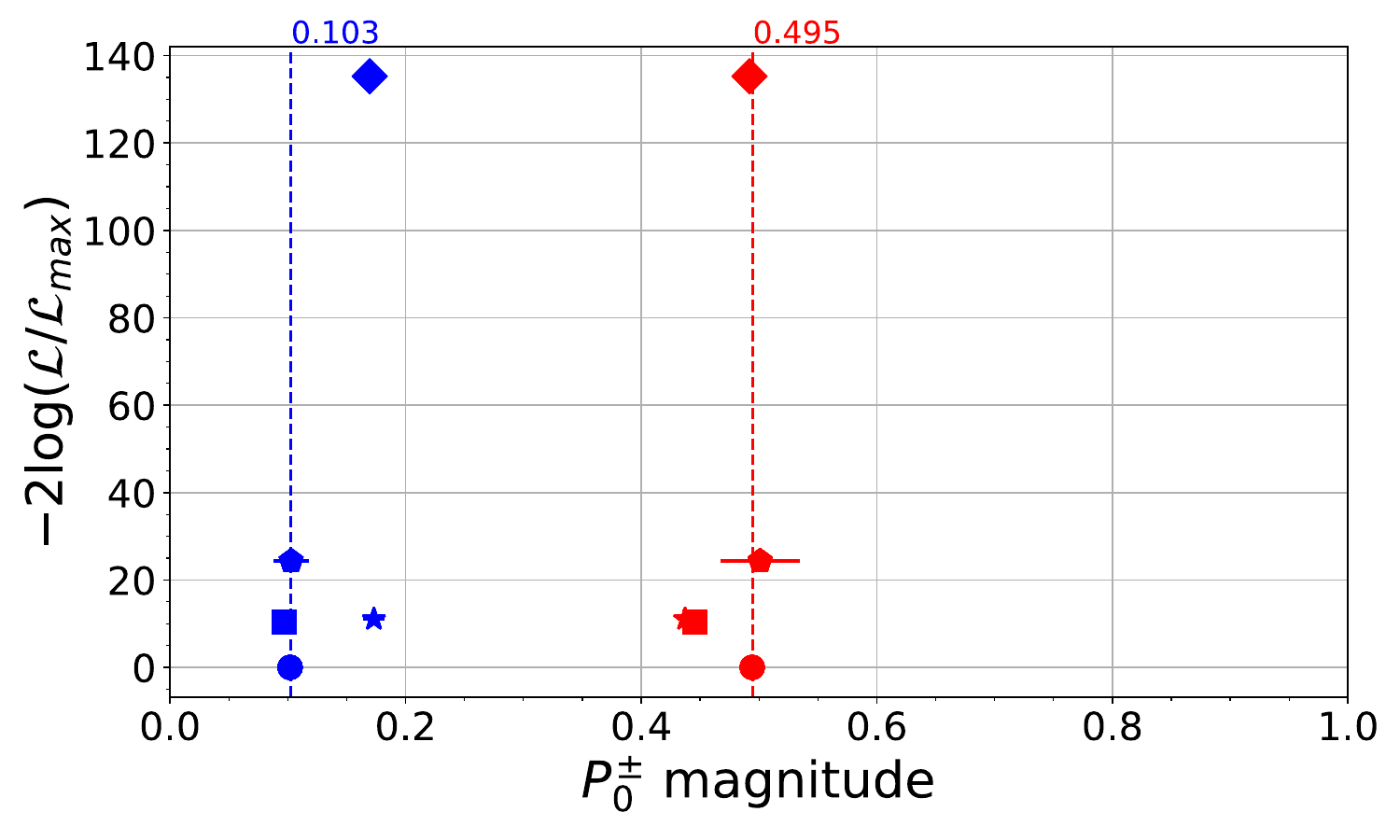}
\includegraphics[width=0.45\textwidth]{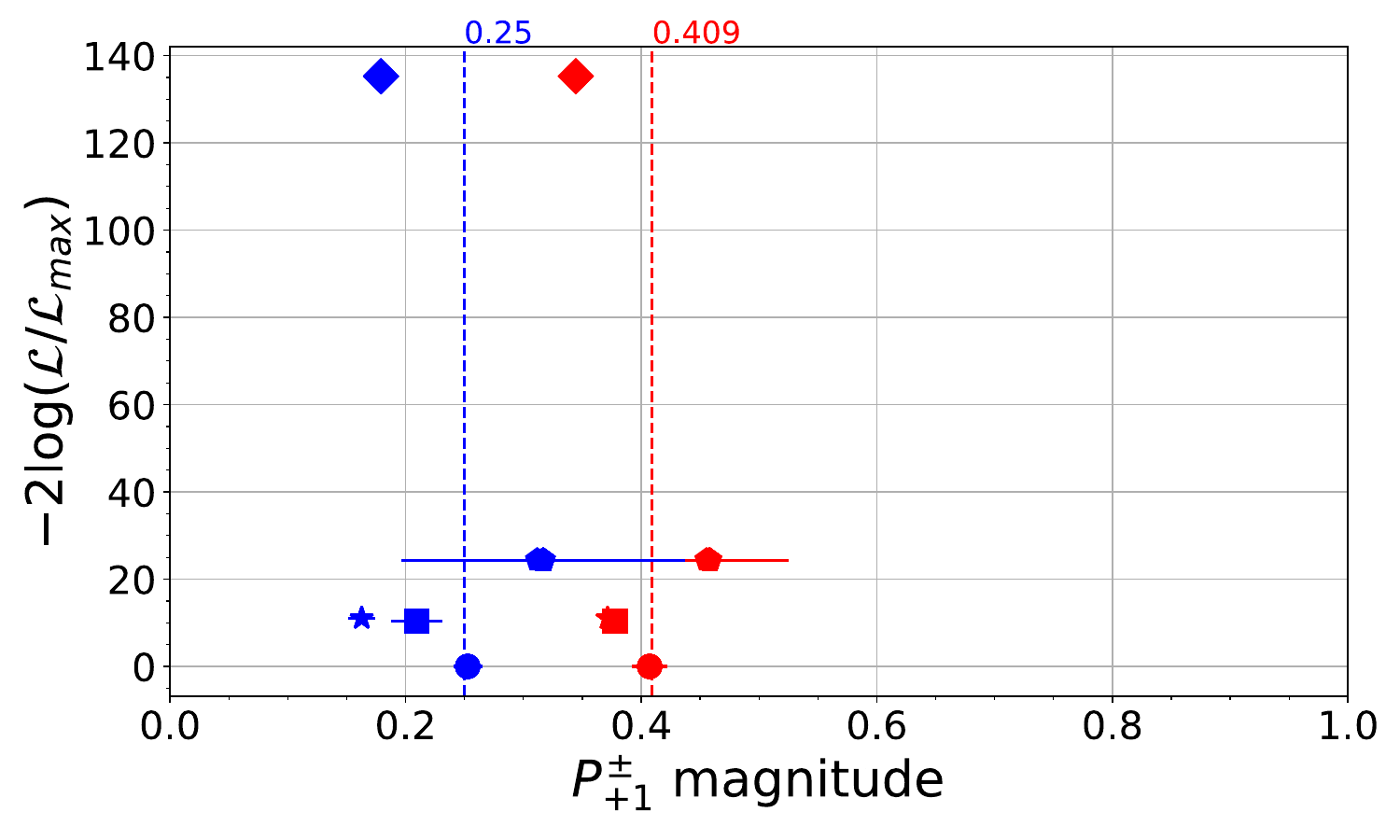}
\includegraphics[width=0.45\textwidth]{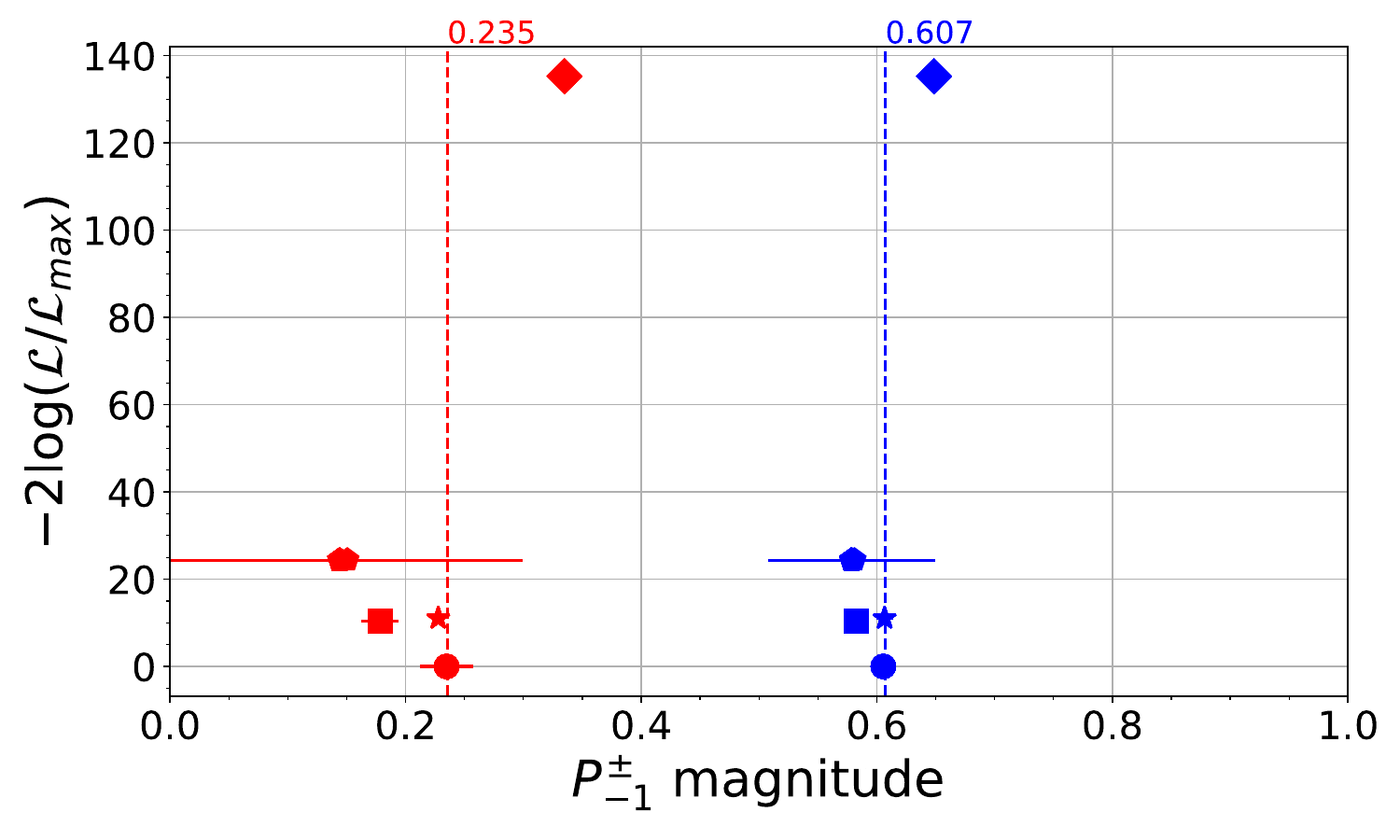}
\includegraphics[width=0.45\textwidth]{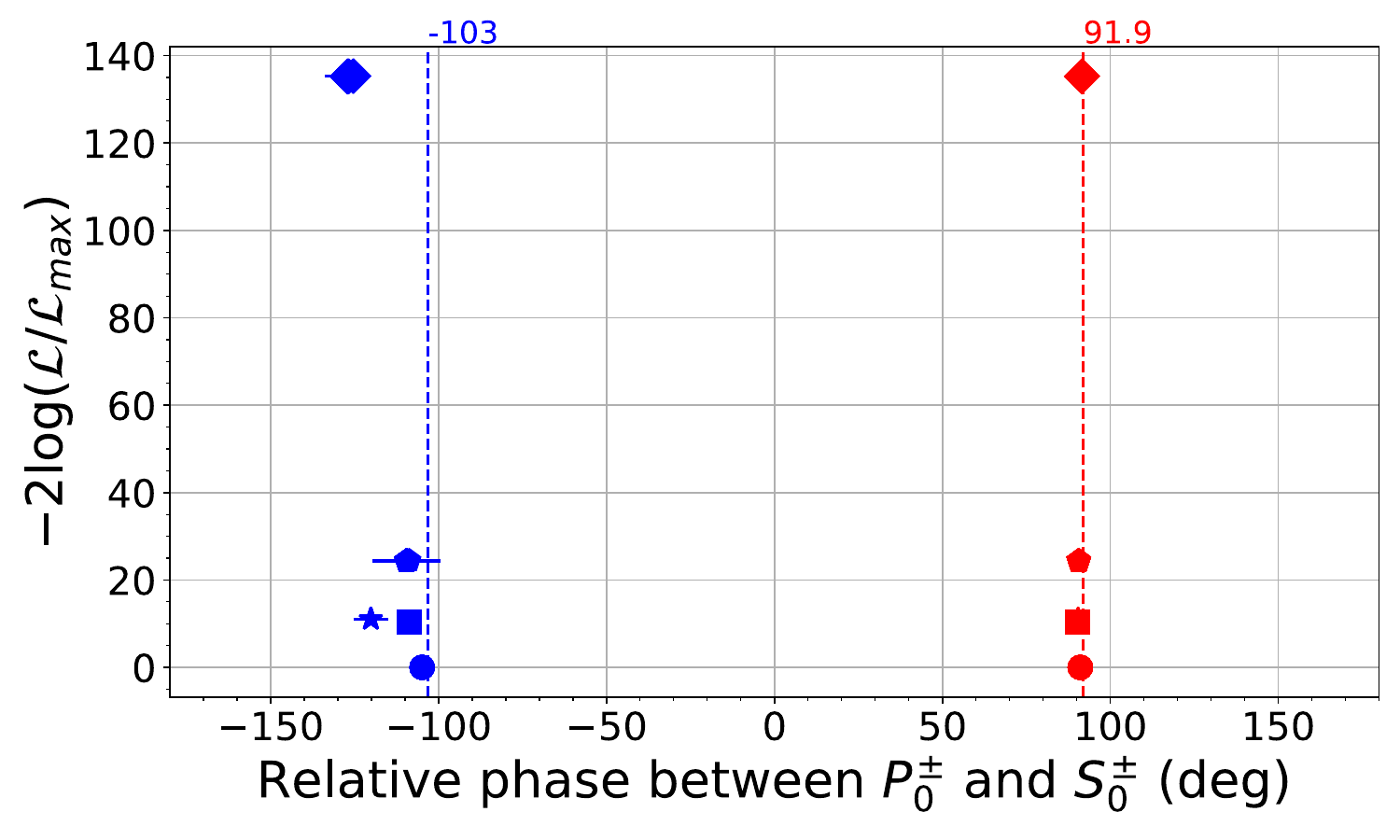}
\includegraphics[width=0.45\textwidth]{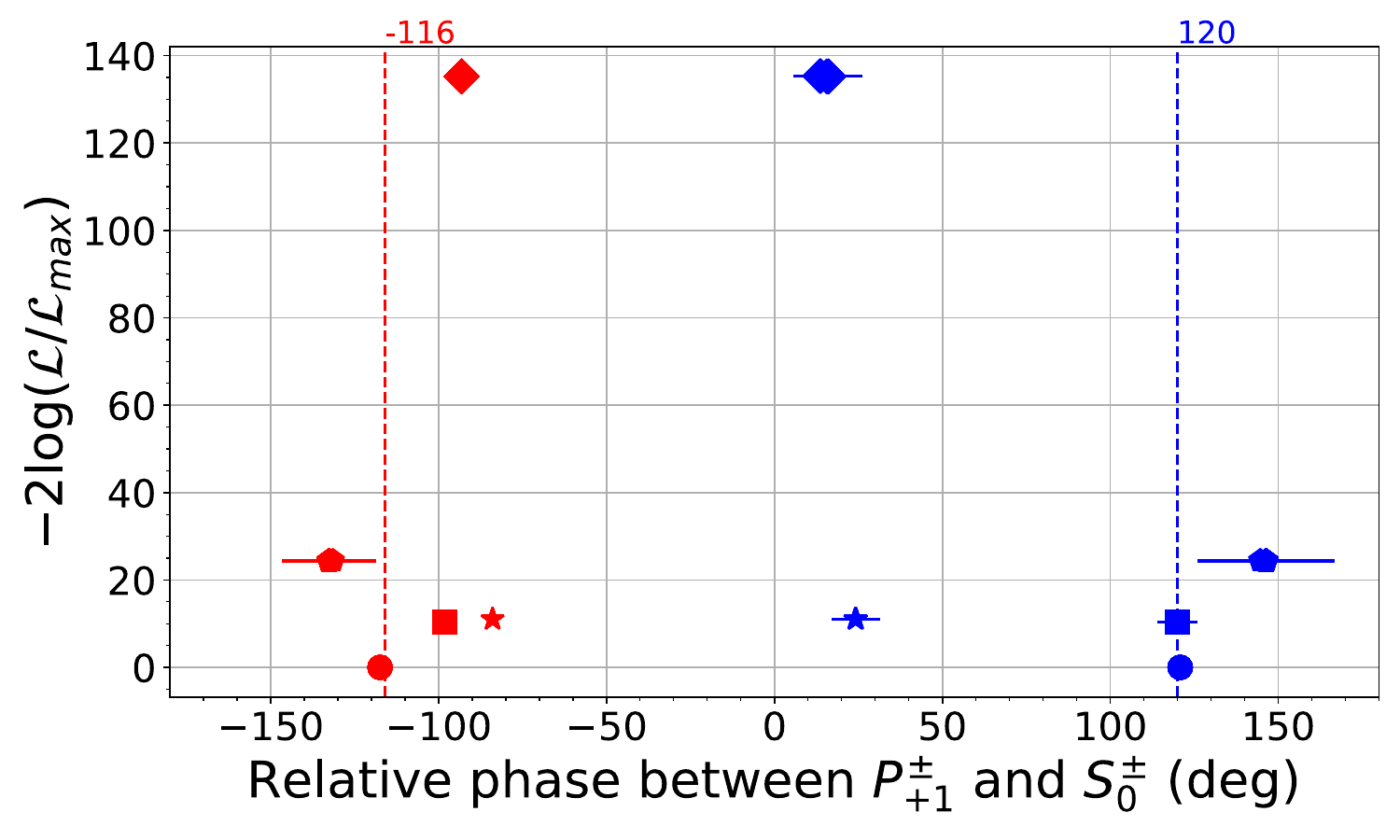}
\includegraphics[width=0.45\textwidth]{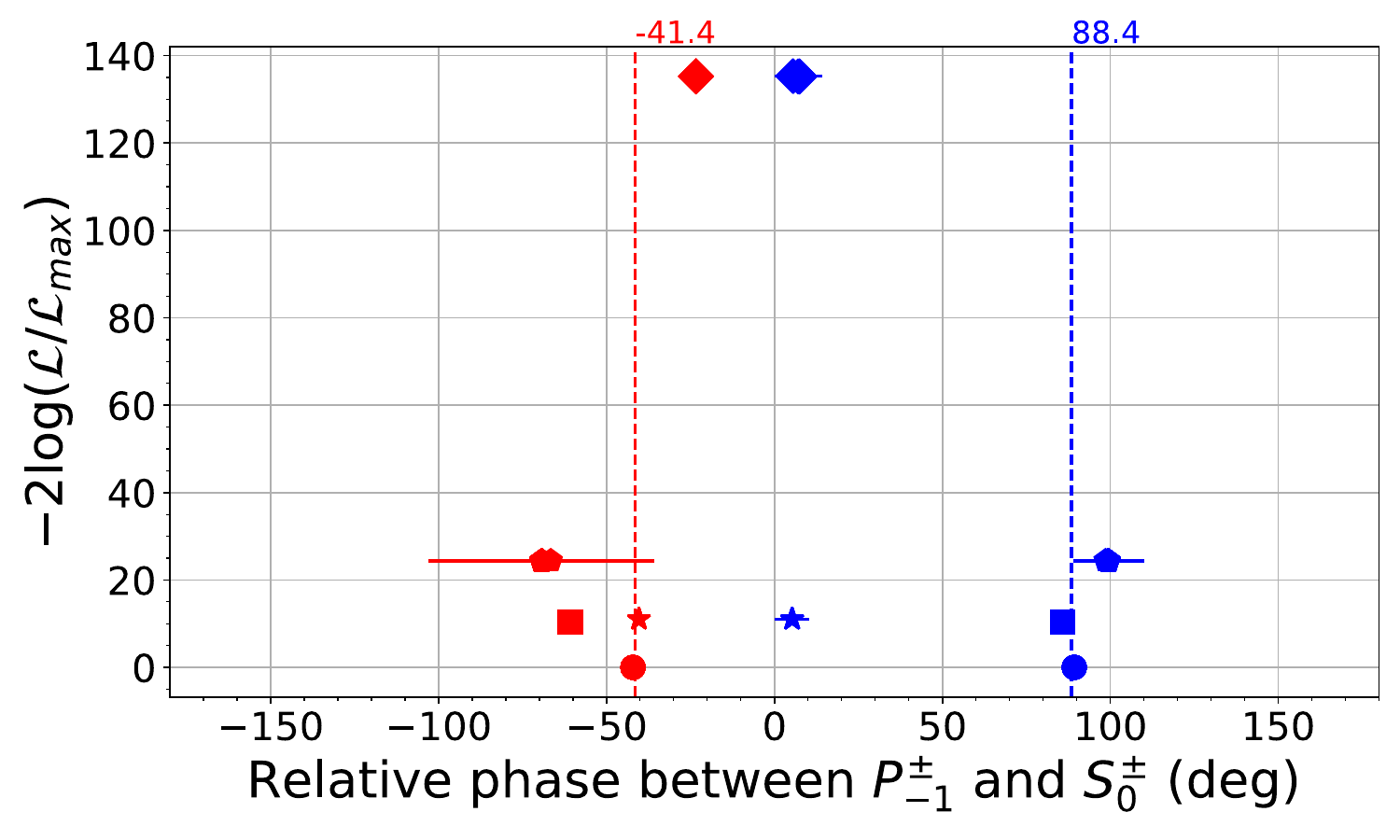}
\caption[]{\label{fig:SP}Results of 76 converged fits of $10^6$ events generated with the wave set \{$S_0^{\pm}, P_{+1}^{\pm}, P_{0}^{\pm}, P_{-1}^{\pm}$\}, performed using random start values. The upper four plots show the NLL differences with respect to the lowest NLL versus the magnitudes of the $S^{\pm}_{0}$, $P^{\pm}_{0}$, $P^{\pm}_{+1}$, and $P^{\pm}_{-1}$ waves. The lower three plots show the NLL differences versus the phases of the $P^{\pm}_{0}$, $P^{\pm}_{+1}$, and $P^{\pm}_{-1}$ waves relative to the $S^{\pm}_0$ wave. Positive-reflectivity waves are shown in red and negative-reflectivity ones in blue. The marker shapes represent different minima of the NLL function. The dashed lines indicate the input values used to generate the pseudo-data.}
\end{figure}

For the Monte Carlo input-output study, we generate pseudo-data using the amplitude values shown as dashed lines in \cref{fig:SP} together with the results of the PWA fits, again flipping the sign of the trivial ambiguous solution in the phases. Out of 100 fit attempts, 76 converge but only $21\%$ of these converged fits correspond to the global minimum with the lowest NLL. The partial-wave amplitudes at the global minimum agree with the corresponding input values. The remaining fit attempts converge to local minima, with some differing from the global NLL value by as little as 10 units. For these local minima, the estimated partial-wave amplitudes deviate significantly from the true values. Although the estimated uncertainties for the fourth-smallest NLL solution are unreliable, further analysis using the \texttt{MINOS} algorithm (see \cref{appen:SP_minos}), which accounts for the non-parabolic NLL shape near the local minimum, confirms that this solution genuinely represents a local minimum.

\begin{figure}[H]
    \centering
    \includegraphics[width=0.45\textwidth]{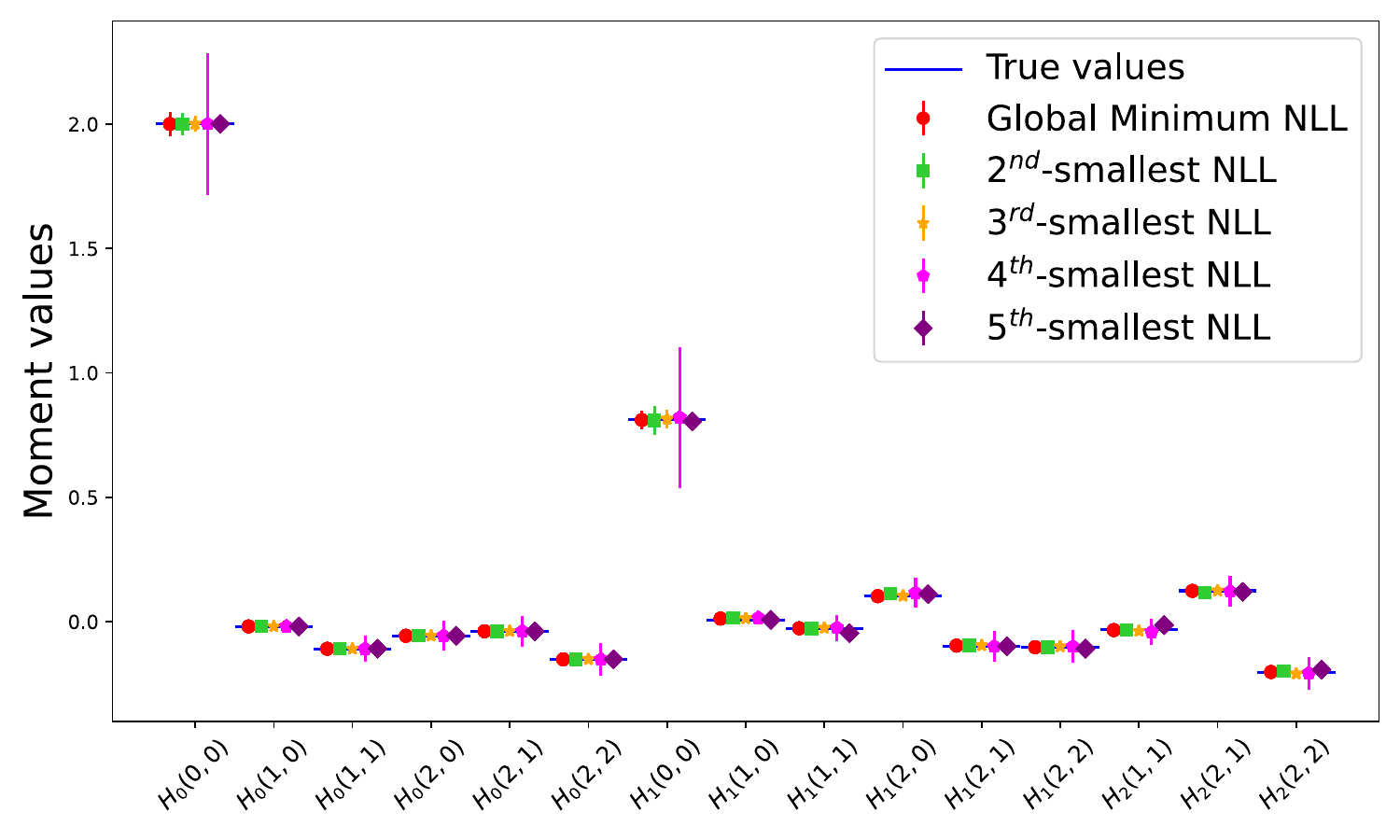}
    \includegraphics[width=0.45\textwidth]{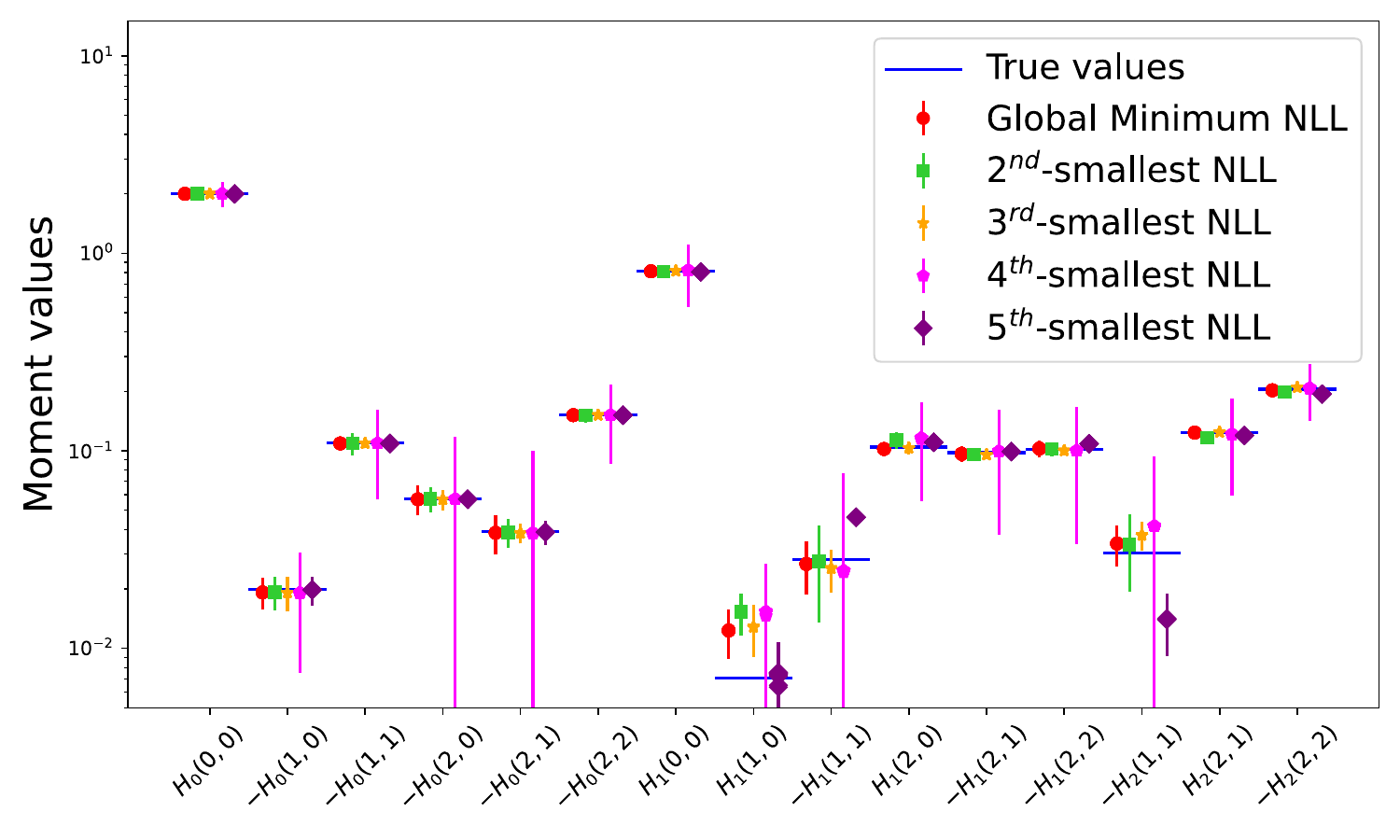}
    \caption{Comparison of the moments calculated from the 76 converged fit results shown in \cref{fig:SP} using \cref{eq:SP_moments} (colored markers with uncertainties) with the expected values (blue lines). In the right-hand plot the signs of all negative moments are flipped and a logarithmic scale is used to enhance the visibility of differences between the minima. As in \cref{fig:SP}, the marker shapes represent different minima of the NLL function.}
    \label{fig:moments_SP}
\end{figure}

In \cref{fig:moments_SP}, we compare the moments calculated from the five different NLL solutions obtained from the fits with the true values used to generate the pseudo-data. As shown in the right plot of \cref{fig:moments_SP}, the moments from all five different solutions are close to the true values, with the global minimum matching the true values more accurately. This indicates that, although the local minima are not mathematically ambiguous solutions, their angular distributions closely resemble the true angular distribution. Given the similarity of the NLL values, it may be challenging to distinguish the global minimum from these local minima in a real-world analysis, especially when the likelihood surface is distorted by detector acceptance and resolution effects. A brief discussion of these potential complications in real analyses is provided in \cref{discussion_challenge}. Overall, the fit results indicate that the \{$S_0^{\pm},P_{+1}^{\pm}, P_0^{\pm}, P_{-1}^{\pm}$\} wave set does not have mathematical ambiguities.

\section{Discussion}
\label{sec:discussion}

\subsection{Extractable information in case of ambiguities}
From the wave sets studied in \cref{sec:case}, we found that continuous mathematical ambiguities appear when the wave set only includes a pair $[\ell]_m^{\pm }$ and $[\ell]_{-m}^{\pm }$ of partial-wave amplitude, where $m \neq 0$, such as the \{$P_{+1,-1}^{\pm}$\} wave set. However, this ambiguity is generally resolved when more partial-wave amplitudes are included in the model, since the resulting interference terms with other amplitudes increase the number of linearly independent equations that determine the amplitudes.

Even though we cannot precisely determine all partial-wave amplitudes when a continuous ambiguity is present, we can still extract some useful information. For a pair $[\ell]_m^{\pm }$ and $[\ell]_{-m}^{\pm }$ of partial waves with $m \neq 0$, we can still get the total intensity of all amplitudes using \cref{eq:h0_appendix}: 
\begin{equation}
    \Big|[\ell]_m^+\Big|^2 + \Big|[\ell]_m^-\Big|^2 + \Big|[\ell]_{-m}^+\Big|^2 + \Big|[\ell]_{-m}^-\Big|^2 = \frac{1}{2}H_0(0,0) .
\end{equation}

In addition, from \cref{eq:h1_appendix,eq:h2_appendix} we have
\begin{align}
    H_1(2\ell, 2m) =& C_{\ell m;\ell-m}^{2\ell \ 2m} (-1)^m \left( \Big|[\ell]_m^+\Big|^2  + \Big|[\ell]_{-m}^+\Big|^2 - \Big|[\ell]_m^-\Big|^2 - \Big|[\ell]_{-m}^-\Big|^2 \right), \\
    H_2(2\ell, 2m) =& i C_{\ell m;\ell-m}^{2\ell \ 2m} (-1)^m \left( -\Big|[\ell]_m^+\Big|^2  + \Big|[\ell]_{-m}^+\Big|^2 + \Big|[\ell]_m^-\Big|^2 - \Big|[\ell]_{-m}^-\Big|^2 \right).
\end{align}

Therefore, we can also get the total intensity for each reflectivity as 
\begin{equation}
\begin{aligned}
    \Big|[\ell]_m^+\Big|^2  + \Big|[\ell]_{-m}^+\Big|^2 =& \frac{1}{4} H_0(0,0) + \frac{(-1)^m}{2C_{\ell m;\ell-m}^{2\ell \ 2m}} H_1(2\ell, 2m), \\
    \Big|[\ell]_m^-\Big|^2  + \Big|[\ell]_{-m}^-\Big|^2 =& \frac{1}{4} H_0(0,0) - \frac{(-1)^m}{2C_{\ell m;\ell-m}^{2\ell \ 2m}} H_1(2\ell, 2m).
\end{aligned}
\end{equation}

Although we would lose information about the phases, it might still be helpful for us to search for resonance signatures in these intensity distributions.

Furthermore, for high-energy photoproduction the reflectivity corresponds to the naturality of the exchanged particle. Therefore, we can learn which naturality of the exchanged particles is dominant from the differences
\begin{equation}
    \begin{aligned}
        \Big|[\ell]_m^+\Big|^2 - \Big|[\ell]_m^-\Big|^2 =&   \frac{(-1)^m}{2C_{\ell m;\ell-m}^{2\ell \ 2m}} \Bigg[H_{1}(2\ell, 2m) - H_{2}(2\ell, 2m)/i \Bigg],\\
        \Big|[\ell]_{-m}^+\Big|^2 - \Big|[\ell]_{-m}^-\Big|^2 =& \frac{(-1)^m}{2C_{\ell m;\ell-m}^{2\ell \ 2m}} \Bigg[H_{1}(2\ell, 2m) + H_{2}(2\ell, 2m)/i\Bigg].
    \end{aligned}
\end{equation}

\subsection{Applying external constraints to resolve ambiguities}

\begin{figure}[H]
    \centering
    \includegraphics[width=0.45\textwidth]{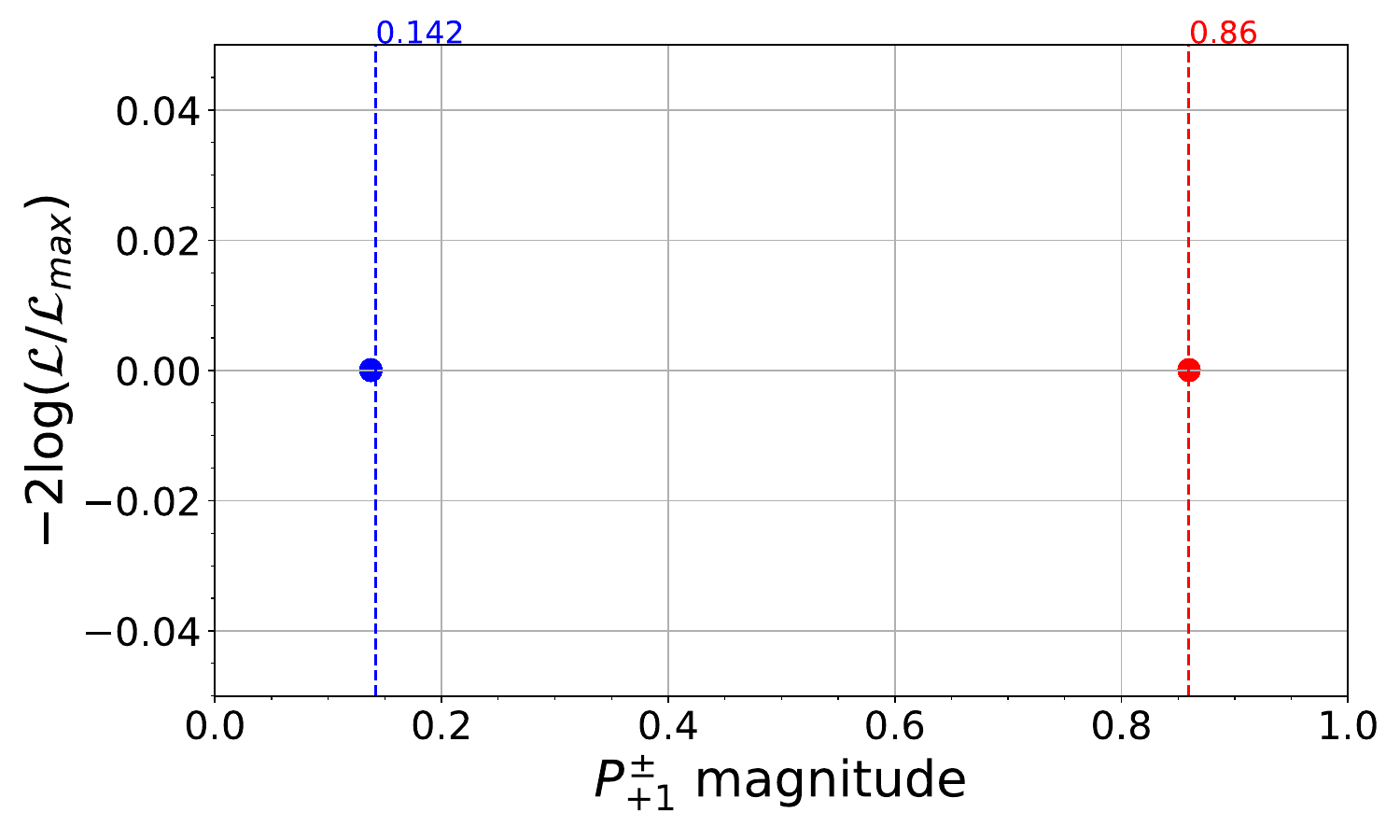}
    \includegraphics[width=0.45\textwidth]{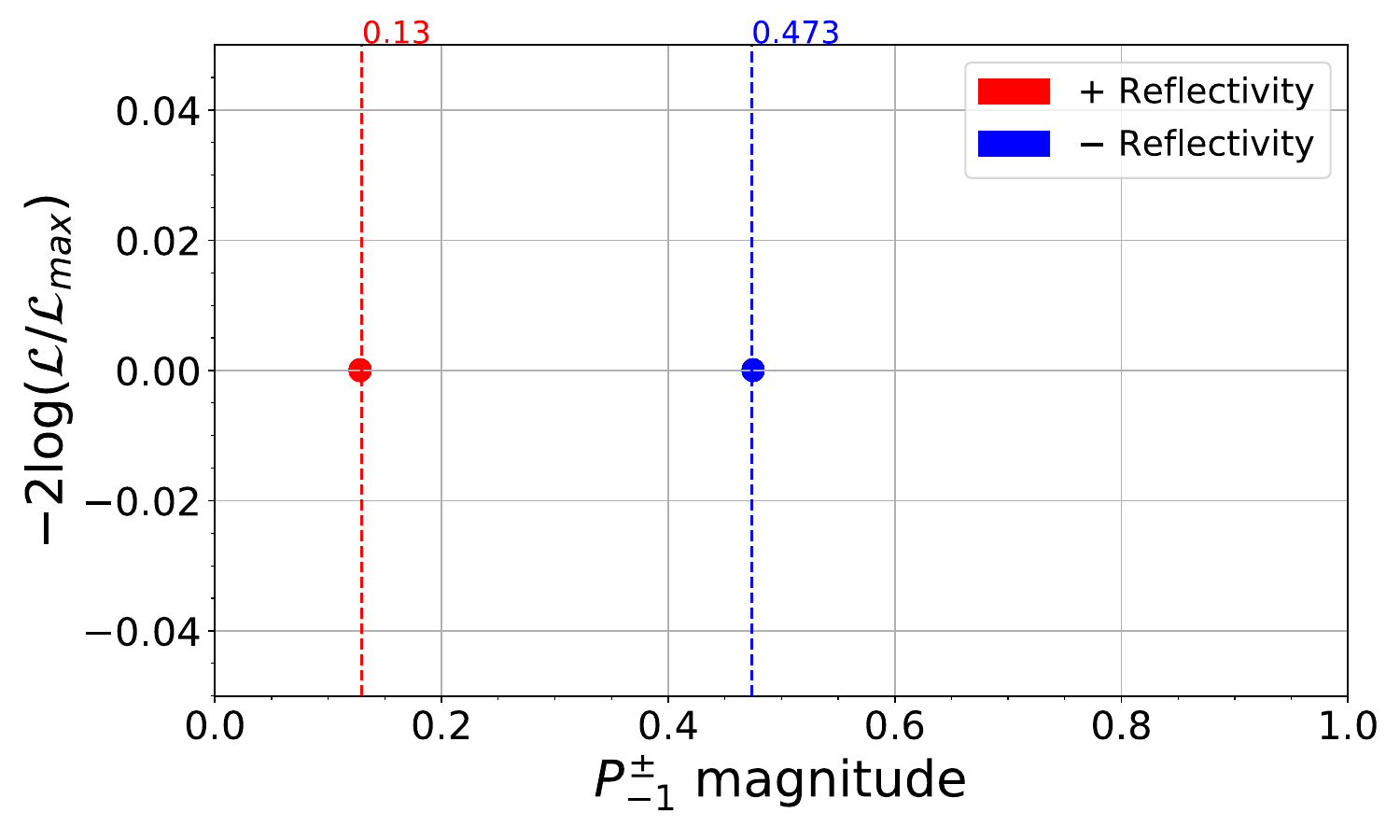}
    \caption{Results of $100$ fits of $10^6$ events generated with wave set \{$P_{+1}^{\pm}, P_{-1}^{\pm}$\}, performed constraining the relative phase differences to the generated values. The two plots show the NLL differences with respect to the lowest NLL versus the partial-wave magnitudes (left for $P_{+1}^{\pm}$ and right for $P_{-1}^{\pm})$. Positive-reflectivity waves are shown in red and negative-reflectivity ones in blue. The dashed lines indicate the input values used to generate the pseudo-data.}
    \label{fig:P+1-1_sharedphase}
\end{figure}

Alternatively, we can try to resolve the ambiguities by introducing external constraints. In some cases, we can constrain the relative phases between waves with the same $\ell$ and $\epsilon$ but different $m$-projections so that the number of unknown parameters is reduced. For example, in the $a_2^0(1320)$ cross-section analysis of GlueX data in Ref.~\cite{GlueX:2025kma}, the relative phases of all $m$ projections of a resonance produced with a given reflectivity are assumed to be the same. We study the effect of this relative-phase constraint for the \{$P_{+1}^{\pm}, P_{-1}^{\pm}$\} wave set by fixing the relative phase differences between $P_{+1}^+, P_{-1}^+$ and $P_{+1}^-, P_{-1}^-$ at the generated values, leaving the four unknown magnitudes as the only free parameters. As shown in \cref{fig:P+1-1_sharedphase}, this constraint resolves the mathematical ambiguity and the likelihood fit converges for all fit attempts and finds always a global minimum with amplitude values that match the true values, as expected.





\subsection{Challenges arising in data analysis}
\label{discussion_challenge}

The first challenge is selecting an optimal wave set, which is a difficult topic in real analyses. In \cref{sec:analysis}, we assumed that all the partial-wave amplitudes in the wave set are nonzero. However, if for a real analysis a wave set is chosen, for which certain partial-wave amplitudes are zero in the data, it will result in zero-value components in the vector $\mathbf{f}$ in \cref{eq:matrix_h=af} and hence, the rank of the matrix $\mathbf{A}$ would shrink. For example, even though we select an in principle ambiguity-free wave set like \{$P_{+1}^{\pm}, P_0^{\pm}, P_{-1}^{\pm}$\}, we may still obtain fit results with continuous ambiguity, if the data contain only \{$P_{+1}^{\pm}, P_{-1}^{\pm}$\} and $P_0^{\pm}\approx 0$ because all the components of $\mathbf{f}$ with $P_0^{\pm}$ become actually zero and the rank of $\mathbf{A}$ decreases from 10 to 5.

The second challenge is the presence of local minima close to the global minimum in more complex wave sets that do not suffer from mathematical ambiguity.  For example, the local minima observed in the fit results of the \{$S_0^{\pm},P_{+1}^{\pm}, P_0^{\pm}, P_{-1}^{\pm}$\} wave set studied in \cref{sec:SP-simulation} have only small NLL differences with respect to the global minimum, ranging from 10 to 140 units as shown in \cref{fig:SP}. Performing a similar study for the more comprehensive wave set $\{S_0^\pm, P_{+1,0,-1}^\pm, D_{+2,+1,0,-1,-2}^\pm\}$, we observe that this issue becomes even more severe. So, although larger wave sets are in most cases ambiguity free, it might become difficult in real analyses to distinguish the global minimum from local minima leading to multiple indistinguishable solutions with amplitude values that may deviate significantly from the true values.

Similarly, a decrease of the precision of the data sample may cause local minima to shift closer in likelihood to the true solution making them eventually indistinguishable, as shown in Ref.~\cite{JointPhysicsAnalysisCenter:2023gku}. Our studies of the \{$S_0^{\pm},P_{+1}^{\pm}, P_0^{\pm}, P_{-1}^{\pm}$\} wave set show that with smaller sample size, the likelihood surface become distorted such that the true parameter values correspond only to a local NLL minimum. The details are discussed in \cref{appen:stats}.

\begin{figure}[!ht]
\centering
\includegraphics[width=0.45\textwidth]{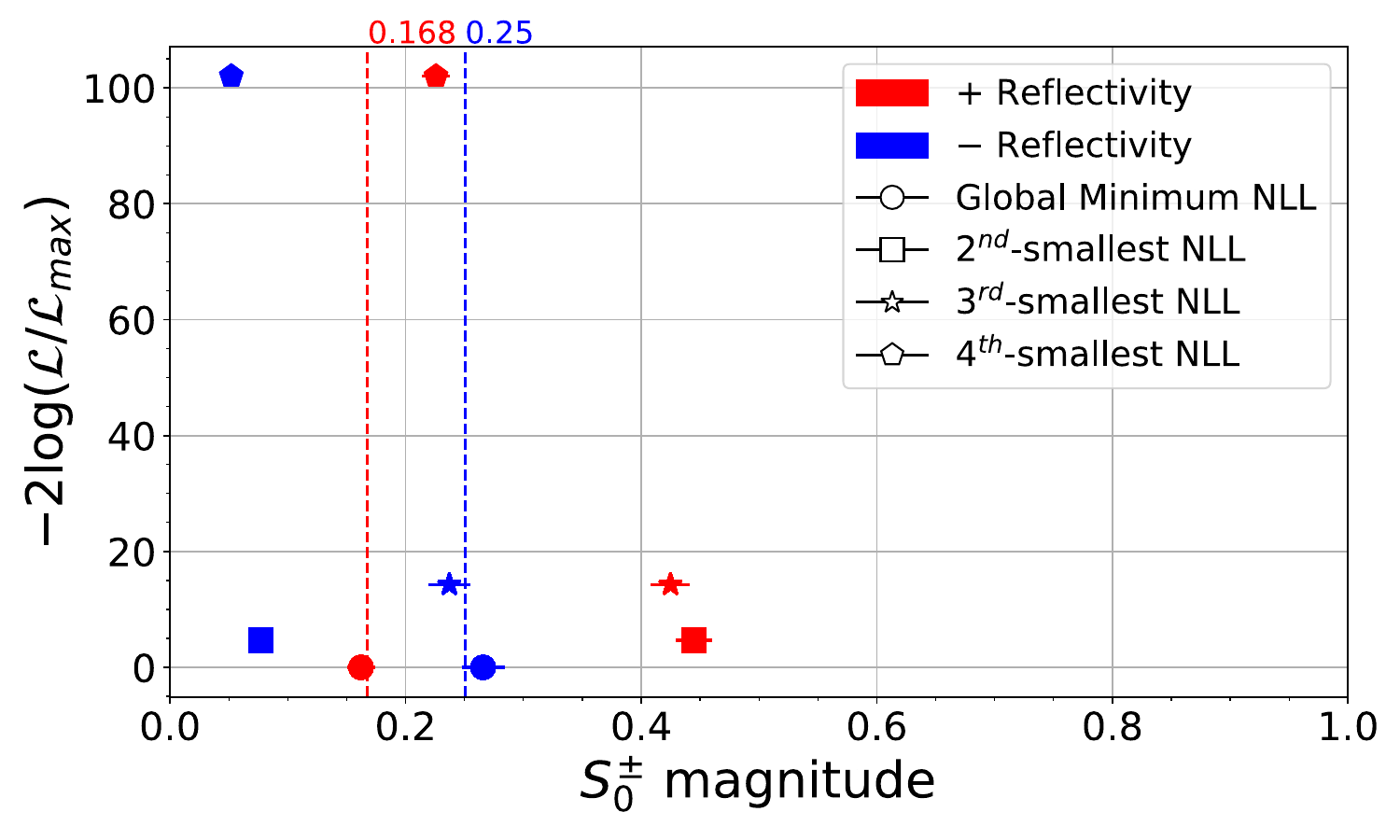}
\includegraphics[width=0.45\textwidth]{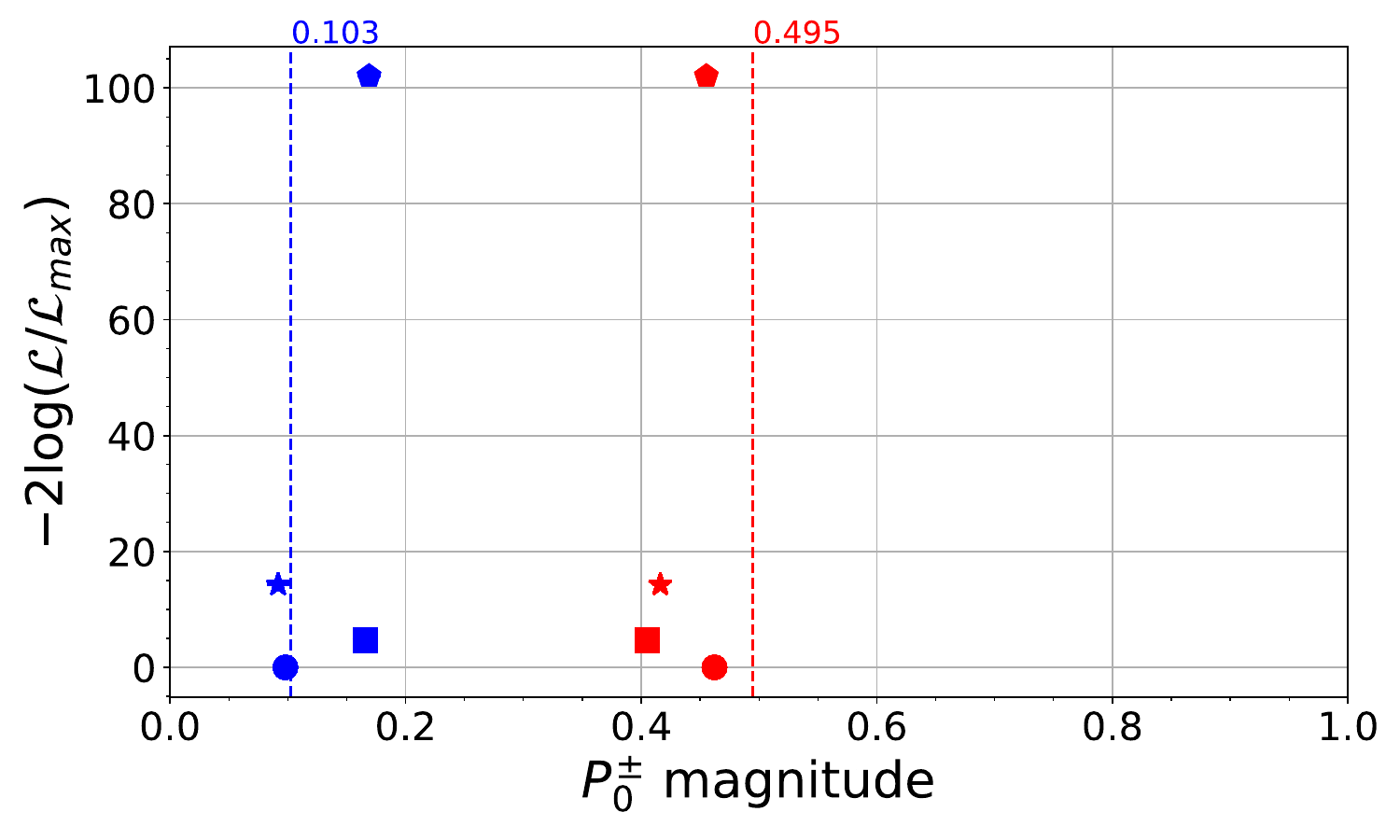}
\includegraphics[width=0.45\textwidth]{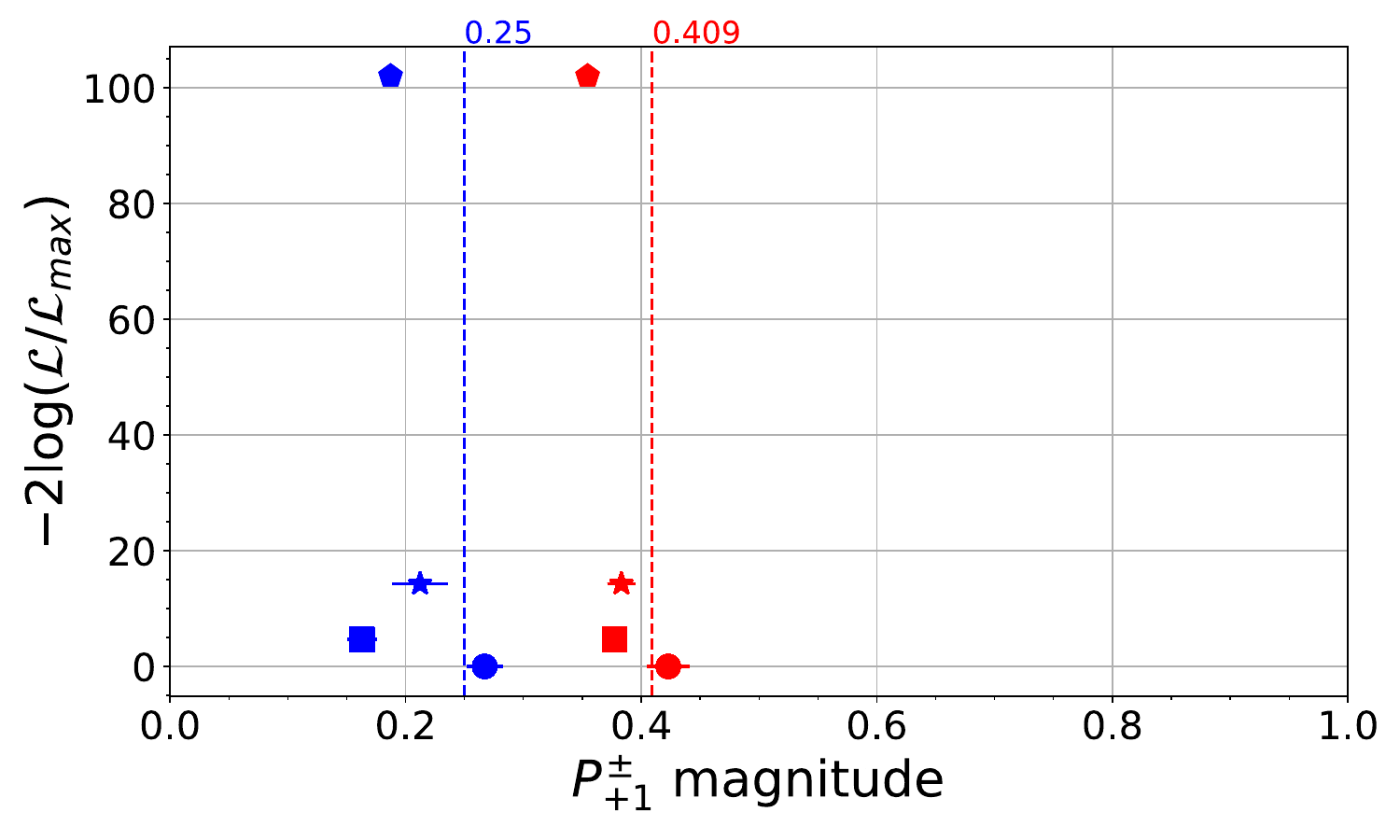}
\includegraphics[width=0.45\textwidth]{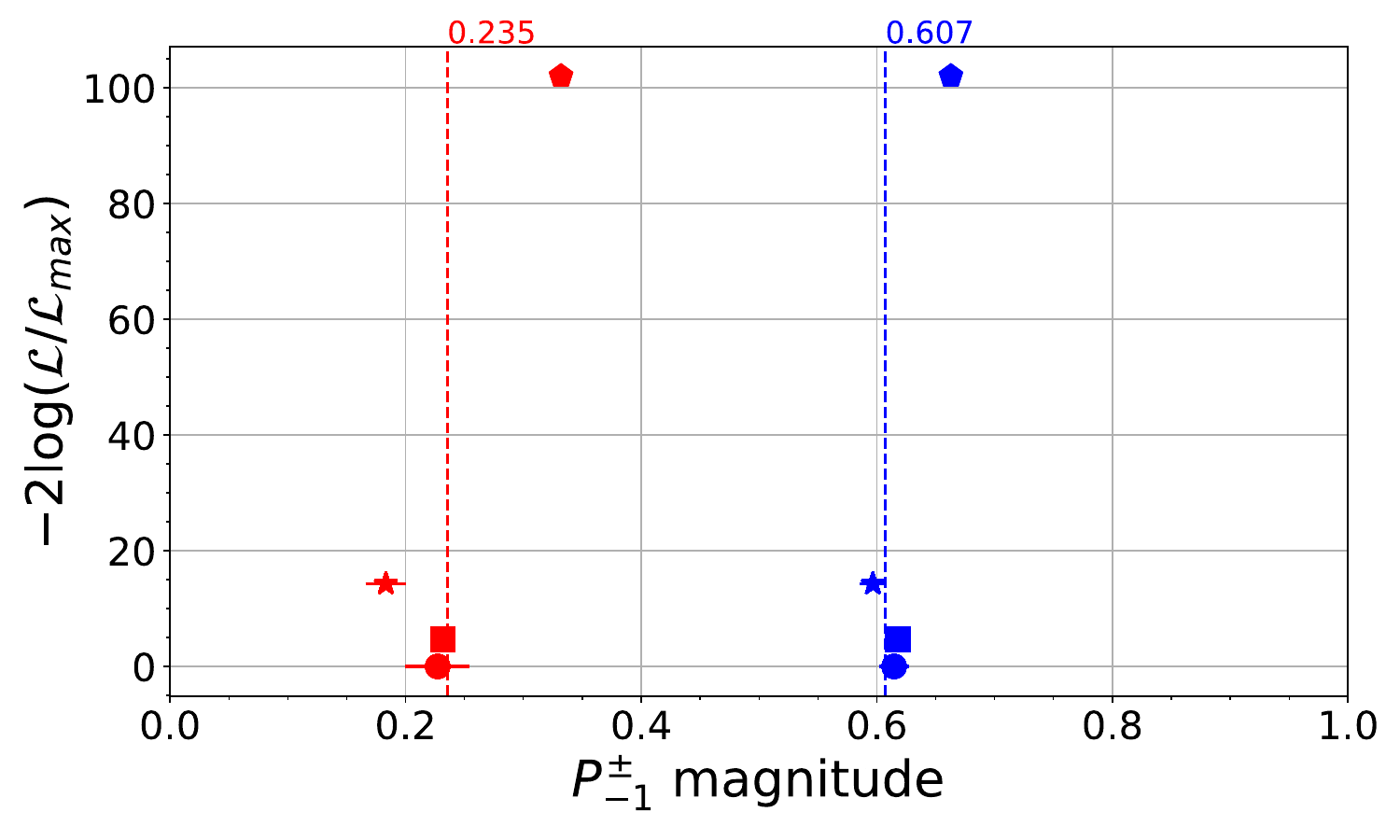}
\includegraphics[width=0.45\textwidth]{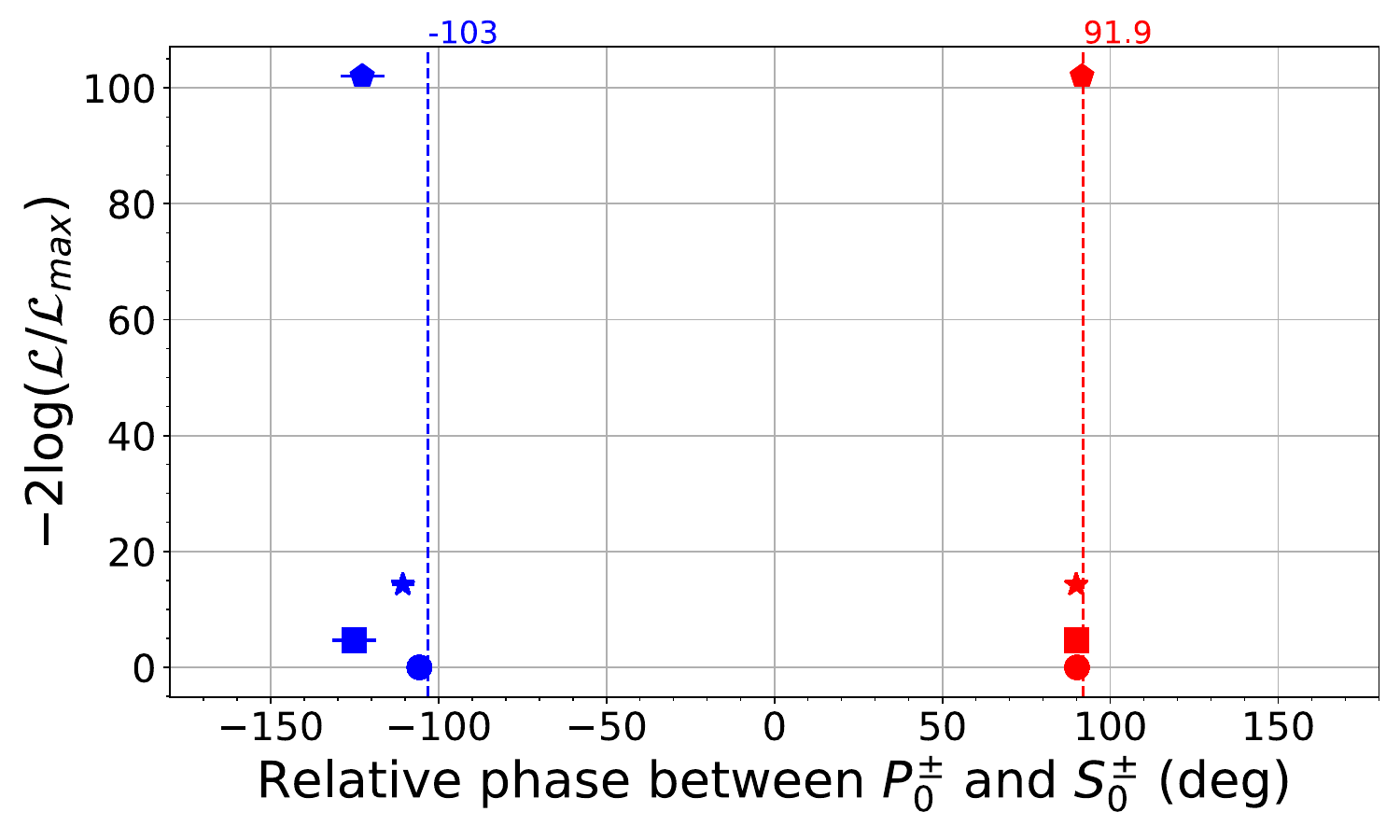}
\includegraphics[width=0.45\textwidth]{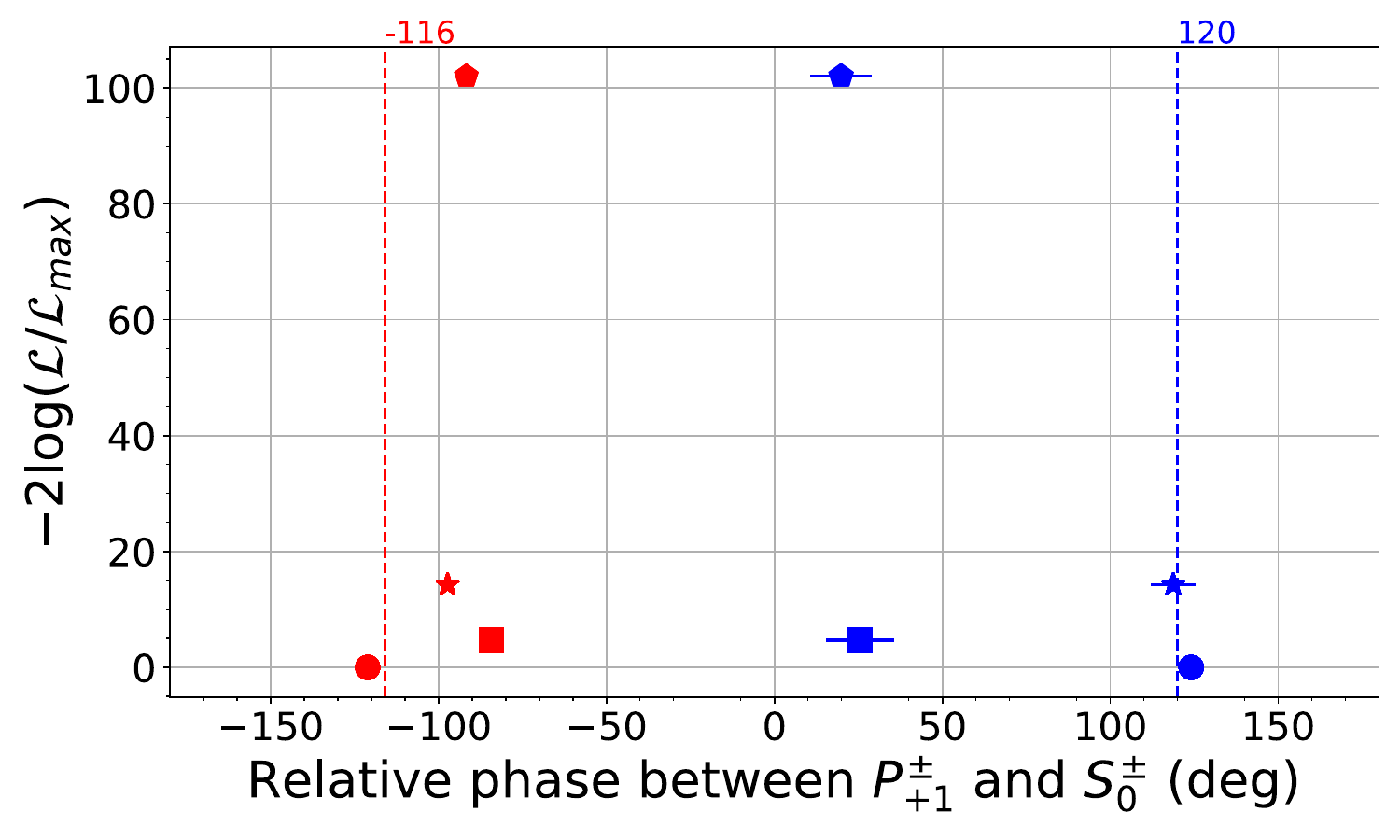}
\includegraphics[width=0.45\textwidth]{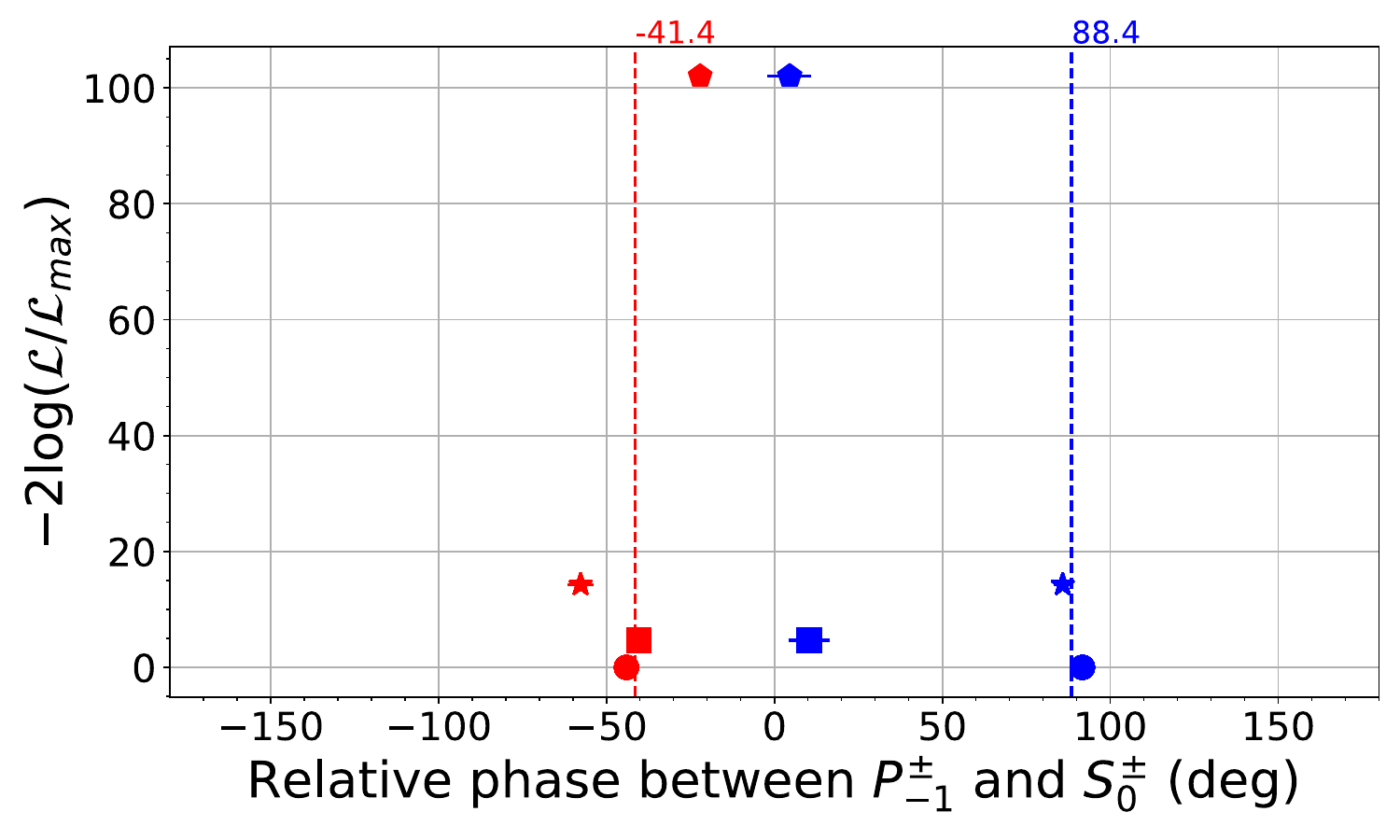}
\caption[]{\label{fig:SP_cut2degree} Similar to \cref{fig:SP}, but showing the 83 converged out of 100 attempted fits of $10^6$ before-cut events generated with the wave set \{$S_0^{\pm}, P_{-1}^{\pm}, P_{0}^{\pm}, P_{+1}^{\pm}$\} and where events with $\theta_{\text{lab}}^{\text{track}}<2^{\circ}$ are removed.}
\end{figure}

Finally, the effect of detector acceptance and resolution in real analyses can also affect the results. To study the potential effect of the detector acceptance, we model the detector hole along the beam line, by removing all events, in which any final-state particle has a polar angle in the lab frame of $\theta_{\text{lab}}^{\text{track}} < 2^{\circ}$, from the pseudo-data used to study the \{$S_0^{\pm},P_{+1}^{\pm}, P_0^{\pm}, P_{-1}^{\pm}$\} wave set in \cref{sec:SP-simulation}. In the log-likelihood fits the acceptance is accounted for using phase-space distributed Monte Carlo events with the same selection cut applied. The 83 converged fit results are shown in \cref{fig:SP_cut2degree}. Again, the sign of trivial ambiguities of the phases is flipped in the plots. Although the partial-wave model is ambiguity-free, the global-minimum NLL solution deviates slightly from the true values, particularly for $P_0^+$. Moreover, the gap in NLL between the global minimum and the local minima is smaller (cf. \cref{fig:SP}). These observations highlight potential challenges in real analyses, where detector effects and limited acceptance can lead to multiple indistinguishable solutions. In such cases, careful modeling of the detector response in the Monte Carlo simulations becomes crucial to mitigate these effects and improve the robustness of partial-wave analyses.

\clearpage
\section{Conclusions}
\label{sec:conclusion}
We studied mathematical ambiguities in the partial-wave analysis of two pseudoscalar mesons produced in linearly polarized photoproduction when both reflectivities are populated. Expressing the moments of the intensity distribution in terms of bilinears of the partial-wave amplitudes, we determined for a given wave set the number of linearly independent equations that constrain the unknown partial-wave amplitudes. We demonstrated that for simple wave sets that include both reflectivities like \{$P_{+1}^{\pm}, P_{-1}^{\pm}$\}, the partial-wave amplitudes are underdetermined resulting in a continuous mathematical ambiguity. Adding more non-zero partial waves into the wave set usually breaks the ambiguity. We performed input-output studies of partial-wave analyses on pseudo-data sets generated for wave sets with and without continuous ambiguity confirming the calculations. In addition, we discussed what information about amplitudes can be obtained when a continuous ambiguity occurs and how external constraints on relative phases could resolve the ambiguity. However, ambiguities may still arise even though an in principle non-ambiguous wave set is used, if certain partial waves are close to zero the data. Furthermore, we find that larger wave sets may exhibit local minima of the negative log-likelihood function. Depending on the precision of the data and the detector acceptance, these local minima may lead to indistinguishable solutions with amplitude values that deviate significantly from the true values.

\begin{acknowledgments}
The authors appreciate a useful discussion with Dr. Vincent Mathieu. This work was supported by the U.S. Department of Energy, Office of Science, Office of Nuclear Physics under contract No.\ DE-FG02-87ER40315, No.\ DE-SC0023978, No.\ DE-FG02-92ER40735, and No. DE-AC05-06OR23177. The work of A.A. was supported by the DOE, Office of Science, Office of Nuclear Physics in the Early Career Program.

\end{acknowledgments}

\appendix
\section{Expression of moments in terms of \cref{eq:PairOfAmplitudes}}
\label{appen:pair_amplitude}
For $H_0(L,M)$, we can reorder the terms in the sum in \cref{eq:moments0}:
\begin{equation}
\begin{aligned}
    H_0(L,M) &= \sum_{\substack{\ell m \\ \ell' m'}}^{\infty} C_{\ell \, m;\ell' \, m'}^{L \, M}\sum_{\epsilon=\pm} [\ell]_m^{(\epsilon)}[\ell']_{m'}^{(\epsilon)*}  \\
    &+\sum_{\substack{\ell m \\ \ell' m'}}^{\infty} C_{\ell \, m;\ell' \, m'}^{L \, M}  \sum_{\epsilon=\pm} (-1)^{m-m'} [\ell]_{-m}^{(\epsilon)}[\ell']_{-m'}^{(\epsilon)*} .
\end{aligned}
\end{equation}
In the second term in on the right-hand side, we interchange $\ell$ and $\ell'$ as well as $m$ and $-m'$, so that this term becomes
\begin{equation}
    \sum_{\substack{\ell m \\ \ell' m'}}^{\infty} C_{\ell' \, -m';\ell \, -m}^{L \, M}  \sum_{\epsilon=\pm} (-1)^{-m+m'} [\ell']_{m'}^{(\epsilon)}[\ell]_{m}^{(\epsilon)*} .
\label{eq:second_term_moment0}
\end{equation}
Applying the symmetry property of the Clebsch-Gordan coefficients, 
\begin{equation}
    \langle j_1 m_1 j_2 m_2 | J M\rangle = (-1)^{j_2+m_2} \sqrt{\frac{2J+1}{2j_1 + 1}} \langle J (-M) j_2 m_2 | j_1 (-m_1) \rangle ,
    \label{eq:cg_relation}
\end{equation}
\cref{eq:second_term_moment0} becomes 

\begin{equation}
    \sum_{\substack{\ell m \\ \ell' m'}}^{\infty}  (-1)^{2L+M} C_{\ell \, m;\ell' \, m'}^{L \, M} \sum_{\epsilon=\pm} (-1)^{-m+m'} [\ell']_{m'}^{(\epsilon)}[\ell]_{m}^{(\epsilon)*} .
\end{equation}
Since the Clebsch-Gordan coefficients in $C_{\ell \, m;\ell' \, m'}^{L \, M}$ vanish unless $M+m' = m$, we get
\begin{equation}
    H_0(L,M) = \sum_{\substack{\ell m \\ \ell' m'}}^{\infty} C_{\ell \, m;\ell' \, m'}^{L \, M} \sum_{\epsilon=\pm}\bigg( [\ell]_m^{(\epsilon)}[\ell']_{m'}^{(\epsilon)*} + [\ell]_{m}^{(\epsilon)*} [\ell']_{m'}^{(\epsilon)}  \bigg) .
\label{eq:h0_appendix}
\end{equation}
The term in parentheses has the same form as the left-hand side of \cref{eq:PairOfAmplitudes}.

Interchanging $\ell$ and $\ell'$ as well as $m$ and $-m'$ in the expression for $H_1(L,M)$ in \cref{eq:moments1} only reorders the terms in the sums and consequently leaves the value of $H_1(L,M)$ unchanged. Hence, we can add the original expression and the one with interchanged indices and compensate by dividing by 2:
\begin{equation}
\begin{aligned}
    H_1(L,M) = \sum_{\substack{\ell m \\ 
    \ell' m'}}^{\infty} \sum_{\epsilon=\pm} \frac{\epsilon}{2} \bigg[
    &C_{\ell \, m;\ell' \, m'}^{L \, M}   \ \bigg( (-1)^m [\ell]_{-m}^{(\epsilon)}[\ell']_{m'}^{(\epsilon)*} + (-1)^{m'} [\ell]_{m}^{(\epsilon)}[\ell']_{-m'}^{(\epsilon)*} \bigg) +\\
    &C_{\ell' \, -m';\ell \, -m}^{L \, M}  \ \bigg( (-1)^{-m'} [\ell']_{m'}^{(\epsilon)}[\ell]_{-m}^{(\epsilon)*} + (-1)^{-m} [\ell']_{-m'}^{(\epsilon)}[\ell]_{m}^{(\epsilon)*} \bigg)\bigg] .
\end{aligned}
\end{equation}
Using \cref{eq:cg_relation} and $M+m' = m$, we obtain
\begin{equation}
\begin{aligned}
    H_1(L,M) = \sum_{\substack{\ell m \\ \ell' m'}}^{\infty} C_{\ell \, m;\ell' \, m'}^{L \, M} \sum_{\epsilon=\pm} \frac{\epsilon}{2} \ \bigg[ &(-1)^m \bigg( [\ell]_{-m}^{(\epsilon)}[\ell']_{m'}^{(\epsilon)*} + [\ell]_{-m}^{(\epsilon)*}[\ell']_{m'}^{(\epsilon)}\bigg)\\ 
    + &(-1)^{m'} \bigg( [\ell]_{m}^{(\epsilon)}[\ell']_{-m'}^{(\epsilon)*} + [\ell]_{m}^{(\epsilon)*}[\ell']_{-m'}^{(\epsilon)}\bigg ) \bigg] ,
\label{eq:h1_appendix}
\end{aligned}
\end{equation}
which is again a linear combination of terms in the form as the left-hand side of \cref{eq:PairOfAmplitudes}.

Since $H_2(L,M)$ has the same structure as $H_1(L,M)$, we can apply the same method and write $H_2(L,M)$ as
\begin{equation}
\begin{aligned}
    H_2(L,M) = i\sum_{\substack{\ell m \\ \ell' m'}}^{\infty} C_{\ell \, m;\ell' \, m'}^{L \, M} \sum_{\epsilon=\pm} \frac{\epsilon}{2} \ \bigg[ &(-1)^m \bigg( [\ell]_{-m}^{(\epsilon)}[\ell']_{m'}^{(\epsilon)*} + [\ell]_{-m}^{(\epsilon)*}[\ell']_{m'}^{(\epsilon)}\bigg) \\ - &(-1)^{m'} \bigg( [\ell]_{m}^{(\epsilon)}[\ell']_{-m'}^{(\epsilon)*} + [\ell]_{m}^{(\epsilon)*}[\ell']_{-m'}^{(\epsilon)}\bigg ) \bigg] .
\label{eq:h2_appendix}
\end{aligned}
\end{equation}

\section{Fit results for the $S_0^\pm$ and $P_{+1,0,-1}^\pm$ wave set with \texttt{MINOS} error estimation}
\label{appen:SP_minos}
In \cref{sec:SP-simulation}, we presented the \texttt{MIGRAD} fit results for the $\{ S_0^\pm, P_{+1}^\pm, P_{0}^\pm, P_{-1}^\pm \}$ wave set. We noted that the fourth-smallest NLL solution exhibits inflated uncertainties compared to others. To further investigate these uncertainties, we employ the \texttt{MINOS} algorithm within \texttt{MINUIT}. \texttt{MINOS} provides a more robust error estimation by accounting for a potential non-parabolic shape of the NLL around the minimum.

\begin{figure}[H]
\centering
\includegraphics[width=0.45\textwidth]{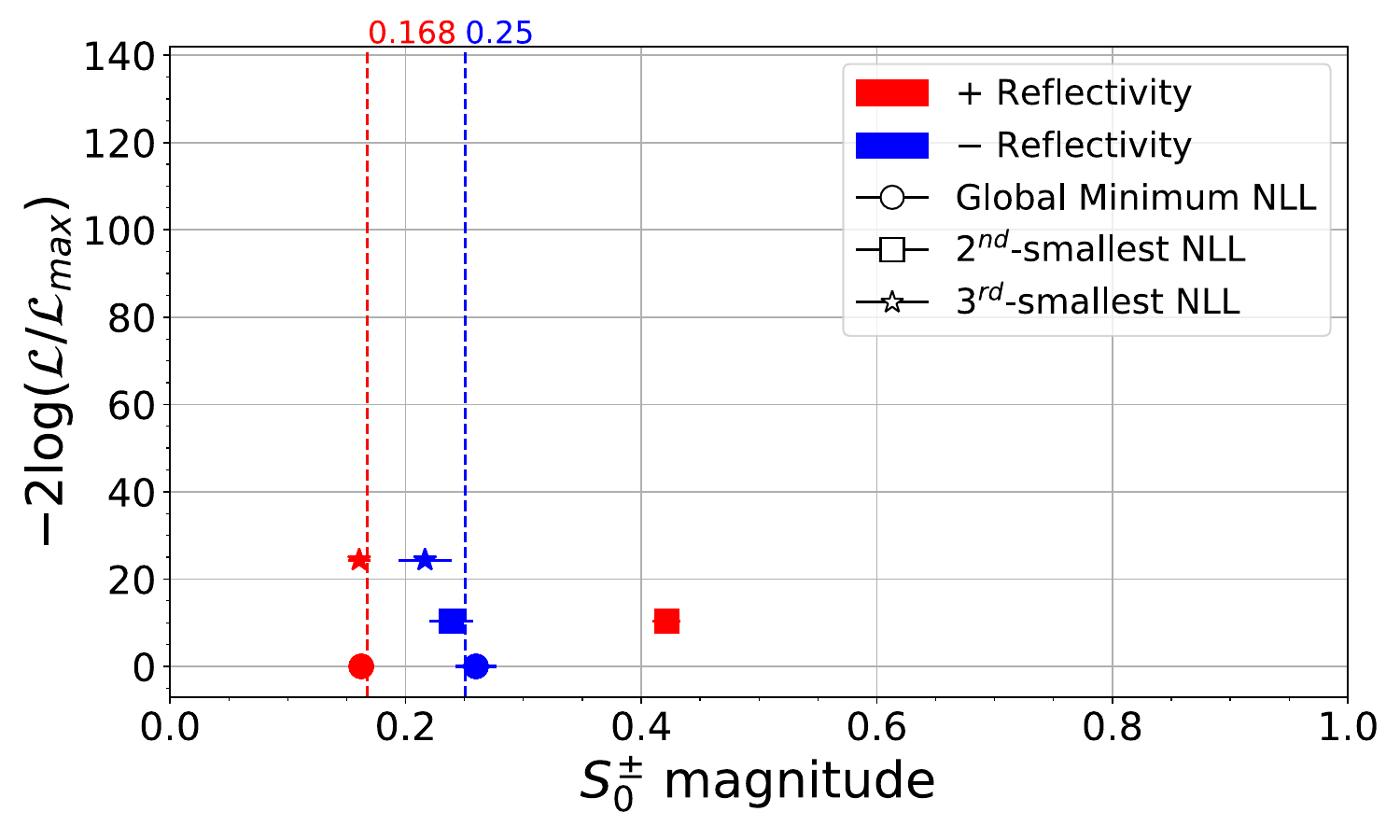}
\includegraphics[width=0.45\textwidth]{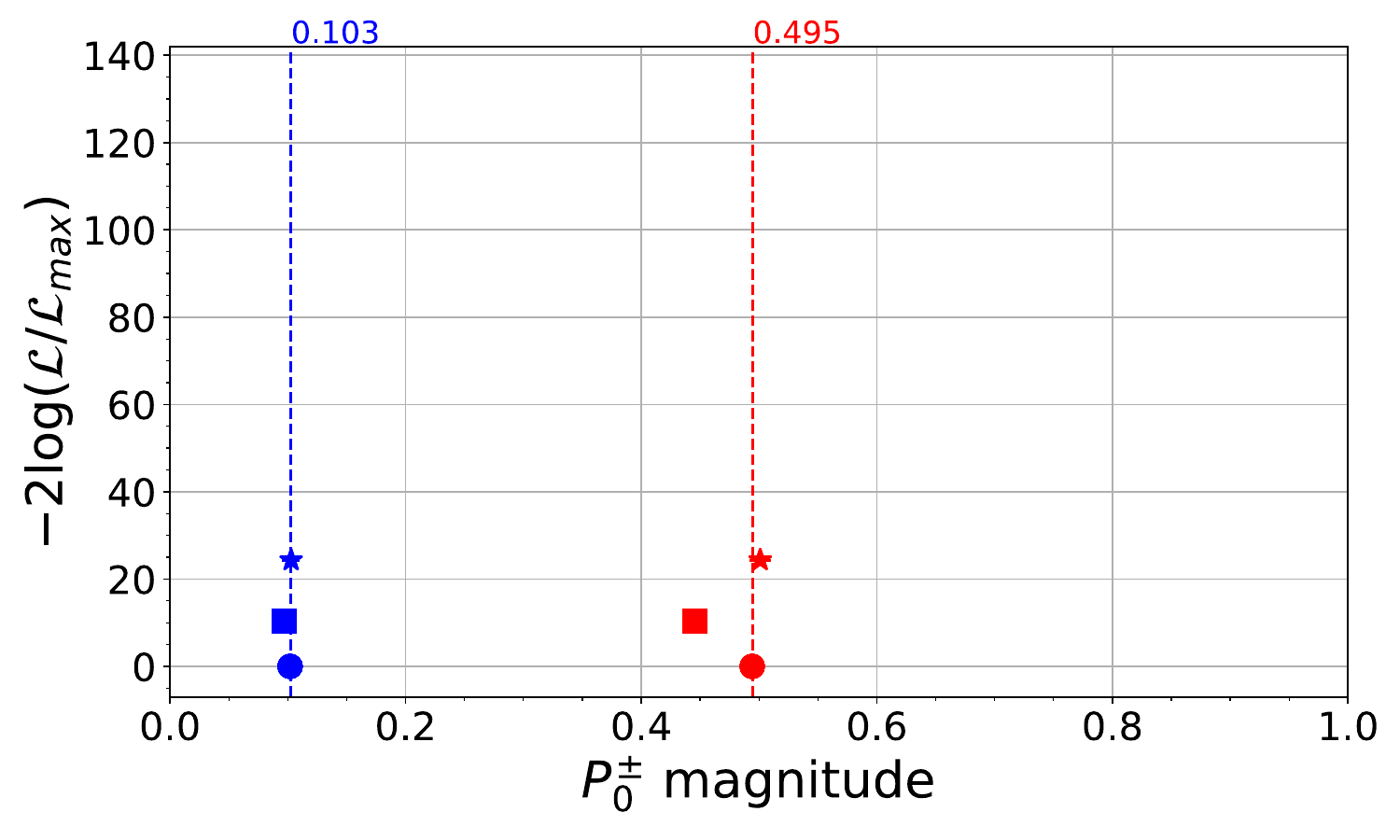}
\includegraphics[width=0.45\textwidth]{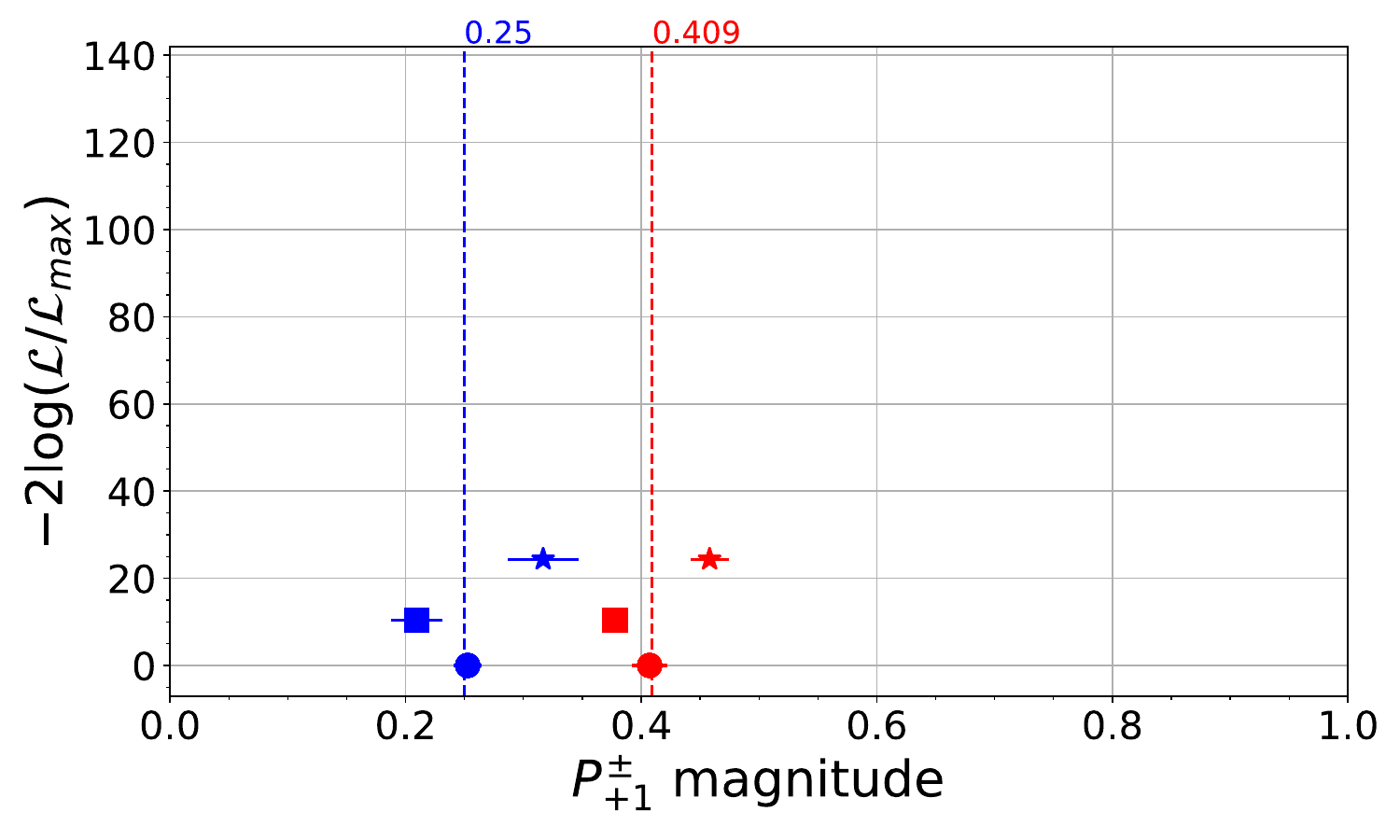}
\includegraphics[width=0.45\textwidth]{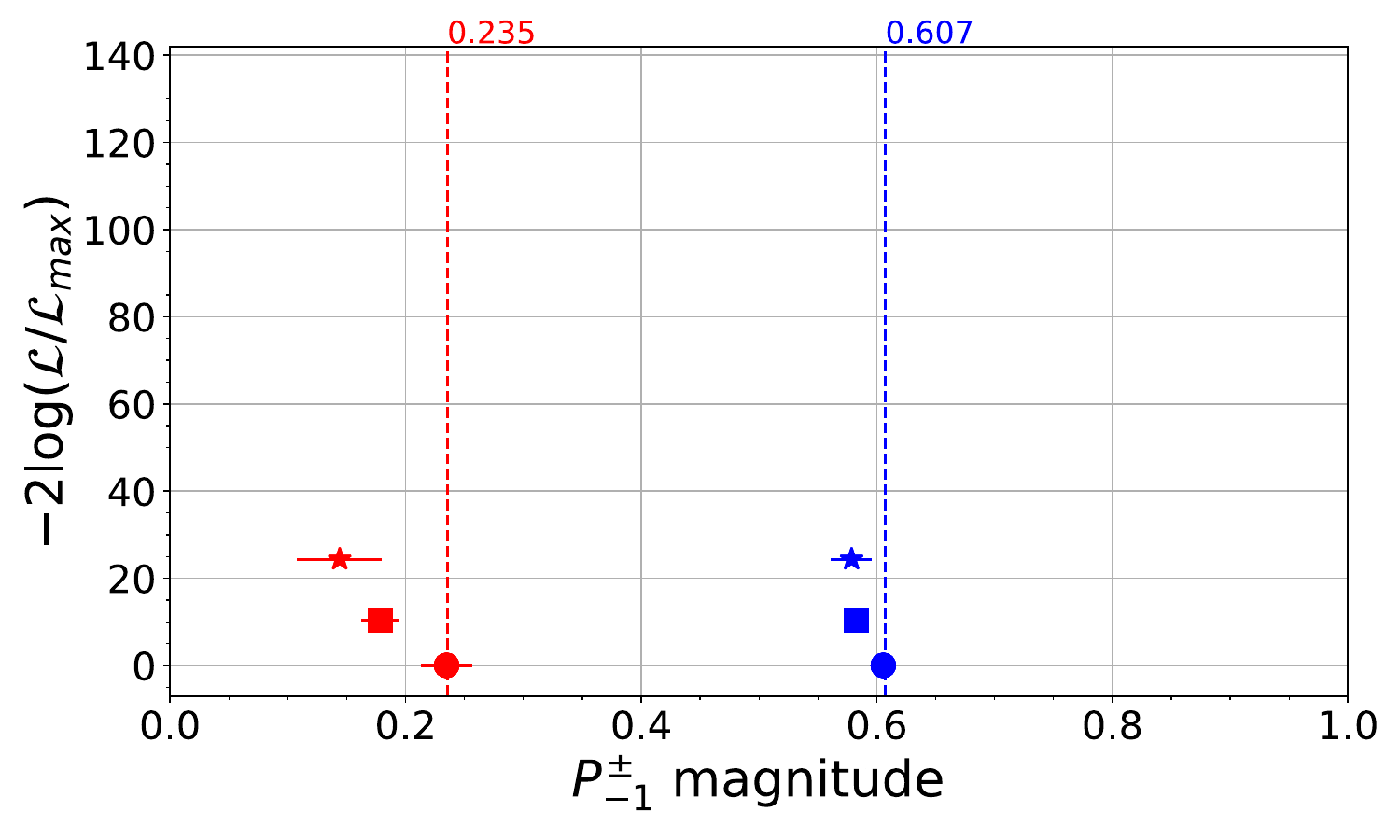}
\includegraphics[width=0.45\textwidth]{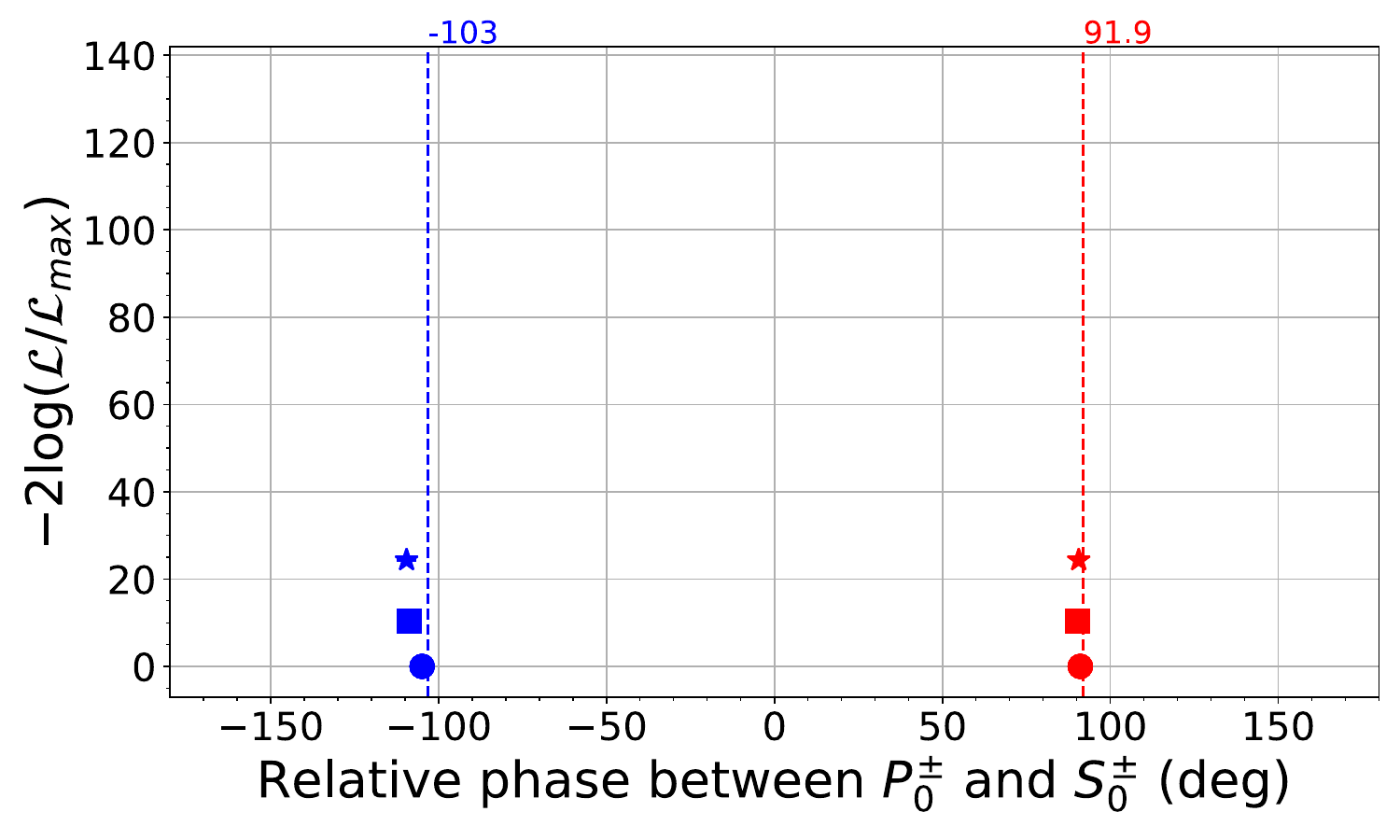}
\includegraphics[width=0.45\textwidth]{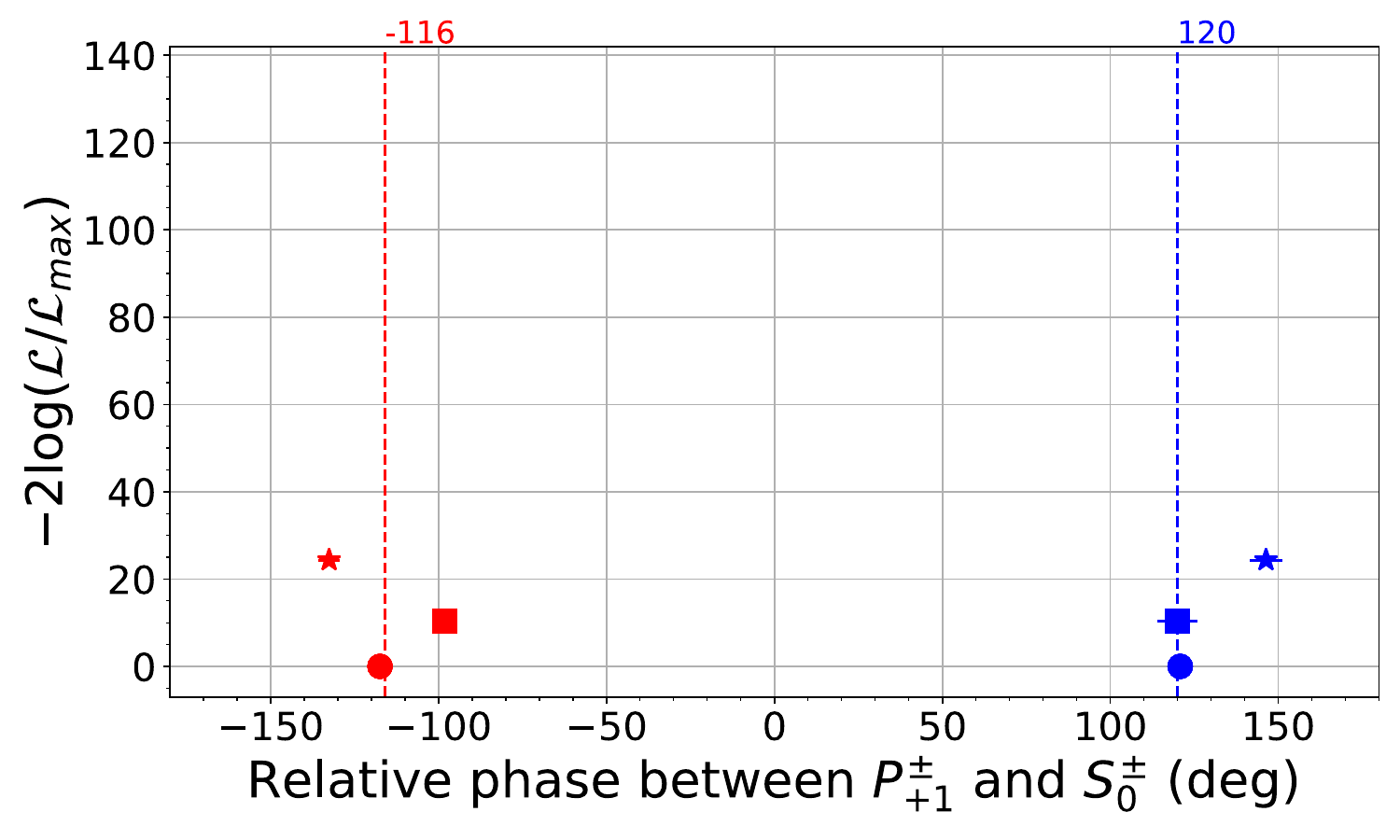}
\includegraphics[width=0.45\textwidth]{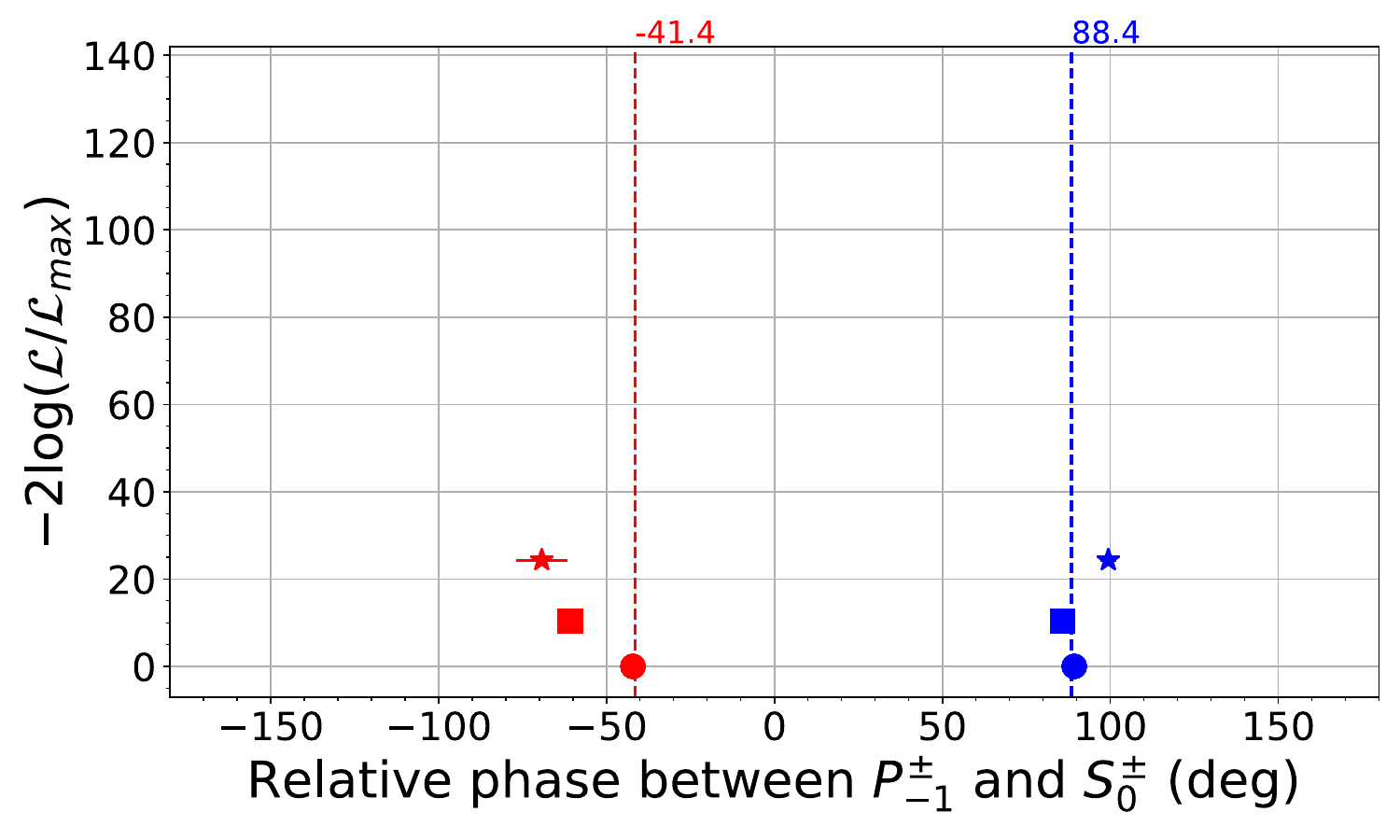}
\caption[]{\label{fig:SP_minos}Similar to \cref{fig:SP}, but showing the 18 converged out of 100 attempted fits of $10^6$ events generated with the wave set \{$S_0^{\pm}, P_{+1}^{\pm}, P_{0}^{\pm}, P_{-1}^{\pm}$\}, performed using random start values and \texttt{MINOS} error estimation.}
\end{figure}

\cref{fig:SP_minos} shows the \texttt{MINOS} fit results for the same pseudo-dataset and wave set, using the same vertical scale as in \cref{fig:SP}. The sign of trivial phase ambiguous solutions has been flipped. In contrast to the \texttt{MIGRAD} results, the \texttt{MINOS} error estimation was successful for only 18 fits out of 100 fit attempts and 50\% of these fits converged to the global minimum identified in \cref{sec:SP-simulation}. Comparing with \cref{fig:SP}, the positions of the minima remain nearly unchanged, while the uncertainties associated with the fourth-smallest NLL are reduced but still significantly larger than for the other solutions. Notably, the third- and the fifth-smallest NLL solutions, vanished, but the remaining local minima are consistent with \cref{fig:SP}. Overall, the \texttt{MINOS} analysis reinforces our earlier conclusion that the $\{ S_0^\pm, P_{+1}^\pm, P_{0}^\pm, P_{-1}^\pm \}$ wave set is free from mathematical ambiguities.

\section{Dependence of fit results on sample size}
\label{appen:stats}
\begin{figure}[!h]
\centering
\includegraphics[width=0.45\textwidth]{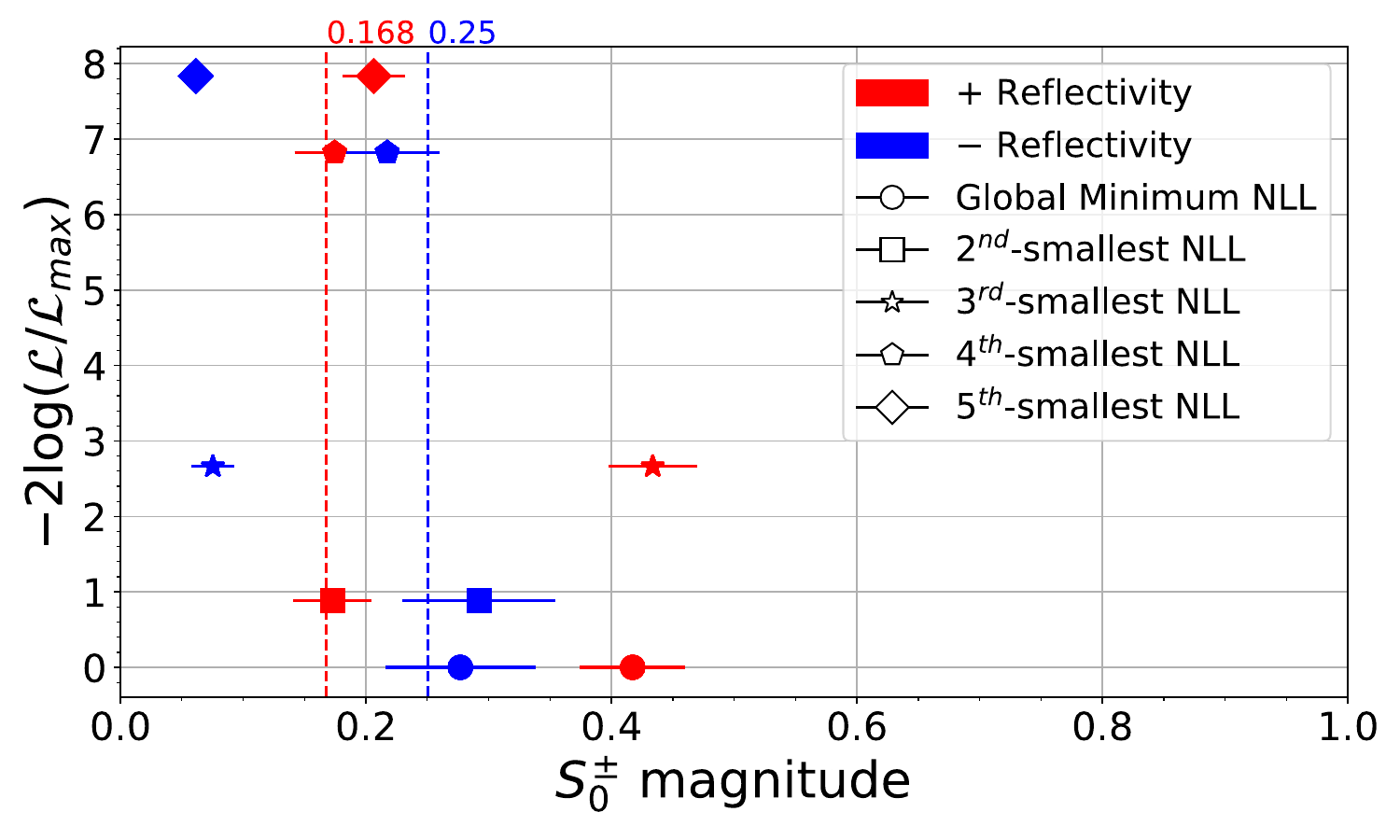}
\includegraphics[width=0.45\textwidth]{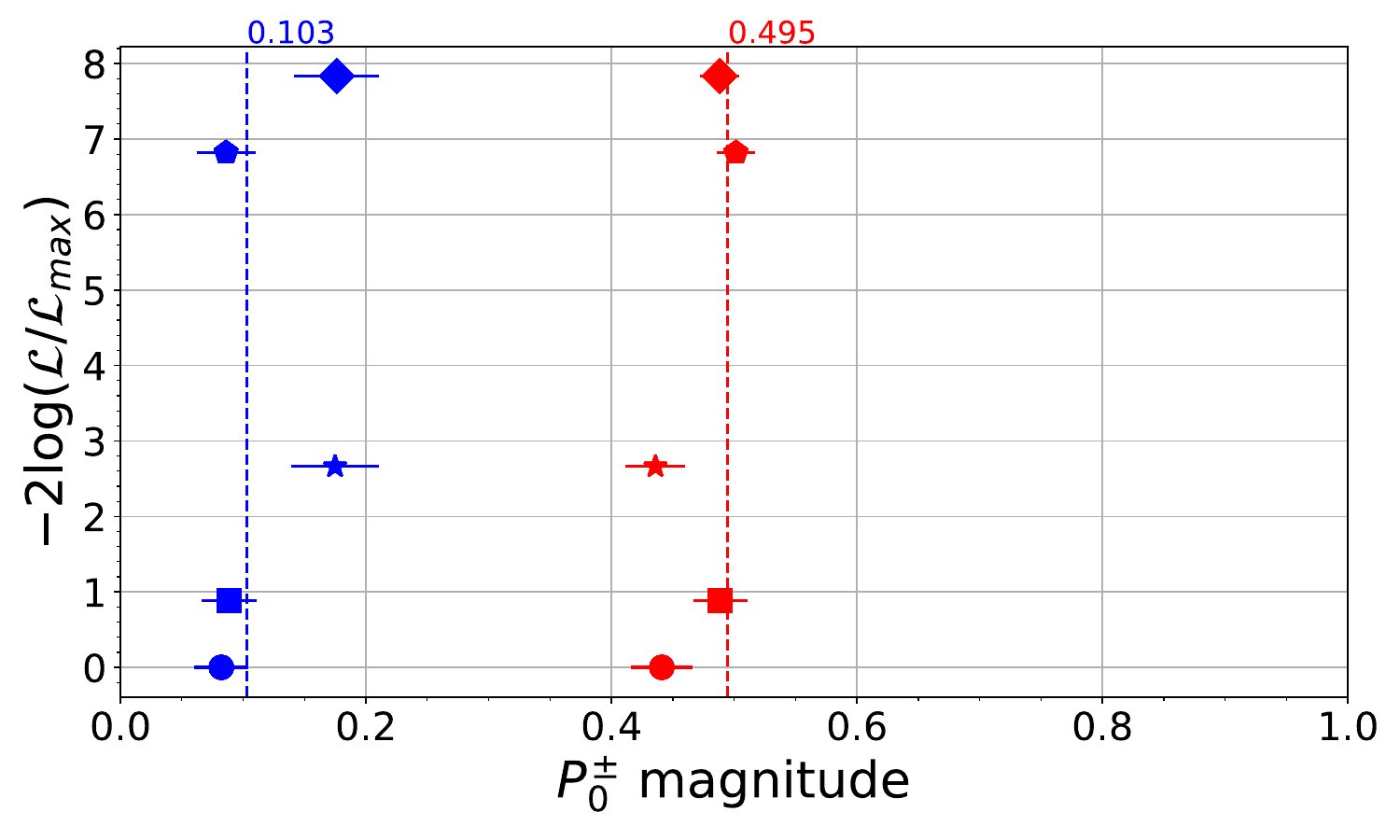}
\includegraphics[width=0.45\textwidth]{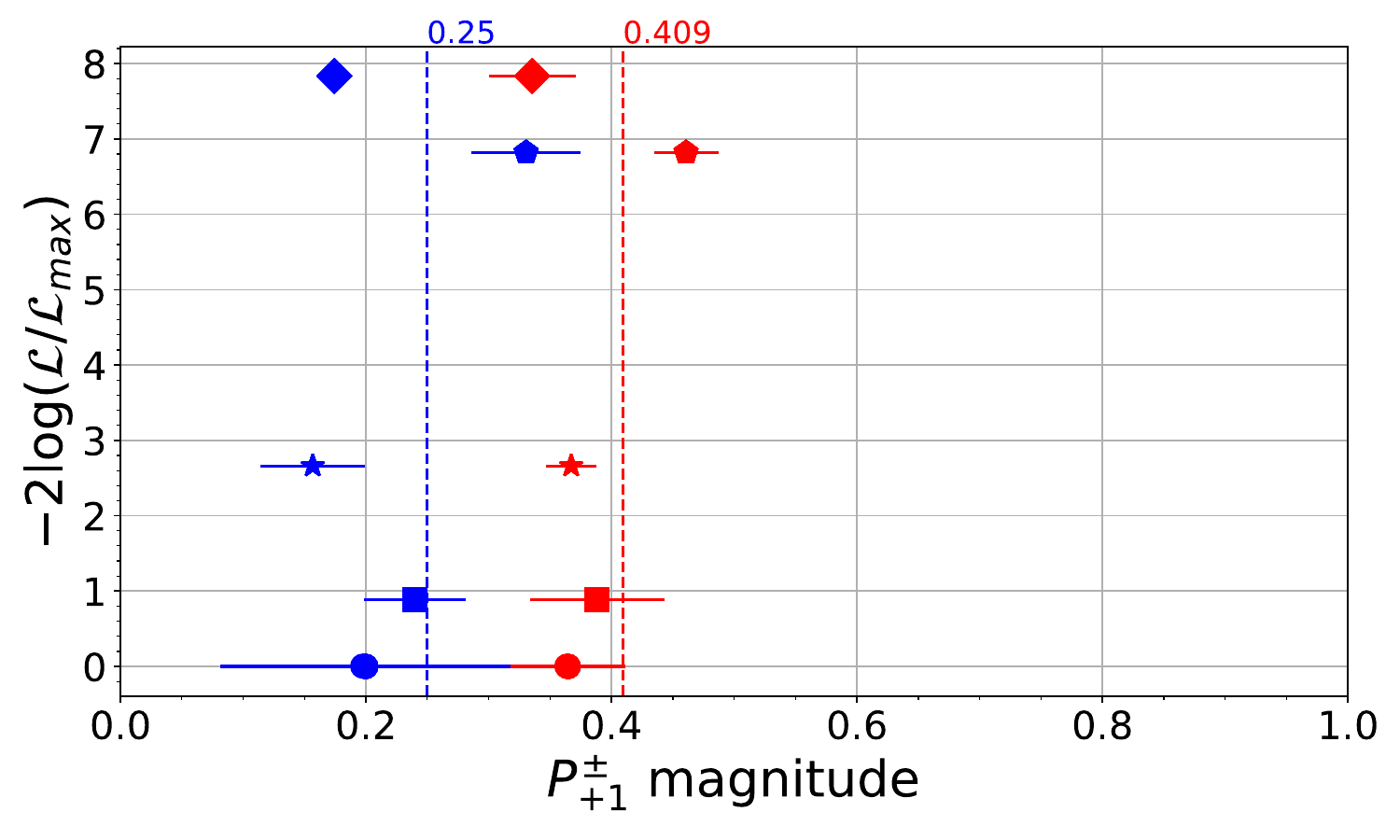}
\includegraphics[width=0.45\textwidth]{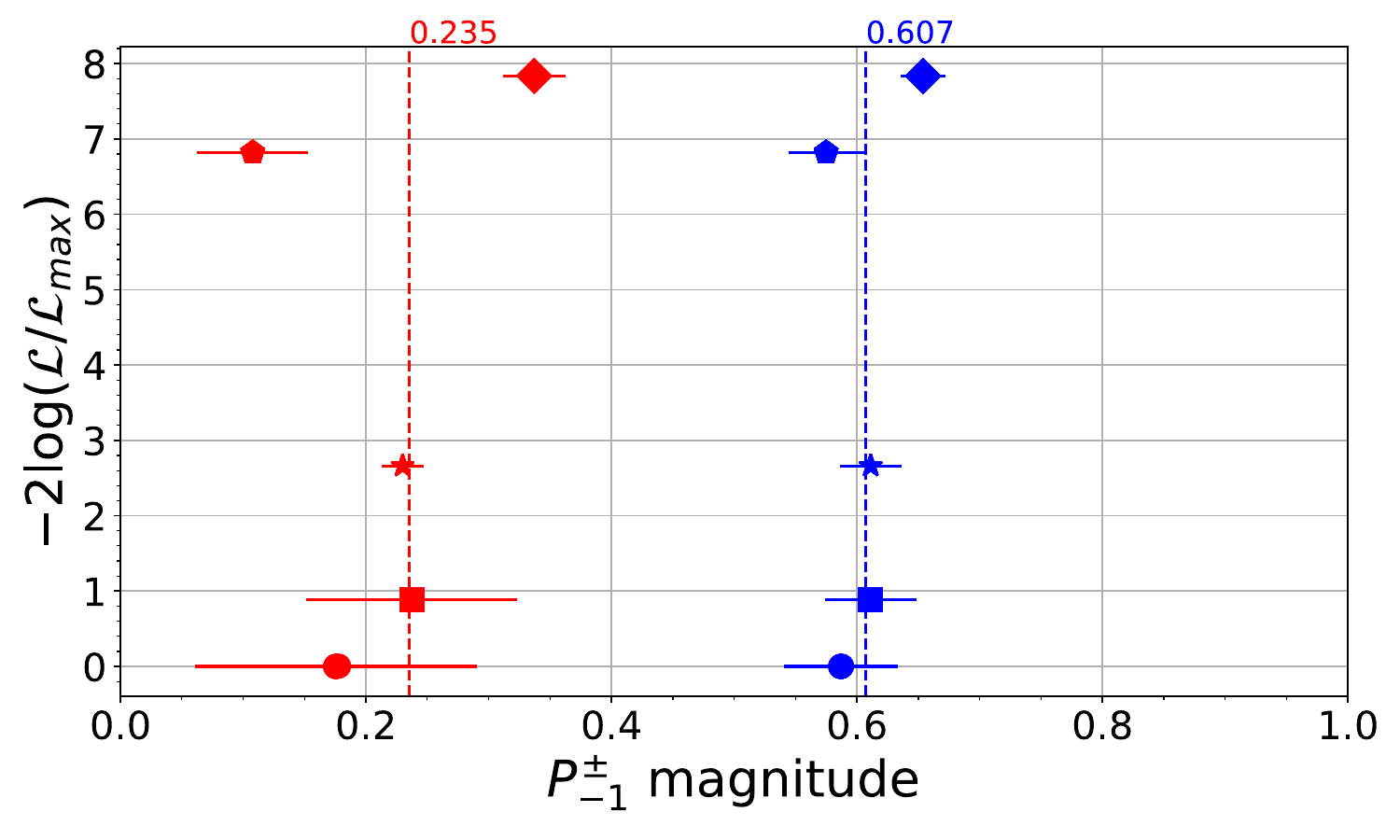}
\includegraphics[width=0.45\textwidth]{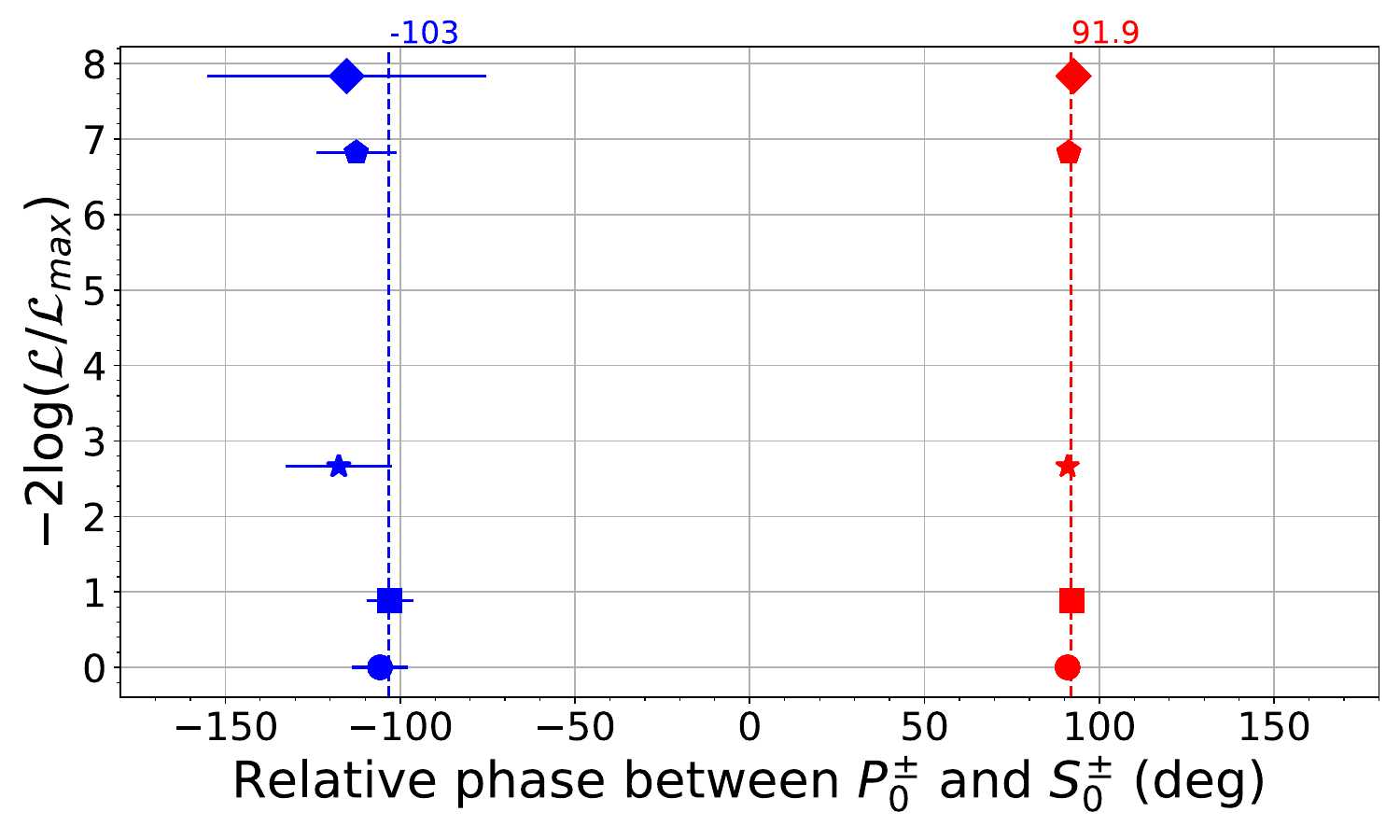}
\includegraphics[width=0.45\textwidth]{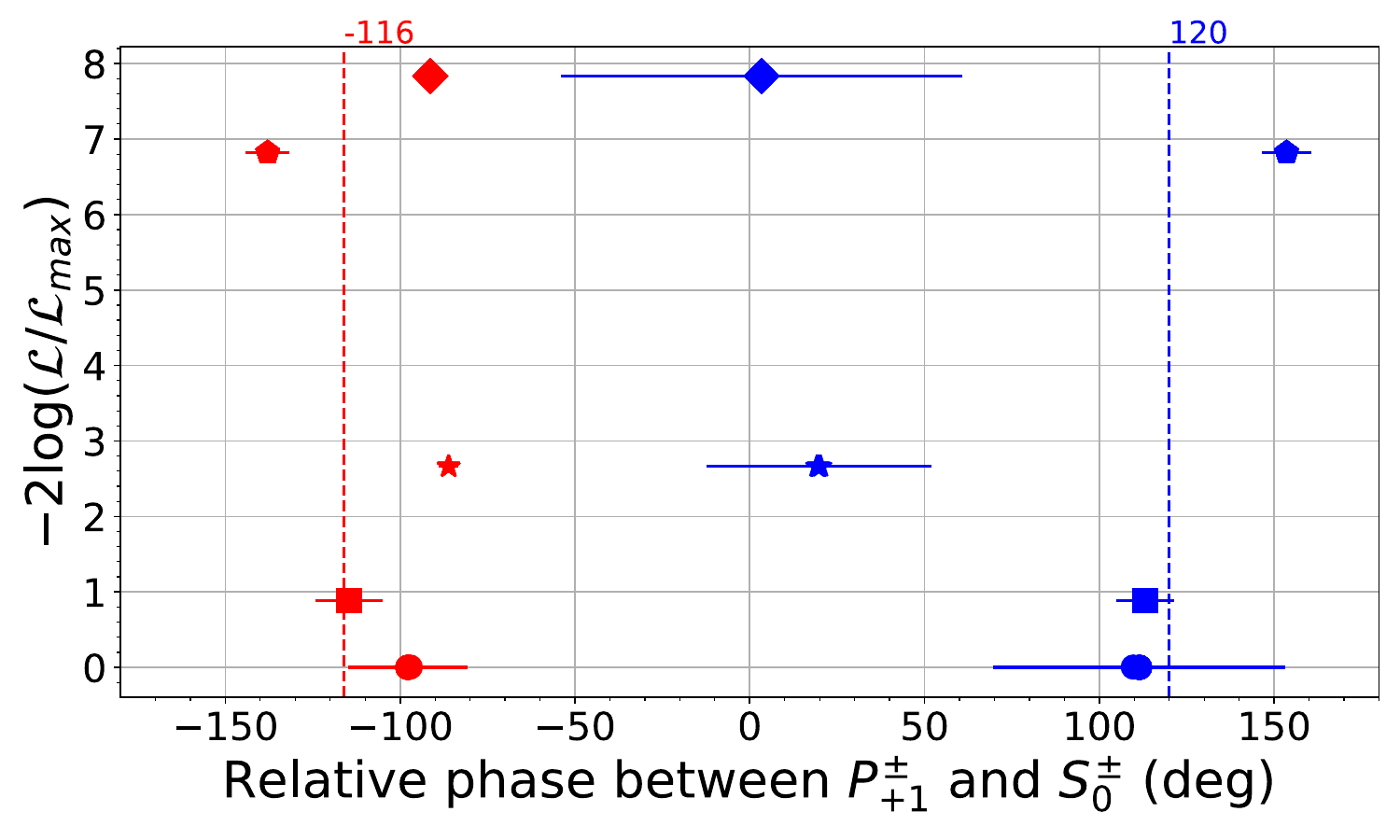}
\includegraphics[width=0.45\textwidth]{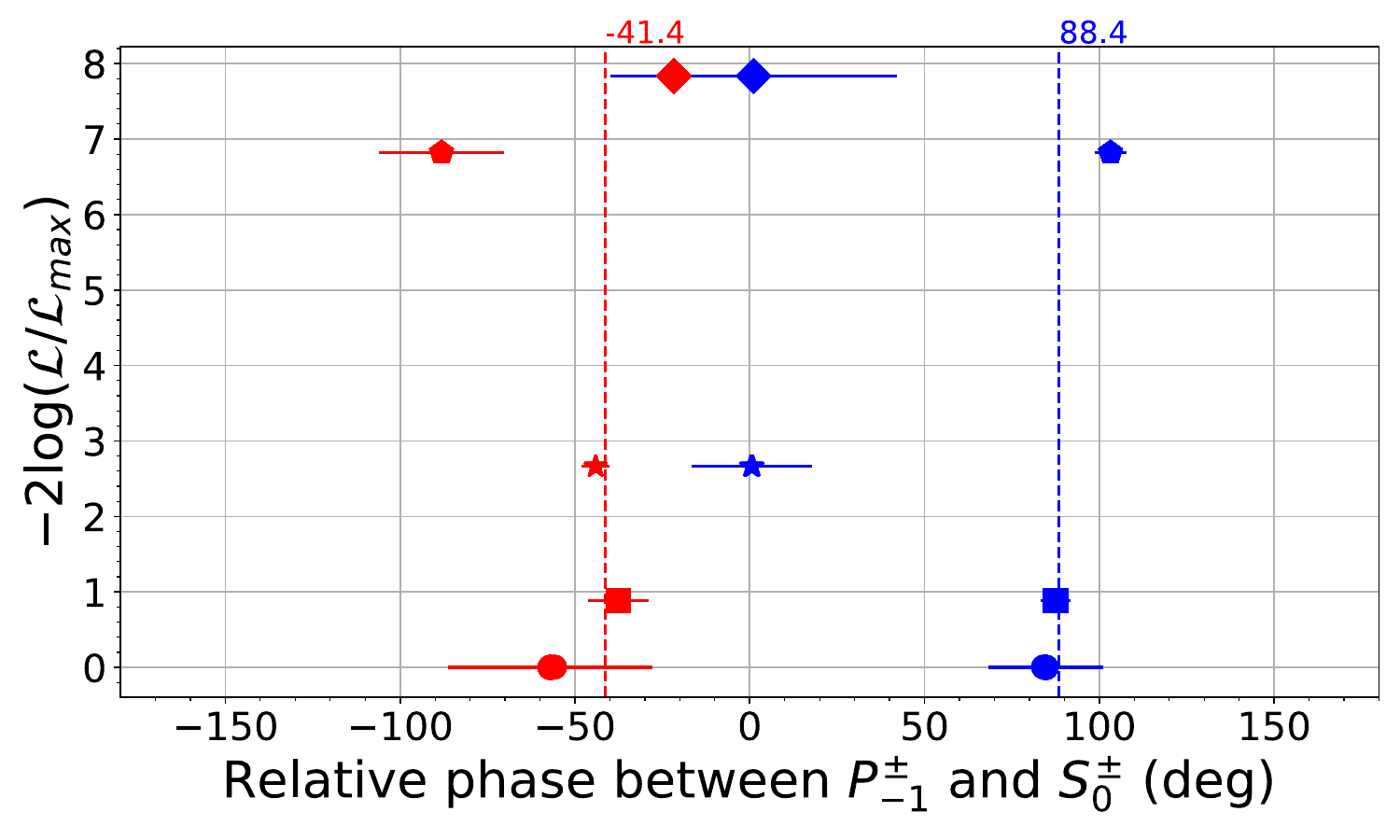}
\caption[]{\label{fig:SP_100k} Similar to \cref{fig:SP}, but showing the 93 converged out of 100 attempted fits of $10^5$ events generated with the wave set \{$S_0^{\pm}, P_{+1}^{\pm}, P_{0}^{\pm}, P_{-1}^{\pm}$\}, performed using random start values.}
\end{figure}

To study the dependence of the fit results on the size of the analyzed data sample, we generate a smaller sample of $10^5$ events using the same amplitude values as in \cref{sec:SP-simulation} for the wave set $\{ S_0^\pm, P_{+1}^\pm, P_0^\pm, P_{-1}^\pm \}$. The results from the 93 converged fits, performed using \texttt{MIGRAD} with randomized starting values, are shown in \cref{fig:SP_100k}. The sign of trivial phase ambiguities is flipped.

Compared to the data sample with $10^6$ events shown in \cref{fig:SP}, the NLL differences between the five distinct solutions are significantly smaller. Notably, the global minimum does not correspond anymore to the true amplitude values. Instead, the second-best NLL solution, with an NLL value approximately 1 unit higher than that of the global minium, more accurately matches the true values and the global minimum found here corresponds to the second-best NLL solutions observed in \cref{fig:SP}, indicating that for smaller sample sizes the likelihood surface may be distorted. Therefore, the risk of obtaining indistinguishable solutions due to local minima increases for smaller data sets and the global minimum is not guaranteed to correspond to a solution close to the true amplitude values.

\clearpage

\section{Moments with $S,P$, and $D$ waves}
\label{appen:SPD_moments}

\begin{equation*}
\begin{aligned}
H_{0}(0,0) = 2 \sum_{\epsilon} \bigg( &|S_{0}^{\epsilon}|^2 + |P_{+1}^{\epsilon}|^2 + |P_{0}^{\epsilon}|^2 + |P_{-1}^{\epsilon}|^2
+|D_{+2}^{\epsilon}|^2 + |D_{+1}^{\epsilon}|^2 + |D_{0}^{\epsilon}|^2 + |D_{-1}^{\epsilon}|^2 + |D_{-2}^{\epsilon}|^2 \bigg)
\end{aligned}
\end{equation*}

\begin{equation*}
\begin{aligned}
H_{1}(0,0) = 2 \sum_{\epsilon} \epsilon\bigg( & |S_{0}^{\epsilon}|^2 + |P_{+1}^{-\epsilon}|^2 + |D_{0}^{\epsilon}|^2 -2 \operatorname{Re}[P_{+1}^{\epsilon} (P_{-1}^{\epsilon})^*] +2 \operatorname{Re}[D_{+2}^{\epsilon} (D_{-2}^{\epsilon})^*] -2 \operatorname{Re}[D_{+1}^{\epsilon} (D_{-1}^{\epsilon})^*] \bigg)
\end{aligned}
\end{equation*}

\begin{equation*}
\begin{aligned}
H_{0}(1,0) =4 \sum_{\epsilon}\bigg( & \frac{\sqrt{3}}{3}\operatorname{Re}[S_{0}^{\epsilon} (P_{0}^{\epsilon})^*] +\frac{\sqrt{5}}{5}\operatorname{Re}[P_{+1}^{\epsilon} (D_{+1}^{\epsilon})^*]+\frac{2 \sqrt{15}}{15}\operatorname{Re}[P_{0}^{\epsilon} (D_{0}^{\epsilon})^*] +\frac{\sqrt{5}}{5}\operatorname{Re}[P_{-1}^{\epsilon} (D_{-1}^{+})^*]\bigg)
\end{aligned}
\end{equation*}

\begin{equation*}
\begin{aligned}
H_{0}(1,1) = 2 \sum_{\epsilon}\bigg( &\frac{\sqrt{3}}{3}\operatorname{Re}[S_{0}^{\epsilon} (P_{+1}^{\epsilon})^*]
- \frac{\sqrt{3}}{3}\operatorname{Re}[S_{0}^{\epsilon} (P_{-1}^{\epsilon})^*]
+\frac{\sqrt{10}}{5}\operatorname{Re}[P_{+1}^{\epsilon} (D_{+2}^{\epsilon})^*]
- \frac{\sqrt{15}}{15}\operatorname{Re}[P_{+1}^{\epsilon} (D_{0}^{\epsilon})^*]\\ 
+&\frac{\sqrt{5}}{5}\operatorname{Re}[P_{0}^{\epsilon} (D_{+1}^{\epsilon})^*]
- \frac{\sqrt{5}}{5}\operatorname{Re}[P_{0}^{\epsilon} (D_{-1}^{\epsilon})^*]+\frac{\sqrt{15}}{15}\operatorname{Re}[P_{-1}^{\epsilon} (D_{0}^{\epsilon})^*]
- \frac{\sqrt{10}}{5}\operatorname{Re}[P_{-1}^{\epsilon} (D_{-2}^{\epsilon})^*]\bigg)
\end{aligned}
\end{equation*}

\begin{equation*}
\begin{aligned}
H_{1}(1,0) =4\sum_{\epsilon} \epsilon \bigg( &\frac{\sqrt{3}}{3}\operatorname{Re}[S_{0}^{\epsilon} (P_{0}^{\epsilon})^*]
- \frac{\sqrt{5}}{5}\operatorname{Re}[P_{+1}^{\epsilon} (D_{-1}^{\epsilon})^*]+\frac{2 \sqrt{15}}{15}\operatorname{Re}[P_{0}^{\epsilon} (D_{0}^{\epsilon})^*]
- \frac{\sqrt{5}}{5}\operatorname{Re}[P_{-1}^{\epsilon} (D_{+1}^{\epsilon})^*]\bigg)
\end{aligned}
\end{equation*}

\begin{equation*}
\begin{aligned}
H_{1}(1,1) = 2 \sum_{\epsilon} \epsilon\bigg(& \frac{\sqrt{3}}{3}\operatorname{Re}[S_{0}^{\epsilon} (P_{+1}^{\epsilon})^*]
- \frac{\sqrt{3}}{3}\operatorname{Re}[S_{0}^{\epsilon} (P_{-1}^{\epsilon})^*]- \frac{\sqrt{15}}{15}\operatorname{Re}[P_{+1}^{\epsilon} (D_{0}^{\epsilon})^*] +\frac{\sqrt{10}}{5}\operatorname{Re}[P_{+1}^{\epsilon} (D_{-2}^{\epsilon})^*]\\ 
+&\frac{\sqrt{5}}{5}\operatorname{Re}[P_{0}^{\epsilon} (D_{+1}^{\epsilon})^*]- \frac{\sqrt{5}}{5}\operatorname{Re}[P_{0}^{\epsilon} (D_{-1}^{\epsilon})^*]- \frac{\sqrt{10}}{5}\operatorname{Re}[P_{-1}^{\epsilon} (D_{+2}^{\epsilon})^*] +\frac{\sqrt{15}}{15}\operatorname{Re}[P_{-1}^{\epsilon} (D_{0}^{\epsilon})^*]\bigg)
\end{aligned}
\end{equation*}

\begin{equation*}
\begin{aligned}
\frac{H_{2}(1,1)}{i} = 2 \sum_{\epsilon} \epsilon\bigg(&-\frac{\sqrt{3}}{3}\operatorname{Re}[S_{0}^{\epsilon} (P_{+1}^{\epsilon})^*]
- \frac{\sqrt{3}}{3}\operatorname{Re}[S_{0}^{\epsilon} (P_{-1}^{\epsilon})^*]+\frac{\sqrt{15}}{15}\operatorname{Re}[P_{+1}^{\epsilon} (D_{0}^{\epsilon})^*]+\frac{\sqrt{10}}{5}\operatorname{Re}[P_{+1}^{\epsilon} (D_{-2}^{\epsilon})^*]\\ 
&- \frac{\sqrt{5}}{5}\operatorname{Re}[P_{0}^{\epsilon} (D_{+1}^{\epsilon})^*]
- \frac{\sqrt{5}}{5}\operatorname{Re}[P_{0}^{\epsilon} (D_{-1}^{\epsilon})^*]+\frac{\sqrt{10}}{5}\operatorname{Re}[P_{-1}^{\epsilon} (D_{+2}^{\epsilon})^*]+\frac{\sqrt{15}}{15}\operatorname{Re}[P_{-1}^{\epsilon} (D_{0}^{\epsilon})^*]\bigg)
\end{aligned}
\end{equation*}

\begin{equation*}
\begin{aligned}
H_{0}(2,0) = 2\sum_{\epsilon} \bigg(&\frac{2\sqrt{5}}{5}\operatorname{Re}[S_{0}^{\epsilon} (D_{0}^{\epsilon})^*]
- \frac{1}{5}|P_{+1}^{\epsilon}|^2
+\frac{2}{5}|P_{0}^{\epsilon}|^2- \frac{1}{5}|P_{-1}^{\epsilon}|^2\\
- &\frac{2}{7}|D_{+2}^{\epsilon}|^2
+\frac{1}{7}|D_{+1}^{\epsilon}|^2
+\frac{2}{7}|D_{0}^{\epsilon}|^2
+\frac{1}{7}|D_{-1}^{\epsilon}|^2
- \frac{2}{7}|D_{-2}^{\epsilon}|^2\bigg)
\end{aligned}
\end{equation*}

\begin{equation*}
\begin{aligned}
H_{0}(2,1) = 2\sum_{\epsilon} \bigg(&\frac{\sqrt{5}}{5}\operatorname{Re}[S_{0}^{\epsilon} (D_{+1}^{\epsilon})^*]
- \frac{\sqrt{5}}{5}\operatorname{Re}[S_{0}^{\epsilon} (D_{-1}^{\epsilon})^*]+\frac{\sqrt{3}}{5}\operatorname{Re}[P_{+1}^{\epsilon} (P_{0}^{\epsilon})^*]
- \frac{\sqrt{3}}{5}\operatorname{Re}[P_{0}^{\epsilon} (P_{-1}^{\epsilon})^*]\\ 
&+\frac{\sqrt{6}}{7}\operatorname{Re}[D_{+2}^{\epsilon} (D_{+1}^{\epsilon})^*]
+\frac{1}{7}\operatorname{Re}[D_{+1}^{\epsilon} (D_{0}^{\epsilon})^*]
- \frac{1}{7}\operatorname{Re}[D_{0}^{\epsilon} (D_{-1}^{\epsilon})^*]
- \frac{\sqrt{6}}{7}\operatorname{Re}[D_{-1}^{\epsilon} (D_{-2}^{\epsilon})^*]\bigg)
\end{aligned}
\end{equation*}

\begin{equation*}
\begin{aligned}
H_{0}(2,2) = 2\sum_{\epsilon} \bigg( &\frac{\sqrt{5}}{5}\operatorname{Re}[S_{0}^{\epsilon} (D_{+2}^{\epsilon})^*]+\frac{\sqrt{5}}{5}\operatorname{Re}[S_{0}^{\epsilon} (D_{-2}^{\epsilon})^*]
- \frac{ \sqrt{6}}{5}\operatorname{Re}[P_{+1}^{\epsilon} (P_{-1}^{\epsilon})^*]
- \frac{2}{7}\operatorname{Re}[D_{+2}^{\epsilon} (D_{0}^{\epsilon})^*] \\
&- \frac{\sqrt{6}}{7}\operatorname{Re}[D_{+1}^{\epsilon} (D_{-1}^{\epsilon})^*] - \frac{2}{7}\operatorname{Re}[D_{0}^{\epsilon} (D_{-2}^{\epsilon})^*] \bigg)
\end{aligned}
\end{equation*}

\begin{equation*}
\begin{aligned}
H_{1}(2,0) = 4\sum_{\epsilon} \epsilon \bigg( &\frac{\sqrt{5}}{5}\operatorname{Re}[S_{0}^{\epsilon} (D_{0}^{\epsilon})^*] +\frac{1}{5}\operatorname{Re}[P_{+1}^{\epsilon} (P_{-1}^{\epsilon})^*] 
+\frac{1}{5}|P_{0}^{\epsilon}|^2 \\
- &\frac{2}{7}\operatorname{Re}[D_{+2}^{\epsilon} (D_{-2}^{\epsilon})^*] - \frac{1}{7}\operatorname{Re}[D_{+1}^{\epsilon} (D_{-1}^{\epsilon})^*] +\frac{1}{7}|D_{0}^{\epsilon}|^2  \bigg)
\end{aligned}
\end{equation*}

\begin{equation*}
\begin{aligned}
H_{1}(2,1) = 2\sum_{\epsilon} \epsilon \bigg( &\frac{\sqrt{5}}{5}\operatorname{Re}[S_{0}^{\epsilon} (D_{+1}^{\epsilon})^*] - \frac{\sqrt{5}}{5}\operatorname{Re}[S_{0}^{\epsilon} (D_{-1}^{\epsilon})^*] +\frac{\sqrt{3}}{5}\operatorname{Re}[P_{+1}^{\epsilon} (P_{0}^{\epsilon})^*] - \frac{\sqrt{3}}{5}\operatorname{Re}[P_{0}^{\epsilon} (P_{-1}^{\epsilon})^*] \\
-& \frac{\sqrt{6}}{7}\operatorname{Re}[D_{+2}^{\epsilon} (D_{-1}^{\epsilon})^*] +\frac{1}{7}\operatorname{Re}[D_{+1}^{\epsilon} (D_{0}^{\epsilon})^*] +\frac{\sqrt{6}}{7}\operatorname{Re}[D_{+1}^{\epsilon} (D_{-2}^{\epsilon})^*] - \frac{1}{7}\operatorname{Re}[D_{0}^{\epsilon} (D_{-1}^{\epsilon})^*] \bigg)
\end{aligned}
\end{equation*}

\begin{equation*}
\begin{aligned}
H_{1}(2,2) = 2\sum_{\epsilon} \epsilon \bigg( &\frac{\sqrt{5}}{5}\operatorname{Re}[S_{0}^{\epsilon} (D_{+2}^{\epsilon})^*] +\frac{\sqrt{5}}{5}\operatorname{Re}[S_{0}^{\epsilon} (D_{-2}^{\epsilon})^*]+\frac{\sqrt{6}}{10}|P_{+1}^{\epsilon}|^2 +\frac{\sqrt{6}}{10}|P_{-1}^{\epsilon}|^2 \\
-& \frac{2}{7}\operatorname{Re}[D_{+2}^{\epsilon} (D_{0}^{\epsilon})^*] +\frac{\sqrt{6}}{14}|D_{+1}^{\epsilon}|^2 - \frac{2}{7}\operatorname{Re}[D_{0}^{\epsilon} (D_{-2}^{\epsilon})^*] +\frac{\sqrt{6}}{14}|D_{-1}^{\epsilon}|^2 \bigg)
\end{aligned}
\end{equation*}

\begin{equation*}
\begin{aligned}
\frac{H_{2}(2,1)}{i} = 2\sum_{\epsilon} \epsilon \bigg( &- \frac{\sqrt{5}}{5}\operatorname{Re}[S_{0}^{\epsilon} (D_{+1}^{\epsilon})^*] - \frac{\sqrt{5}}{5}\operatorname{Re}[S_{0}^{\epsilon} (D_{-1}^{\epsilon})^*]
- \frac{\sqrt{3}}{5}\operatorname{Re}[P_{+1}^{\epsilon} (P_{0}^{\epsilon})^*] - \frac{\sqrt{3}}{5}\operatorname{Re}[P_{0}^{\epsilon} (P_{-1}^{\epsilon})^*] \\
&+ \frac{\sqrt{6}}{7}\operatorname{Re}[D_{+2}^{\epsilon} (D_{-1}^{\epsilon})^*] - \frac{1}{7}\operatorname{Re}[D_{+1}^{\epsilon} (D_{0}^{\epsilon})^*] + \frac{\sqrt{6}}{7}\operatorname{Re}[D_{+1}^{\epsilon} (D_{-2}^{\epsilon})^*] - \frac{1}{7}\operatorname{Re}[D_{0}^{\epsilon} (D_{-1}^{\epsilon})^*] \bigg)
\end{aligned}
\end{equation*}

\begin{equation*}
\begin{aligned}
\frac{H_{2}(2,2)}{i} = 2 \sum_{\epsilon} \epsilon \bigg( &- \frac{\sqrt{5}}{5}\operatorname{Re}[S_{0}^{\epsilon} (D_{+2}^{\epsilon})^*] + \frac{\sqrt{5}}{5}\operatorname{Re}[S_{0}^{\epsilon} (D_{-2}^{\epsilon})^*] - \frac{\sqrt{6}}{10}|P_{+1}^{\epsilon}|^2 + \frac{\sqrt{6}}{10}|P_{-1}^{\epsilon}|^2 \\
&+ \frac{2}{7}\operatorname{Re}[D_{+2}^{\epsilon} (D_{0}^{\epsilon})^*] - \frac{\sqrt{6}}{14}|D_{+1}^{\epsilon}|^2 - \frac{2}{7}\operatorname{Re}[D_{0}^{\epsilon} (D_{-2}^{\epsilon})^*] + \frac{\sqrt{6}}{14}|D_{-1}^{\epsilon}|^2 \bigg)
\end{aligned}
\end{equation*}

\begin{equation*}
\begin{aligned}
H_{0}(3,0) = \frac{12\sqrt{5}}{35} \sum_{\epsilon}  \bigg( &- \operatorname{Re}[P_{+1}^{\epsilon} (D_{+1}^{\epsilon})^*] -\operatorname{Re}[P_{-1}^{\epsilon} (D_{-1}^{\epsilon})^*]
+ \sqrt{3}\operatorname{Re}[P_{0}^{\epsilon} (D_{0}^{\epsilon})^*] \bigg)
\end{aligned}
\end{equation*}

\begin{equation*}
\begin{aligned}
H_{0}(3,1) =  \frac{2 \sqrt{5}}{35} \sum_{\epsilon}  \bigg( &- \sqrt{3}\operatorname{Re}[P_{+1}^{\epsilon} (D_{+2}^{\epsilon})^*] + 3 \sqrt{2}\operatorname{Re}[P_{+1}^{\epsilon} (D_{0}^{\epsilon})^*] + 2 \sqrt{6}\operatorname{Re}[P_{0}^{\epsilon} (D_{+1}^{\epsilon})^*]\\
&- 2 \sqrt{6}\operatorname{Re}[P_{0}^{\epsilon} (D_{-1}^{\epsilon})^*] 
- 3 \sqrt{2}\operatorname{Re}[P_{-1}^{\epsilon} (D_{0}^{\epsilon})^*] + \sqrt{3}\operatorname{Re}[P_{-1}^{\epsilon} (D_{-2}^{\epsilon})^*] \bigg)
\end{aligned}
\end{equation*}

\begin{equation*}
\begin{aligned}
H_{0}(3,2) =  \frac{2\sqrt{3}}{7} \sum_{\epsilon} \bigg( &- \sqrt{2}\operatorname{Re}[P_{+1}^{\epsilon} (D_{-1}^{\epsilon})^*] + \operatorname{Re}[P_{0}^{\epsilon} (D_{+2}^{\epsilon})^*] + \operatorname{Re}[P_{0}^{\epsilon} (D_{-2}^{\epsilon})^*] - \sqrt{2}\operatorname{Re}[P_{-1}^{\epsilon} (D_{+1}^{\epsilon})^*] \bigg)
\end{aligned}
\end{equation*}

\begin{equation*}
\begin{aligned}
H_{0}(3,3) = \frac{6}{7} \sum_{\epsilon} \bigg( &\operatorname{Re}[P_{+1}^{\epsilon} (D_{-2}^{\epsilon})^*] - \operatorname{Re}[P_{-1}^{\epsilon} (D_{+2}^{\epsilon})^*] \bigg)
\end{aligned}
\end{equation*}

\begin{equation*}
\begin{aligned}
H_{1}(3,0) = \frac{12 \sqrt{5}}{35} \sum_{\epsilon}  \epsilon \bigg( &\operatorname{Re}[P_{+1}^{\epsilon} (D_{-1}^{\epsilon})^*] + \sqrt{3}\operatorname{Re}[P_{0}^{\epsilon} (D_{0}^{\epsilon})^*] + \operatorname{Re}[P_{-1}^{\epsilon} (D_{+1}^{\epsilon})^*] \bigg)
\end{aligned}
\end{equation*}

\begin{equation*}
\begin{aligned}
H_{1}(3,1) = \frac{2 \sqrt{5}}{35} \sum_{\epsilon} \epsilon \bigg( &3 \sqrt{2}\operatorname{Re}[P_{+1}^{\epsilon} (D_{0}^{\epsilon})^*] - \sqrt{3}\operatorname{Re}[P_{+1}^{\epsilon} (D_{-2}^{\epsilon})^*] + 2 \sqrt{6}\operatorname{Re}[P_{0}^{\epsilon} (D_{+1}^{\epsilon})^*] \\
- &2 \sqrt{6}\operatorname{Re}[P_{0}^{\epsilon} (D_{-1}^{\epsilon})^*]+ \sqrt{3}\operatorname{Re}[P_{-1}^{\epsilon} (D_{+2}^{\epsilon})^*] - 3 \sqrt{2}\operatorname{Re}[P_{-1}^{\epsilon} (D_{0}^{\epsilon})^*] \bigg)
\end{aligned}
\end{equation*}

\begin{equation*}
\begin{aligned}
H_{1}(3,2) = \frac{2}{7}\sum_{\epsilon} \epsilon \bigg( &\sqrt{6}\operatorname{Re}[P_{+1}^{\epsilon} (D_{+1}^{\epsilon})^*] + \sqrt{3}\operatorname{Re}[P_{0}^{\epsilon} (D_{+2}^{\epsilon})^*] + \sqrt{3}\operatorname{Re}[P_{0}^{\epsilon} (D_{-2}^{\epsilon})^*] + \sqrt{6}\operatorname{Re}[P_{-1}^{\epsilon} (D_{-1}^{\epsilon})^*] \bigg)
\end{aligned}
\end{equation*}

\begin{equation*}
\begin{aligned}
H_{1}(3,3) = \frac{6}{7}\sum_{\epsilon} \epsilon \bigg( &\operatorname{Re}[P_{+1}^{\epsilon} (D_{+2}^{\epsilon})^*] - \operatorname{Re}[P_{-1}^{\epsilon} (D_{-2}^{\epsilon})^*] \bigg)
\end{aligned}
\end{equation*}

\begin{equation*}
\begin{aligned}
\frac{H_{2}(3,1)}{i} = \frac{2 \sqrt{5}}{35} \sum_{\epsilon} \epsilon \bigg( &- 3 \sqrt{2}\operatorname{Re}[P_{+1}^{\epsilon} (D_{0}^{\epsilon})^*] - \sqrt{3}\operatorname{Re}[P_{+1}^{\epsilon} (D_{-2}^{\epsilon})^*] - 2 \sqrt{6}\operatorname{Re}[P_{0}^{\epsilon} (D_{+1}^{\epsilon})^*] \\
&- 2 \sqrt{6}\operatorname{Re}[P_{0}^{\epsilon} (D_{-1}^{\epsilon})^*] - \sqrt{3}\operatorname{Re}[P_{-1}^{\epsilon} (D_{+2}^{\epsilon})^*] - 3 \sqrt{2}\operatorname{Re}[P_{-1}^{\epsilon} (D_{0}^{\epsilon})^*] \bigg)
\end{aligned}
\end{equation*}

\begin{equation*}
\begin{aligned}
\frac{H_{2}(3,2)}{i} = \frac{2\sqrt{3}}{7} \sum_{\epsilon} \epsilon \bigg( &- \sqrt{2}\operatorname{Re}[P_{+1}^{\epsilon} (D_{+1}^{\epsilon})^*] - \operatorname{Re}[P_{0}^{\epsilon} (D_{+2}^{\epsilon})^*] + \operatorname{Re}[P_{0}^{\epsilon} (D_{-2}^{\epsilon})^*] + \sqrt{2}\operatorname{Re}[P_{-1}^{\epsilon} (D_{-1}^{\epsilon})^*] \bigg)
\end{aligned}
\end{equation*}

\begin{equation*}
\begin{aligned}
H_{2}(3,3)/i = \frac{6}{7} \sum_{\epsilon} \epsilon \bigg( &- \operatorname{Re}[P_{+1}^{\epsilon} (D_{+2}^{\epsilon})^*] - \operatorname{Re}[P_{-1}^{\epsilon} (D_{-2}^{\epsilon})^*] \bigg)
\end{aligned}
\end{equation*}

\begin{equation*}
\begin{aligned}
H_{0}(4,0) = \frac{2}{21} \sum_{\epsilon} \bigg( &|D_{+2}^{\epsilon}|^2 - 4|D_{+1}^{\epsilon}|^2 + 6|D_{0}^{\epsilon}|^2 - 4|D_{-1}^{\epsilon}|^2 + |D_{-2}^{\epsilon}|^2 \bigg)
\end{aligned}
\end{equation*}

\begin{equation*}
\begin{aligned}
H_{0}(4,1) = \frac{2 \sqrt{5}}{21} \sum_{\epsilon} \bigg( &- \operatorname{Re}[D_{+2}^{\epsilon} (D_{+1}^{\epsilon})^*] + \sqrt{6}\operatorname{Re}[D_{+1}^{\epsilon} (D_{0}^{\epsilon})^*] - \sqrt{6}\operatorname{Re}[D_{0}^{\epsilon} (D_{-1}^{\epsilon})^*] + \operatorname{Re}[D_{-1}^{\epsilon} (D_{-2}^{\epsilon})^*] \bigg)
\end{aligned}
\end{equation*}

\begin{equation*}
\begin{aligned}
H_{0}(4,2) = \frac{2 \sqrt{5}}{21} \sum_{\epsilon} \bigg( &\sqrt{3}\operatorname{Re}[D_{+2}^{\epsilon} (D_{0}^{\epsilon})^*] - 2 \sqrt{2}\operatorname{Re}[D_{+1}^{\epsilon} (D_{-1}^{\epsilon})^*] + \sqrt{3}\operatorname{Re}[D_{0}^{\epsilon} (D_{-2}^{\epsilon})^*] \bigg)
\end{aligned}
\end{equation*}

\begin{equation*}
\begin{aligned}
H_{0}(4,3) = \frac{2 \sqrt{35}}{21} \sum_{\epsilon} \bigg( &- \operatorname{Re}[D_{+2}^{\epsilon} (D_{-1}^{\epsilon})^*] + \operatorname{Re}[D_{+1}^{\epsilon} (D_{-2}^{\epsilon})^*] \bigg)
\end{aligned}
\end{equation*}

\begin{equation*}
\begin{aligned}
H_{0}(4,4) = \frac{2 \sqrt{70}}{21} \sum_{\epsilon} \bigg( &\operatorname{Re}[D_{+2}^{\epsilon} (D_{-2}^{\epsilon})^*] \bigg)
\end{aligned}
\end{equation*}

\begin{equation*}
\begin{aligned}
H_{1}(4,0) = \frac{4}{21} \sum_{\epsilon} \epsilon \bigg( &+\operatorname{Re}[D_{+2}^{\epsilon} (D_{-2}^{\epsilon})^*] + 4\operatorname{Re}[D_{+1}^{\epsilon} (D_{-1}^{\epsilon})^*] + 3|D_{0}^{\epsilon}|^2 \bigg)
\end{aligned}
\end{equation*}

\begin{equation*}
\begin{aligned}
H_{1}(4,1) = \frac{2 \sqrt{5}}{21} \sum_{\epsilon} \epsilon \bigg( &\operatorname{Re}[D_{+2}^{\epsilon} (D_{-1}^{\epsilon})^*] - \sqrt{6}\operatorname{Re}[D_{+1}^{\epsilon} (D_{0}^{\epsilon})^*] + \operatorname{Re}[D_{+1}^{\epsilon} (D_{-2}^{\epsilon})^*] + \sqrt{6}\operatorname{Re}[D_{0}^{\epsilon} (D_{-1}^{\epsilon})^*] \bigg)
\end{aligned}
\end{equation*}

\begin{equation*}
\begin{aligned}
H_{1}(4,2) = \frac{2 \sqrt{5}}{21} \sum_{\epsilon} \epsilon \bigg( &\sqrt{3}\operatorname{Re}[D_{+2}^{\epsilon} (D_{0}^{\epsilon})^*] + \sqrt{2}|D_{+1}^{\epsilon}|^2+ \sqrt{3}\operatorname{Re}[D_{0}^{\epsilon} (D_{-2}^{\epsilon})^*] + \sqrt{2}|D_{-1}^{\epsilon}|^2 \bigg)
\end{aligned}
\end{equation*}

\begin{equation*}
\begin{aligned}
H_{1}(4,3) = \frac{2 \sqrt{35}}{21} \sum_{\epsilon} \epsilon \bigg( &\operatorname{Re}[D_{+2}^{\epsilon} (D_{+1}^{\epsilon})^*] - \operatorname{Re}[D_{-1}^{\epsilon} (D_{-2}^{\epsilon})^*] \bigg)
\end{aligned}
\end{equation*}

\begin{equation*}
\begin{aligned}
H_{1}(4,4) = \frac{\sqrt{70}}{21} \sum_{\epsilon} \epsilon \bigg( &|D_{+2}^{\epsilon}|^2 + |D_{-2}^{\epsilon}|^2 \bigg)
\end{aligned}
\end{equation*}

\begin{equation*}
\begin{aligned}
\frac{H_{2}(4,1)}{i} = \frac{2 \sqrt{5}}{21} \sum_{\epsilon} \epsilon \bigg( &- \operatorname{Re}[D_{+2}^{\epsilon} (D_{-1}^{\epsilon})^*] - \sqrt{6}\operatorname{Re}[D_{+1}^{\epsilon} (D_{0}^{\epsilon})^*] - \operatorname{Re}[D_{+1}^{\epsilon} (D_{-2}^{\epsilon})^*] - \sqrt{6}\operatorname{Re}[D_{0}^{\epsilon} (D_{-1}^{\epsilon})^*] \bigg)
\end{aligned}
\end{equation*}

\begin{equation*}
\begin{aligned}
\frac{H_{2}(4,2)}{i} = \frac{2 \sqrt{5}}{21} \sum_{\epsilon} \epsilon \bigg( &- \sqrt{3}\operatorname{Re}[D_{+2}^{\epsilon} (D_{0}^{\epsilon})^*] - \sqrt{2}|D_{+1}^{\epsilon}|^2 + \sqrt{3}\operatorname{Re}[D_{0}^{\epsilon} (D_{-2}^{\epsilon})^*] - \sqrt{2}|D_{-1}^{\epsilon}|^2 \bigg)
\end{aligned}
\end{equation*}

\begin{equation*}
\begin{aligned}
\frac{H_{2}(4,3)}{i} = \frac{2 \sqrt{35}}{21} \sum_{\epsilon} \epsilon \bigg( &- \operatorname{Re}[D_{+2}^{\epsilon} (D_{+1}^{\epsilon})^*] - \operatorname{Re}[D_{-1}^{\epsilon} (D_{-2}^{\epsilon})^*] \bigg)
\end{aligned}
\end{equation*}

\begin{equation*}
\begin{aligned}
\frac{H_{2}(4,4)}{i} = \frac{\sqrt{70}}{21} \sum_{\epsilon} \epsilon \bigg( &- |D_{+2}^{\epsilon}|^2 + |D_{-2}^{\epsilon}|^2 \bigg)
\end{aligned}
\end{equation*}

%
\bibliographystyle{unsrtnat}
\bibliography{Ambiguities}
\end{document}